**Perspective**

# Computing at the Edge Enabled by Indoor Photovoltaics and Ferroelectrics

Robert L. Z. Hoye,[1,*] Markus Hellenbrand,[2,*] Nuno Estrócio,[3,4] Austin M. Kay,[5] Alice Scardina,[6] Alessandro Mezzetti,[6] Ignasi Fina,[7] Jorge Íñiguez-González,[8,9] Luís S. Marques,[3,4] Francesco Matteucci,[10,11] Quanxi Jia,[12] Judith L. MacManus-Driscoll,[2] Bert Offrein,[13] Florencio Sánchez,[7] Beatriz Noheda,[14,15] George Koutsourakis,[16] Marina Freitag,[17] Gregory Burwell,[5] Giulia Grancini,[6,*] Jose P. B. Silva[3,4,*]

[1] Inorganic Chemistry Laboratory, University of Oxford, South Parks Road, Oxford OX1 3QR, United Kingdom

[2] Department of Materials Science and Metallurgy, 27 Charles Babbage Road, Cambridge CB3 0FS, United Kingdom

[3] Physics Center of Minho and Porto Universities (CF-UM-UP), University of Minho, Campus de Gualtar, 4710-057 Braga, Portugal

[4] Laboratory of Physics for Materials and Emergent Technologies, LapMET, University of Minho, 4710-057 Braga, Portugal

[5] Sustainable Advanced Materials (Sêr-SAM), Centre for Integrative Semiconductor Materials (CISM), Department of Physics, Swansea University Bay Campus, Swansea SA1 8EN, UK

[6] Department of Chemistry & INSTM, University of Pavia, Via T. Taramelli 14, 27100, Pavia, Italy

[7] Institut de Ciència de Materials de Barcelona (ICMAB-CSIC), Carrer dels Til.lers, 08193 Cerdanyola del Vallès, Spain

[8] Smart Materials Unit, Luxembourg Institute of Science and Technology (LIST), Avenue des Hauts-Fourneaux 5, L-4362, Esch/Alzette, Luxembourg

[9] Department of Physics and Materials Science, University of Luxembourg, Rue du Brill 41, L-4422 Belvaux, Luxembourg

[10] Nanoshuttle, via. G. Desa 12, Lecce, Italy

[11] Fondazione Bruno Kessler, via Sommariva 18, Povo, Italy

[12] Department of Materials Design and Innovation, University at Buffalo – The State University of New York, Buffalo, NY 14260, USA

[13] IBM Research GmbH – Zürich Research laboratory, CH-8803, Ruschlikon, Switzerland

[14.] CongiGron – Groningen Cognitive Systems and Materials Center, University of Groningen, Nijenborgh 4, 9747AG, Groningen, The Netherlands

[15.] Zernike Institute for Advanced Materials, University of Groningen, Nijenborgh 4, 9747AG Groningen, The Netherlands

[16.] National Physical Laboratory, Hampton Road, Teddington TW11 0LW, UK

[17.] School of Natural and Environmental Science, Newcastle University, Newcastle upon Tyne NE1 7RU, UK

* e-mail: robert.hoye@chem.ox.ac.uk (R. L. Z. H.), mkhh2@cam.ac.uk (M. K. H. H.), giulia.grancini@unipv.it (G. G.), josesilva@fisica.uminho.pt (J. P. B. S.)

## Abstract

The exponential, pervasive rise of artificial intelligence (AI), hosted in large data centres, will test the limits of the underpinning infrastructure, including electricity and water supplies, which are becoming increasingly precious resources. Rather than performing all AI computations in centralized servers, for many functionalities, these can be performed locally across a network of billions of small, autonomous nodes, saving substantial energy costs associated with communication. This alternative paradigm is known as edge computing, or distributed intelligence (DI), but has been held back by the lack of availability of reliable local energy supplies matching the energy requirements of the computational infrastructure. Opportunities to address this challenge are emerging with rapid advances in high-performance indoor photovoltaics (IPVs) for local energy harvesting, as well as reductions in computational cost through neuromorphic or in-memory computing devices. In this perspective, we discuss the requirements of IPVs for DI, the extent to which emerging materials fulfil these requirements, and how current gaps could be addressed. We make the case that (anti)ferroelectric materials can surpass bottlenecks for both electrostatic energy storage and low-power in-memory neuromorphic computing. The core theme of this perspective is that co-creation between energy harvesting, storage and computing is essential for making DI a reliable and more sustainable alternative to centralized AI.

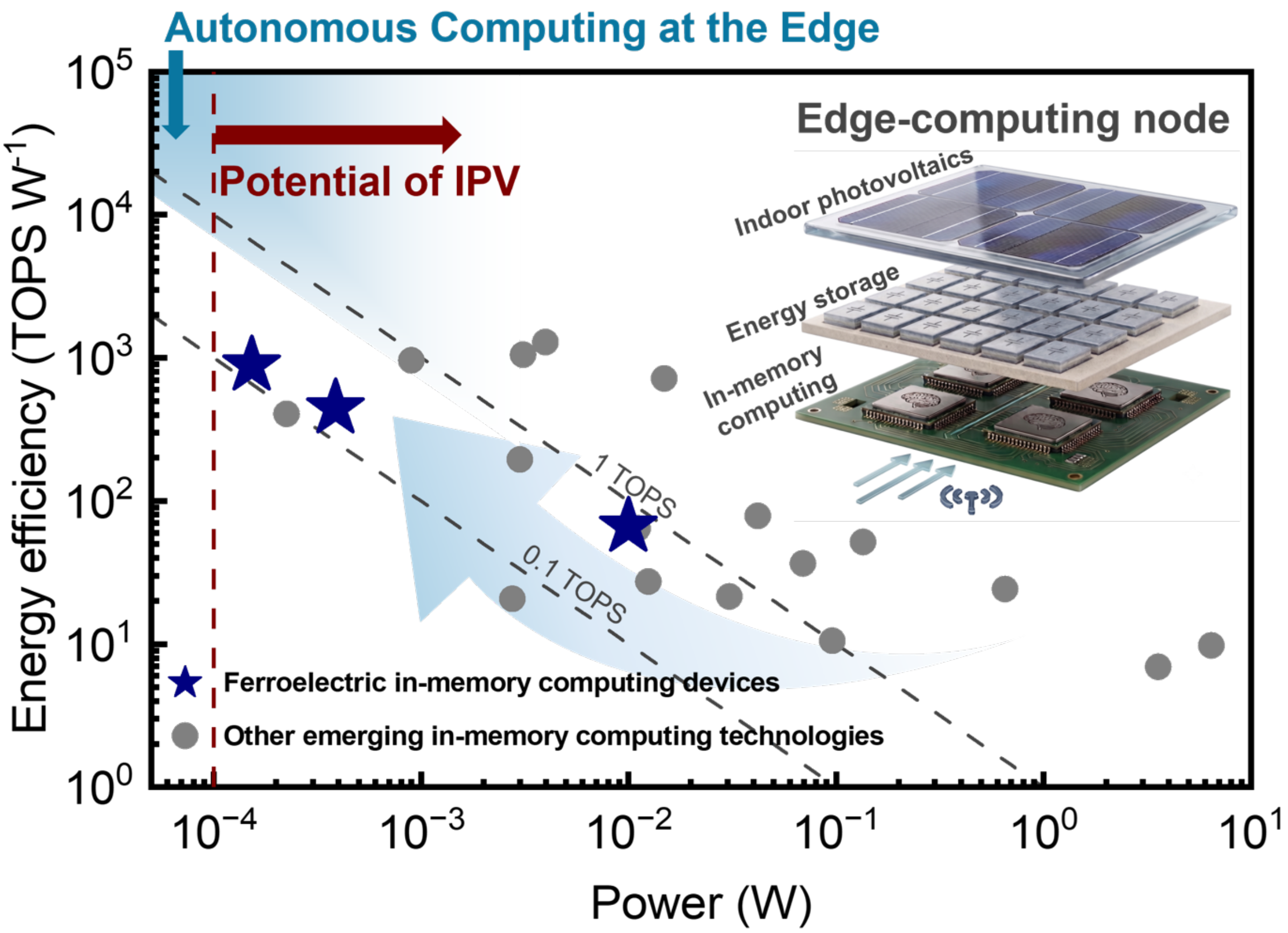

## [H1] Introduction

The coming decade will test the limits of the digital infrastructure underpinning artificial intelligence (AI)[1,2]. Data centre demand, already constituting 1.5% of global electricity consumption in 2024, is projected to grow to 3–4.4% by 2035, surpassing the energy consumption of heavy industry[3]. Growth is especially concentrated in the US and China, which are projected to experience >100% growth in data centre energy demand from 2024 to 2035 [3], driven by the exponential growth of AI workloads. The prevailing centralized model, which funnels data from local devices to resource-intensive hubs, is unsustainable[1,4]. Half of the near-term surge in electricity demand to 2030 is set to be met predominantly with fossil fuels (natural gas and coal), leading to increased $CO_2$ equivalent footprint[3]. The data centres core to the centralized model of AI also require water and cooling during operation, and it is projected that as early as 2027, global water use for data centres will exceed water usage in Denmark by a factor of 6 [4,5]. The proliferation of data centres will also strain the world's resources of critical minerals required for their production, as well as contribute to substantial increases in electronic waste[4].

Distributed intelligence (DI), or edge computing, offers a necessary alternative. In this paradigm, perception, low-latency inference, and local decision making are executed at the sensor node, dramatically reducing backhaul traffic and the energy-delay product associated with the von Neumann bottleneck, where data movement can consume orders of magnitude more energy than computation itself[6]. Field trials show that moving workloads to the edge can lower the energy per inference by 65–80%[7]. Local processing substantially decreases the need for cloud communication, which is often the dominant energy cost in Internet of Things (IoT) nodes due to wireless data transmission. By minimizing Wi-Fi or other wireless communication, the average power consumption of IoT devices can be reduced by orders of magnitude, from the 10-100 mW to 10-100 μW regime[8]. Moreover, DI enables communication over shorter distances using lower transmit power, further reducing the energy required for wireless connectivity. The transition from centralized cloud architectures to DI represents a fundamental shift in how information is processed, stored, and acted upon in the physical world. This transformation is not merely computational but fundamentally materials-driven: to engineer systems that lower the energy cost of on-node computational workflows, while maximising energy provision from local sources and their reliability.

Currently, batteries are commonly used for local on-node energy supply because of their high energy storage density (ESD)[9]. However, batteries have limited lifetime, and expanding the DI ecosystem towards the multi-billion level will create practicality and sustainability challenges: practicality challenges because of the need to frequently recharge or replace a large volume of batteries, which could constrain the growth in the number of DI nodes[10]; sustainability challenges because of the increase in demand on critical raw materials for battery manufacture, as well as the significant waste from battery disposal at end-of-life [11,12]. Harvesting energy from local, ambient energy sources to power each sensor node, as well as recharging the energy storage device, is therefore a highly desired solution, enabling reliable power supply over the entire life of the sensor node without requiring maintenance. Many edge computing devices are expected to operate in indoor environments, where smart buildings, home automation, healthcare, and industrial monitoring represent major application areas. Of the variety of ambient energy sources available in an indoor environment, ambient lighting (particularly from artificial light sources) is appealing because of its high availability, predictability, ubiquity and deployability[11–13]. Energy from ambient lighting can be harvested using indoor photovoltaics (IPVs)[11], but hydrogenated amorphous silicon (a-Si:H) IPVs – historically the most common IPV technology – are typically inefficient (7–16%)[11,13–15], making them not well suited to DI. More recently, an emerging materials portfolio has been developing, and IPV devices made from these novel materials can reach >30% indoor power conversion efficiency ($PCE_i$), and in some cases, >40% $PCE_i$ [13,16–18]. Simultaneously, the development of materials for in-memory neuromorphic computing, which co-locate memory and logic to circumvent the von Neumann bottleneck, are lowering the energy requirements for inference[1,2]. The gap between energy supply and demand from IPVs and computations, respectively, is approaching the level where it could be bridged, putting IPV-driven DI within reach.

At the same time, the requirements for edge computing are increasingly demanding and sometimes conflicting, with powerful real-time data processing, AI functionality, and device interconnectivity required simultaneously with low power consumption. As discussed later, advanced DI functionalities, such as real-time object detection, can require at least a trillion operations per second (*i.e.*, >1 TOPS[19]), which, given a power budget of 0.1–1 mW from IPV, will easily require compute efficiencies of $10^3$–$10^4$ TOPS $W^{-1}$. Current low-power AI accelerators based on complementary metal-oxide-semiconductor (CMOS) technology typically operate in the range of 1–10 TOPS $W^{-1}$ (refer to Box 1), whereas emerging memory technologies have shown promise to improve on this by orders of magnitude. Furthermore, all ambient energy sources, including indoor lighting, are intermittent, while certain computational workflows for inference require short bursts of power. This makes energy storage a crucial part of DI for buffering the power and voltage provided to the computing unit, as well as ensuring consistent operation when the variable ambient energy source is absent (*e.g.*, in the dark). Whilst electrochemical energy storage devices (*e.g.*, batteries) are commonly used for IoT nodes, they face challenges in edge computing, particularly from the sharp spikes in energy required. All of these challenges call for new materials to be developed.

(Anti)ferroelectric materials have recently emerged with potential to address these challenges in both low-power in-memory computing, as well as in energy storage, offering an unparalleled combination of high endurance, energy density and switching speeds. Such materials enable in-memory compute functionality, removing the von Neumann bottleneck that dominates energy consumption in conventional computing[20]. These technologies enable AI inference at sub-millijoule energy budgets, while maintaining the plasticity and adaptability required for edge learning applications. Compared to other diverse materials classes for in-memory computing that rely on mass transport (*e.g.*, of vacancies in transition-metal oxides and solid electrolytes[21–23]), ferroelectrics are particularly compelling because information can be encoded in their polarization state, which can be switched purely electrostatically and without mass transport[24]. Since switching is field-driven, and does not intrinsically require long-range ionic transport or Joule heating, ferroelectric devices offer fast, non-volatile and potentially very low energy operation[24,25]. Moreover, gradual domain nucleation and growth naturally produce multiple history-dependent resistance states in each memory cell, making ferroelectrics especially attractive for analogue memory, adaptive electronics and neuromorphic computing[26]. Furthermore, these (anti)ferroelectric materials can be used for energy storage in the form of electrostatic capacitors. These devices have charge/discharge rates four orders of magnitude higher than batteries, providing the instantaneous current needed for edge computing workloads[27,28]. Their main drawback compared to electrochemical energy storage is their orders of magnitude lower energy storage density (ESD)[29,30], but this is less of a limitation if there is regular recharging from ambient energy harvesting with IPV. Importantly, these capacitors are compatible with fabrication on CMOS, which is critical for on-chip integration. While conventional dielectric materials (*e.g.*, biaxially oriented polypropylene, polyvinylidene fluoride or polyethylene terephthalate polymers) are limited by low energy storage capacities[28,31], (anti)ferroelectric materials stand out because of their giant polarization[32,33] and unique field-induced phase transitions[34–36]. There is therefore significant promise in exploiting the multifunctionality of (anti)ferroelectric materials for both energy storage and low-power in-memory computing for DI, whereas many other materials systems would only serve one of these functions.

Until now, the IPV, ferroelectric, energy storage, and neuromorphic computing communities have been separate fields. But bringing DI from the cusp of realization to a serious commercial alternative to centralized AI requires these communities to be brought together, with the co-evolution of the materials and DI system architectures. This Perspective seeks to fill this gap. We discuss the requirements of IPV for DI, and the light harvesting materials that could fulfil these requirements, as well as the materials challenges that need to be addressed to improve the power that could be harvested by IPV from 100 μW to 1 mW and beyond. We next discuss the potential, challenges and solutions for engineering (anti)ferroelectric materials for energy storage and in-memory computing, including pathways to improving the performance efficiency beyond $10^4$ TOPS $W^{-1}$. This interlinked focus contrasts with prior reviews, which examined each area in isolation, and without consideration of the central thesis of how to make IPV-driven edge computing commercially viable[2,12,13,29]. By bringing to the fore the opportunities recent materials developments with IPV and (anti)ferroelectrics have enabled in energy harvesting, energy storage and neuromorphic computing, and discussing these developments together in the context of the entire DI node, we identify core targets for each community to make IPV-powered DI a strong contender to conventional AI in the near-future.

**Box 1 | Opportunities and Hardware Requirements of Distributed Intelligence**

Distributed intelligence (DI) enables the deployment of AI algorithms and computational power across multiple, interconnected nodes, such as Internet of Things (IoT) devices, edge servers, and the cloud, allowing data to be processed locally and autonomously rather than relying solely on a central server[37,38]. The main components of DI on each node, illustrated in the figure below, are comprised of 1) energy harvesters (*e.g.*, IPVs) and sensors, 2) energy storage devices, and 3) computing devices. By keeping data local, these systems significantly improve privacy and reduce bandwidth costs, as only the necessary information is transmitted to the cloud[39,40]. This allows for near-instantaneous decision-making, which is being used, for example, in autonomous building management systems[41–43]. Traditional cloud-based AI might require continuously streaming thousands of raw data points to perform analyses similar to those performed locally by DI on the floor-level controller, while keeping the energy required for data transmission near zero.

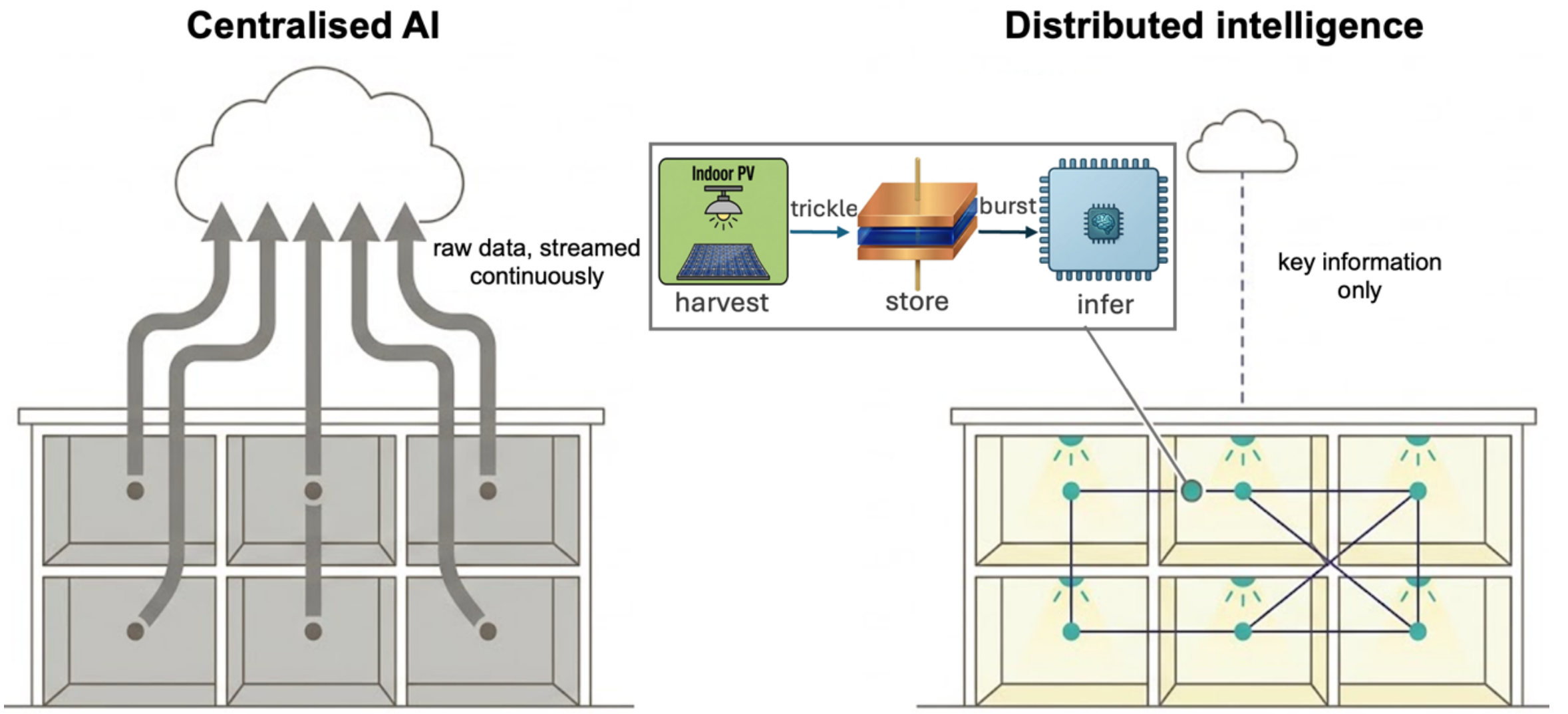

To illustrate the advantage of DI over conventional AI, we can consider the example of a conference room. When the room empties, the floor controller immediately shuts down active airflow. This eliminates the 30 min lag of wasted energy common in centralized systems. Also, localized controllers communicate peer-to-peer, which eliminates electrical current spikes, which can reduce overall system energy costs by 20%[42,43]. At the individual nodes used in DI, local sensors run polynomial regression for temperature extrapolation, Fast Fourier Transforms for mechanical vibration analysis, and low-power deadband logic gating to prevent unnecessary processor wake-ups. Simultaneously, the floor controllers execute Model Predictive Control quadratic programming matrices to balance airflow alongside multivariable regression to dynamically offset room heat from occupancy shifts. Finally, the central hub processes localized reinforcement learning policies to optimize daily chiller and boiler schedules.

Current DI utilizes GPUs (Graphics Processing Units, *e.g.*, NVIDIA Jetson series)—highly specialized processors featuring thousands of computing cores designed to handle millions of mathematical operations simultaneously—to run complex machine learning models and process multi-camera feeds right where data is collected[44]. However, deploying GPUs at the edge introduces severe physical and operational limitations. These chips suffer from high power consumption that drains local energy supplies, alongside extreme heat generation that requires bulky, hard-to-maintain active cooling systems in compact outdoor enclosures[45]. Furthermore, their high financial cost restricts large-scale deployment across thousands of nodes, while their physical size (with a footprint similar to the size of a credit card) limits integration into small sensors, and high "idle power" draw creates significant energy waste during data downtime. To truly excel, future computational hardware must achieve greater energy efficiency by achieving higher performance-per-watt, ensuring devices can run for extended periods with the energy supplies available locally. Heat dissipation remains a critical challenge, as intense local processing in small, fanless enclosures often leads to thermal throttling that affects performance. Additionally, in-memory computing device improvements are needed to eliminate data bottlenecks between the RAM and the GPU cores. By combining energy harvesting with In-Memory Computing (IMC), we can bridge the gap between "low power" and "high performance." In this model, represented in the figure above, the energy storage acts as the heart of the system, while the IMC core serves as the brain, enabling AI "thinking" without the massive power drain of traditional processors[44].

The hardware landscape for edge AI is captured in the figure below, where representative computing platforms are mapped according to their power consumption and computational throughput. The coloured regions denote three broad operating regimes for distributed intelligence, while diagonal lines indicate constant computational efficiency (0.1, 1 and 10 TOPS $W^{-1}$). At the lower end of the power and performance spectrum, the Extreme Edge AI region comprises platforms operating on the order of $10^{-3}$–1 W designed for always-on sensing and lightweight inference, with representative examples including the Syntiant NDP series, GAP processors and Eyeriss. These platforms target applications such as basic sensor signal processing, wake-word detection, inertial sensing and environmental monitoring. The Micro Edge AI region encompasses embedded AI accelerators capable of supporting lightweight computer vision, local audio processing, object detection and object counting, with representative platforms including Cambricon-X, DianNao, Tianjic and STM32N6x7. Finally, the Mid-tier to Heavy Edge AI region includes high-performance edge computing platforms, ranging from NVIDIA Jetson systems and Hailo accelerators to enterprise-scale GPUs, tensor processors and emerging manycore architectures, enabling real-time multi-camera computer vision, facial biometrics, high-resolution object tracking, industrial robotics and autonomous driving. Nevertheless, the current hardware landscape exposes a significant architectural gap when aligning stringent power constraints of edge computing with IPV. Although processors can provide the computational power necessary for complex AI workloads, their energy requirements far surpass the micro-to-milliwatt energy harvesting potential of IPV. As a result, present commercial architectures do not achieve the ultra-high hardware efficiency (*i.e.*, TOPS $W^{-1}$) essential for maintaining uninterrupted, autonomous Edge AI functions exclusively under ambient indoor lighting. Thus, developing high-performing materials across the main components of DI on each node is strictly necessary.

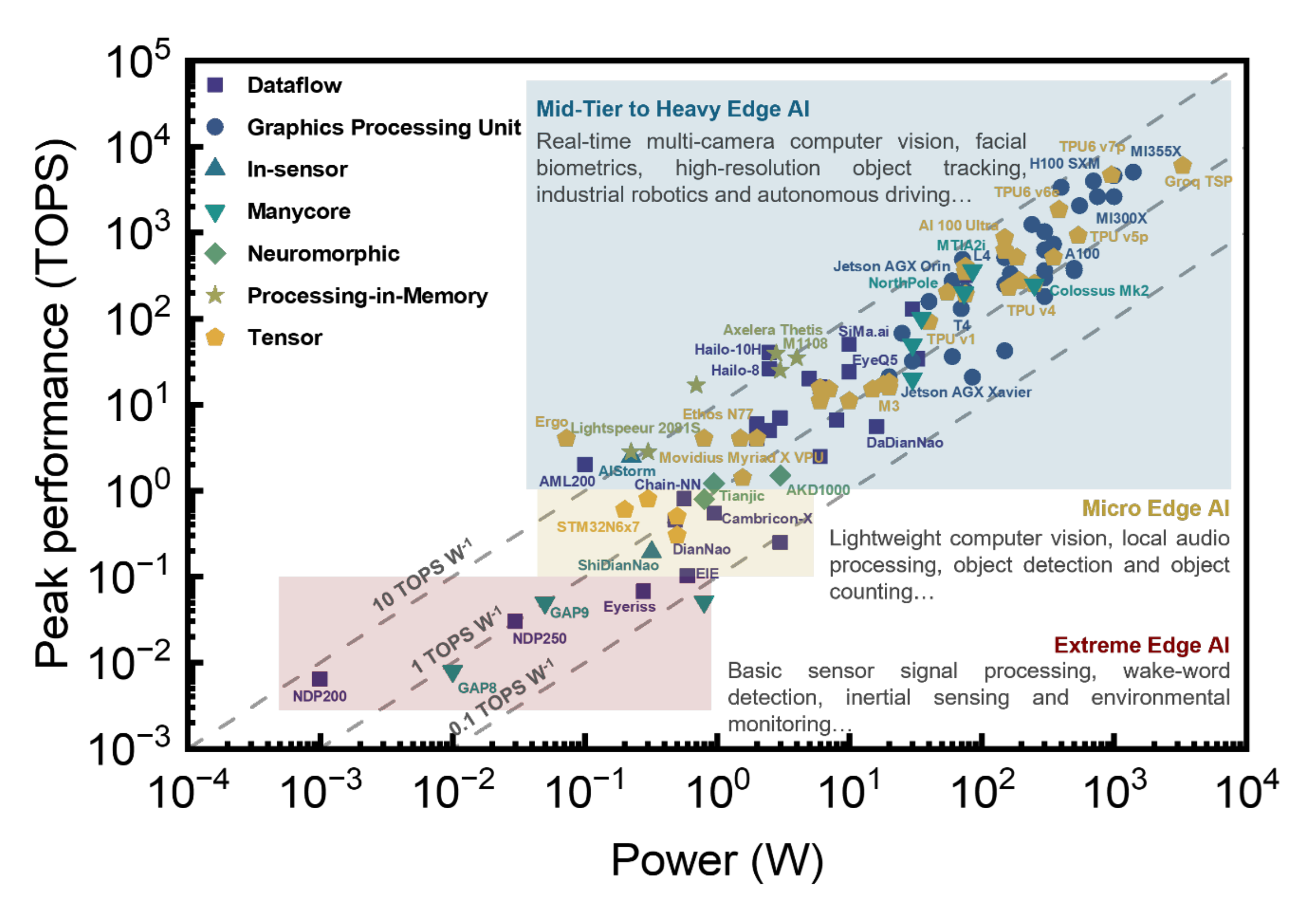

# [H1] Energy Challenges of Edge Computing

Modern DI architectures are built on a hierarchical hardware stack that transitions from low-power microcontrollers (MCUs) with integrated micro-neural processing units (micro-NPUs) at the sensor edge to advanced System-on-Chip accelerators at the coordination layer, and finally to high-bandwidth enterprise edge servers at local facility hubs[46,47]. These processing elements must be coupled with advanced sensor arrays that span environmental monitoring (1–10 μW), and wireless communication modules employing Bluetooth Low Energy or Zigbee protocols that can transmit data bursts at tens of milliwatts but average to microwatt levels through aggressive duty cycling reduction[8,48].

Although indoor lighting presents a higher energy density compared with other sources of ambient energy (*e.g.*, radiofrequency waves, local heat sources[48]), the energy density is orders of magnitude lower than outdoor 1-sun illumination (where the irradiance is 100 mW cm$^{-2}$).

Furthermore, the deployment environments for DI systems indoors span diverse settings with highly variable irradiances (from 15 μW $cm^{-2}$ in shadowed corners to >1 mW $cm^{-2}$ next to windows) and incident light spectra (from narrow white light emitting diodes to broad diffuse sunlight)[13,49,50]. There is therefore no typical power density that could be expected from IPVs, but 100 μW power output can be taken as a minimum benchmark for commercial DI applications[51], and can be provided by state-of-the-art IPV under a typical indoor illuminance of 200 lx (Fig. 1a). A power output of 100 μW is compatible with commercial microwatt-power neural decision processors, such as the Syntiant NDP100, which consumes <140 μW to perform continuous keyword spotting and acoustic event classification. However, with such restrictions in power supply, more advanced processes cannot be operated with conventional DI architectures. For example, image classification or object detection in real-time (*e.g.*, at 30 frames per second; FPS) by convolutional neural networks (CNNs) require at least an order of magnitude higher energy budget. With a power budget of 1 mW, a Syntiant NDP200 has demonstrated learning to play a simple video game as an example of computer-vision-based functionality[52]. The power requirements for more advanced image classification or object detection in real-time by CNNs (*e.g.*, MobileNetV4[53] or YOLO models[54]) are increasing rapidly. For example, the 'smallest' YOLO26n algorithm requires 5.4 billion FLOPs per frame [54], resulting in a minimum of 0.162 TOPS for 30 FPS. Even at the more generous power budget of 1 mW, this would require a minimum hardware efficiency of 162 TOPS $W^{-1}$, whereas the most efficient accelerators to date can at best provide values in the range of 1–10 TOPS $W^{-1}$ at such low power (refer to Box 1). Applications are therefore limited by power efficiency. In order to reach a target computational performance of 1 TOPS, it is necessary to improve both the power conversion efficiency of IPVs ($PCE_i$), coupled with improvements in the performance efficiency (TOPS $W^{-1}$) of IMCs. For example, 1 TOPS can be reached from 100 μW power supply from the IPV if the IMC has $10^4$ TOPS $W^{-1}$ efficiency. But the highest-performing IMC systems using multi-level ferroelectric field-effect transistor (FeFET) cells currently reach only 885.4 TOPS $W^{-1}$ at 153.6 μW[55]. The improvements in the IPVs and IMCs will need to come from the materials level in addition to engineering and device improvements.

Furthermore, peripheral circuits introduce an independent architectural bottleneck. Specifically, the analogue-to-digital converters (ADCs) can consume up to 85% of the total power budget[56]. Recent design paradigms introducing ADC-free peripheral architectures offer a promising path[57], which, when combined with materials engineering at the device level, can allow the desired computational efficiency (1 TOPS) to be achieved at the power provided by IPV.

Thus, whilst operating DI using the resources available indoors presents energy challenges, recent and realistic near-future materials advances in IPV and IMC make these addressable. The communities working on PV and memory are at the cusp of developing edge computing nodes capable of meeting the outlined efficiency requirements. The next sections will discuss the materials challenges and progress towards addressing those challenges to fully bridge the gap between energy supply, demand and reliability for edge computing.

## [H1] Energy Harvesting with Indoor Photovoltaics

There are a wide range of devices capable of harvesting the variety of ambient energy sources available indoors. These include thermoelectrics for harvesting waste heat[58], piezoelectrics for mechanical vibrations (*e.g.*, from footfall or body movements)[59,60], triboelectric nanogenerators for electrostatic charge formed between surfaces brought in contact together[61], Schottky diodes for ambient radiofrequency (RF) energy harvesting[62], and IPVs for indoor energy harvesting[11]. Of these options available, IPVs are particularly appealing. This is because of the ubiquity of lighting in indoor environments, as well as its higher energy density compared to ambient RF energy sources[11]. Mechanical vibrations and triboelectric energy sources may, in some instances, have higher energy density than indoor lighting, but they are less predictable or ubiquitous (*e.g.*, mechanical parts will have to be moving, or surfaces will have to be slid against each other), and also less deployable, in that the energy harvesting device will need to be held against the energy source[11]. IPVs, by contrast, can be deployed at a variety of distances from the energy source, and can readily be scaled from sub-$mm^2$ to >$cm^2$, depending on the application[11]. It is because of these advantages that the IPV field has witnessed a surge in the pace of development over the past decade, with a wide range of novel light-harvesting materials being introduced that have rapidly risen in performance to surpass incumbent technologies[11–13,15,63]. These include lead-halide perovskite photovoltaics (PPV; now with reported $PCE_i$ > 40%[18,64–66]), organic photovoltaics (OPV; $PCE_i$ > 35%[17]), chalcogenides ($PCE_i$ ~ 20% for $Sb_2S_3$ [67] and Se[68]), and dye-sensitized photovoltaics (DSPV; $PCE_i$ >38%[16,69]). This section discusses the requirements for IPVs to be used in DI applications, and evaluates the potential of current and future materials to fulfil these requirements.

## [H2] Materials and device requirements

***Power***. As discussed earlier, 100 μW is taken as the benchmark power output from IPV for edge computing, but improvements in the power that can be produced from IPVs across the diverse set of illumination conditions found indoors can help to meet the target computational performance of 1 TOPS. Improvements in the operating voltage of IPVs, on the other hand, are more relaxed, given that state-of-the-art PPV, OPV, and DSPV IPVs can produce >0.8 V at maximum power point at 200 lx or higher[16,18,70], and CMOS-compatible IMCs require <4 V.

The footprint available for IPV devices varies by application, from <1 $cm^2$ for 'smart dust' sensors[71] to >10 $cm^2$ for higher-power applications[16], but for the purposes of benchmarking, we take 10 $cm^2$ as a representative footprint. The required $PCE_i$ for IPV to deliver a specific power output across a wide range of irradiances is shown in Fig. 1a, with typical indoor irradiances spanning from ~15 μW $cm^{-2}$ to ~300 μW $cm^{-2}$ (or 50 to 1000 lx in the photometric units of illuminance[50,72]). From Fig. 1a, it can be seen that the $PCE_i$ required to produce the benchmark value of 100 μW at 50 lx exceeds the radiative limit for single-junction IPV, and would require a dual- or triple-junction tandem device, which is costlier and more complex to process. However, the most efficient IPVs (44.7% $PCE_i$ from PPV)[73] are capable of producing 100 μW at 90 lx (Fig. 1a). At an average indoor illuminance of 200 lx[74], Fig. 1a shows that a 16%-efficient IPV would just produce sufficient power. To provide a greater tolerance for reductions in efficiency and extend the operable range to lower illuminances, it is desirable for IPVs to have >30% $PCE_i$. As of 2026, the main technologies fulfilling these requirements are PPV, DSPV, OPV, and compound semiconductors based on elements from groups 13 and 15 of the periodic table (*i.e.*, III-V compounds)[15]. But apart from performance, the cost of the PV device (encompassing product cost and capital cost for manufacturing) also needs to be low. PPV, OPV, and DSPV could all meet this requirement, but III-Vs have high production costs due to the requirements of careful epitaxial growth to minimize defect densities[75]. Thus, we primarily focus on PPV, DSPV, and OPV as illustrative examples in our discussion here.

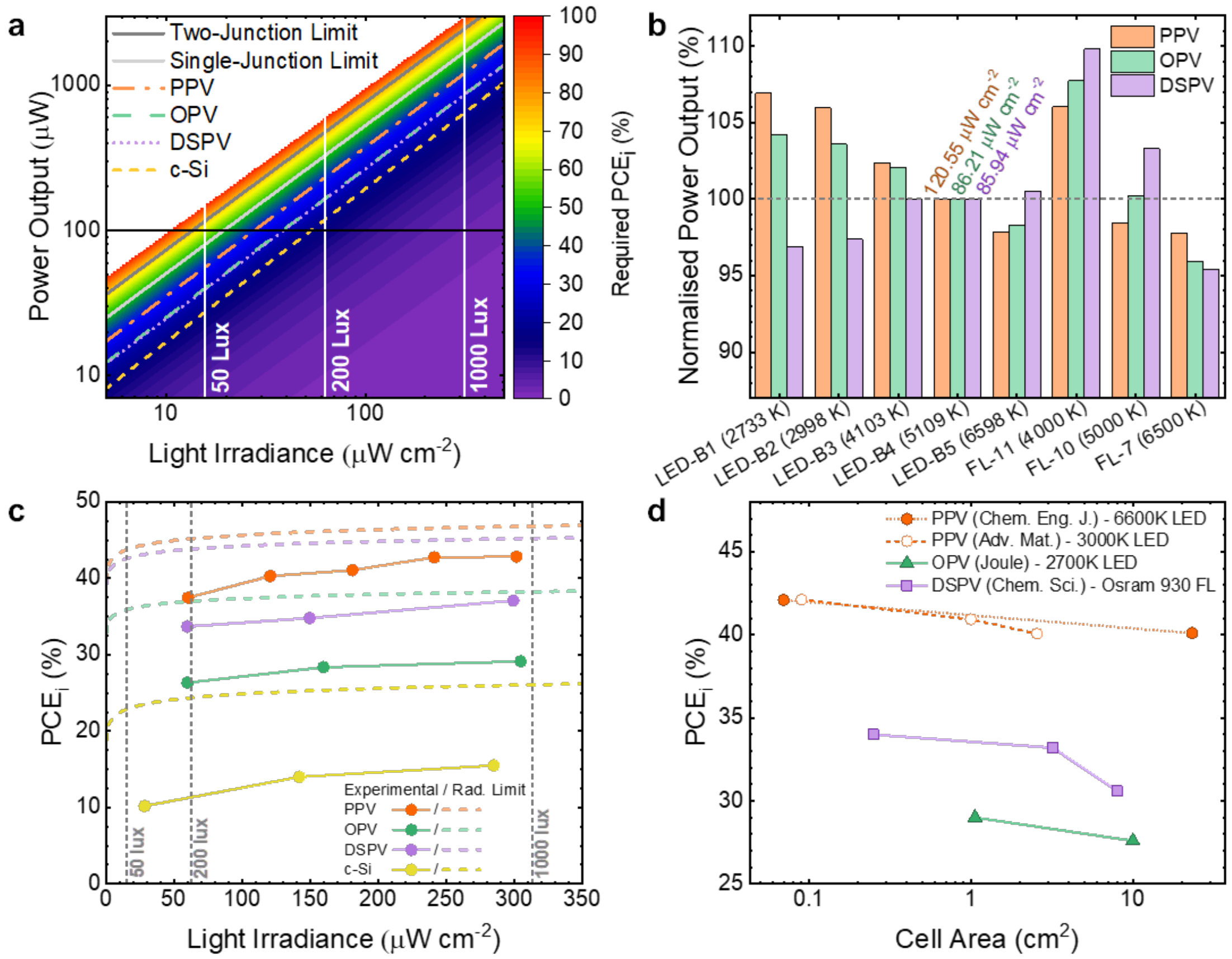

**Fig. 1 | Requirements of IPVs for powering DI, and analysis of current performance of select IPV materials. a,** Heatmap of the indoor power conversion efficiency ($PCE_i$) required to reach selected output electrical power across a range of incident light irradiances. The device area is set to 10 $cm^2$, and the radiative limit for single-junction[49] and two-junction tandem IPV devices[76] under CIE LED-B4 (WLED with 5000 K correlated colour temperature) illumination are indicated. The current state-of-the-art performance of lead-halide perovskite (PPV)[18], organic (OPV)[70], dye-sensitized (DSPV)[16], and crystalline silicon (c-Si) PV are also indicated. **b,** Calculated variation in the power output from PPV, OPV, and DSPV across a set of common indoor artificial light sources with different correlated colour temperatures (CCTs); details of calculations are provided in Supplementary Notes 1-4. Reported variation in the $PCE_i$ of PPV, OPV, and DSPV with **c,** irradiance of incident light compared to the calculated variation in the radiative limit, and **d,** as the device area is increased. In panel **d**, the data originates from Ref. [66] and [77] [66](PPV), [78] (OPV), and [79] (DSPV).

***Consistency.*** Whilst it is currently commonplace to report IPV performance under the most ideal illumination conditions for the materials/devices considered, reliable operation of edge computing devices requires the performance of the IPV modules to be consistent across the wide range of irradiances and spectra experienced in real-world contexts. This consistency depends on several properties: shunt resistance ($R_{sh}$), interfacial recombination, parasitic optical losses, and spectral matching between the incident light spectrum and the device spectral response. A parallel resistance inferred from the slope of an $I-V$ curve can also reflect collection and recombination losses rather than a conducting leakage pathway (detailed discussion in Supplementary Note 5).

Given the low irradiances under indoor lighting compared to outdoor solar illumination (1000 lx corresponds to 0.3 mW cm$^{-2}$, compared to 100 mW cm$^{-2}$ for 1-sun), it is more important for IPV devices to exhibit a high shunt resistance than a low series resistance to minimize the dark current compared to photocurrent, thus avoiding severe reductions in open-circuit voltage ($V_{OC}$) and fill factor as illuminance is reduced. The shunt resistance needed to limit leakage losses depends on the operating illumination, device characteristics and tolerated loss (Supplementary Note 5 and Supplementary Table 2)[12,80–82]. [16,7018]The selected devices shown in Fig. 1c exhibit different illumination dependencies, which should be interpreted in terms of recombination, charge-carrier collection and parasitic conduction. For example, a c-Si IPV device exhibits similar variation in $PCE_i$ with irradiance as the PPV device, but this is due to trade-offs: the c-Si devices has higher $R_{sh}$ than the PPV (Supplementary Table 3), but its lower bandgap results in a higher reverse-bias dark saturation current density ($J_{00}$). This illustrates the advantage of having stable wide-bandgap light harvesters for IPV, with gaps in the range 1.7 to 2.0 being ideal for most artificial light sources[49,83].

In terms of materials and device optimization, interface recombination can also alter the apparent parallel resistance inferred from the $I–V$ curve slope, and should therefore be minimized to improve the consistency of IPVs. For example, in OPV, the higher trap density at interfaces, and between donor and acceptor moieties, in bulk heterojunction devices led to a lower apparent parallel resistance than the bilayer devices studied, and a greater reduction in $V_{OC}$ with reduced illuminance[84]. In contrast, OPVs with non-fullerene acceptors engineered to lower trap densities have high $R_{sh}$ (>$10^5$ Ω cm$^2$), low $J_0$ ~ $10^{-24}$ mA cm$^{-2}$ (Supplementary Table 3), and <10% relative decrease in $PCE_i$ with only an 80 mV decrease in $V_{OC}$ when reducing illuminance from 1000 to 200 lx[70].

Consistency across indoor light spectra requires tuning the device spectral response and reducing parasitic optical absorption in the charge transport layers. Correlated colour temperature (CCT) does not uniquely specify the spectrum of an indoor light source. For example, CdS is widely used as a buffer layer in many of the highest-performing chalcogenide solar cells (*e.g.*, $Sb_2S_3$ [67] and kesterites[85]). But the low bandgap of CdS causes parasitic optical losses in the blue part of the incident light spectrum, such that IPV performance reduces for higher CCT light sources[67,85]. Molecular absorbers (*e.g.*, OPV and dyes in DSPV) can also have their spectral response at specific wavelength ranges tuned, unlike inorganic semiconductors. For example, the XY1:L1 absorption band overlaps well with the FL11 spectrum (Supplementary Fig. 2), hence the DSPV having the highest $PCE_i$ under FL11, despite its $PCE_i$ being lower than the OPV and PPV under the WLED spectra (Fig. 1b).

Moving forward, reporting the $PCE_i$, power density, and voltage of IPVs under the most favourable illumination conditions will be inadequate for edge computing applications. It will be essential to develop and adopt rigorous practices to quantify the consistency of IPVs across a range of lighting conditions (*i.e.*, energy rating standards) in order for manufacturers to specify to DI developers the range of conditions in which the IPVs can provide the requisite power and voltage for operating the edge computing node[13].

***Stability.*** IPVs should be durable over the entire lifetime of the edge computing node, which is expected to be 3–10 years[86]. Currently, there are no international standards for measuring the long-term stability of IPVs[87] or predicting it based on their operative figures-of-merit; alas, the same holds true for accelerated outdoor stability testing of emerging PV materials, such as PPV. Noting this lack of measurement standards, the most responsible choice is to adopt stricter measurement protocols, in order to deliberately but safely underestimate the stability of IPV devices. Given the less harsh conditions of the indoor environments (narrower range of temperature and humidity, reduced UV irradiance), it may be assumed that IPV devices passing outdoor damp heat tests (85 ºC, 85% relative humidity) will have indoor lifetimes >10 years. However, this assumption may overlook degradation pathways that are specifically tied to indoor operation, or underestimate the indoor lifetime of some technologies. For the development of effective stability protocols tailored to IPVs, it is therefore essential to account for all of these factors, using the protocols for outdoor PV only as a starting point. Given the current lack of experimental lifetime testing data, the IPV community should focus on investigating these alternative degradation pathways and understanding the mechanisms by which they affect device stability. The development of international standards frequently stems from laboratory best practices and internal protocols, as well as from industry needs, and the same will likely be true for indoor stability testing, for example by expanding upon the existing ISOS protocols[88]. Cases of extreme values in environmental exposure (irradiance, temperature) should also be considered to cover cases such as devices next to a window facing direct sunlight, or an unoccupied space that is not heated for a long period in winter. IPV devices should be able to tolerate such conditions for short periods, not only for performance stability but for safety reasons (fire hazards).

***Scalability.*** Beyond improvements in efficiency and stability, it is essential to address the scaleup of IPV devices from the sub-centimetre scale typical of fundamental research to areas of around 10 $cm^2$, sufficient to deliver the power required for edge computing applications. As mentioned earlier in the consistency section shunt resistance is critical for IPV[80]. Defects (*e.g.*, pinholes), non-uniform grain growth, or discontinuities at charge-transport layers can all lower the *specific* shunt resistance (typically expressed in $\Omega$ cm²)[89]. Moreover, the $R_{sh}$ values reported for small-area research devices are often not representative of those in larger-area cells, as scaling amplifies the effects of localized film defects and process nonuniformities. Segmenting IPV modules into serially interconnected sub-cells can mitigate these defect-driven shunts while simultaneously raising the output voltage to levels more compatible with power-management electronics. However, methods like laser scribing when creating interconnections need fine tuning for specific materials, as they can create shunt pathways and introduce defects at the edges of sub-cells. Importantly, while the $PCE_i$ of an IPV module is mostly independent of the number of sub-cells (for small interconnect widths)[89], the total efficiency of an IPV-powered energy-harvesting system closely relates to the number of sub-cells, as well as the chosen PV material and its power output under typical illumination conditions (details in Supplementary Notes 3 and 4).

Beyond these essential performance and manufacturing factors, materials considerations for IPV also encompass the capital cost of production, supply chain sustainability, and resource criticality. Important challenges include the widespread use of indium – an element with high criticality – in transparent conducting oxides[90]. Efforts with improving IPV should therefore focus on developing In-free transparent conducing electrodes, or approaches for recycling PV to re-use these elements in new devices as part of a closed-loop manufacturing process. The development of earth-abundant alternatives—such as copper-based DSC mediators, lead-free perovskites, and organic semiconductors—are also important challenges to ensure that scaling DI does not shift the environmental burden from energy consumption to materials depletion.

## [H2] Current and future materials

***Incumbent technologies.*** As discussed earlier, there are currently only a handful of IPV materials that have reached >30% $PCE_i$ and are cost-effective to produce: DSPV, OPV and PPV. Here, we discuss key improvements that enabled them to reach these high $PCE_i$ values, as well as the extent to which they fulfil the requirements of IPV for DI.

With DSPV, the $PCE_i$ at 1000 lx improved from 28.9% in 2017[91] to 38% today, with a 1.0 V open-circuit voltage[16] (and 40% $PCE_i$ at 4000 lx[69]). Three molecular developments drove these improvements in performance. First, one-electron Cu(II/I) coordination complexes replaced the $I^-/I_3^-$ couple, cutting the dye-regeneration driving force from ~0.5 eV to ~0.1 eV and raising photovoltages to above 1 V[91,92]. Second, co-sensitization of donor–π–acceptor dyes matched absorption to the 400–650 nm emission of indoor lighting, with external quantum efficiencies above 90%[79,91]. Third, interface control, hydroxamic acid pre-adsorbents ordering the dye monolayer[93] and ligand engineering of the Cu(II) coordination sphere[94] suppressed recombination. These DSPV devices retain $PCE_i$ above 31% at 200 lx[16,79], although this performance cannot be attributed solely to an estimated shunt resistance (Supplementary Note 5). For state-of-the-art DSPV, fill factors are now limited by Cu(II) mass transport in viscous electrolytes and by counter-electrode kinetics. Quasi-solid and solid-state Cu(II/I) hole conductors (30.7% $PCE_i$ at 1000 lx[95,96]) address both limitations. DSPV are already used commercially for powering maintenance-free sensors and consumer products[97,98], and have already been used in edge computing demonstrations[16,79].

For OPVs, the $PCE_i$ increased from approximately 10–15% in the first demonstrations to over 35% as of 2026[17,99]. These improvements primarily stem from the development of wide-bandgap non-fullerene acceptors (NFAs) with absorption spectra tailored to artificial illumination, alongside advances in donor–acceptor molecular design and interface engineering that improve molecular packing, energy-level alignment, charge selectivity, and charge extraction[17,100]. Despite these advances, an enduring challenge for engineering high-performance indoor OPV is the sub-gap electronic landscape. Residual energetic disorder and the associated sub-gap electronic states contribute to non-radiative voltage losses that limit the achievable open-circuit voltage, while morphological heterogeneity and imperfect interfaces increase recombination losses and reduce the fill factor. Consequently, precise control over factors such as molecular packing, phase morphology, and interfacial energetics is vital for minimising voltage losses and maintaining high efficiencies under indoor operating conditions[49,101,102].

With PPVs, the $PCE_i$ improved from 27.4% under 1000 lx in 2015[103] to over 44% as of 2024[18]. This advance stems from the development of wide bandgap (WBG) perovskite absorbers with improved spectral match to indoor light sources, together with crystallization-control, defect-passivation and interface engineering strategies that suppress non-radiative recombination[63,104–107]. The current record device, based on a 1.71 eV bandgap perovskite, reached 79.3% of its radiative limit, with a $PCE_i$ of 44.72%[18]. One major limitation is that WBG perovskites are typically obtained using Br-rich mixed-halide or fully inorganic compositions. These materials face challenges, including phase segregation, high formation energies and increased defect densities[108,109], which promote energetic disorder and limit the $V_{oc}$, FF, operational stability and device consistency. Further progress will require integrated strategies specific to WBG PPVs, spanning both active-layer engineering and device-level optimisation[110,111], coupled with a deeper understanding of the dominant loss mechanisms[112]. Further gains at low light conditions require distinguishing parasitic conduction from recombination and collection losses, then targeting film quality, charge transport layers and electrodes accordingly (Supplementary Note 5). Concomitantly, exploring alternative structural motifs within the perovskite family (*e.g.*, low-dimensional perovskites[113,114]) may open new avenues for next-generation IPV technologies, enabling intrinsically WBGs while opening new design possibility. Addressing toxicity challenges from the use of Pb in PPVs will also be important, whether through development of Pb-free alternative materials (see next section) or through development of Pb-sequestration strategies[115].

Well-optimized DSPV, OPV and PPV have all demonstrated high stability under indoor operation. PPV have already demonstrated promising early results, with nearly unchanged performance under 200–1000 lx for 1 year. This has been largely enabled by the production of high-quality perovskite films with low defect densities, as achieved through optimized vacuum-deposition process[116]. The dominant indoor failure mode for DSPV is physical rather than photochemical, solvent permeation through the seal, which solid-state Cu(II/I) hole conductors eliminate[92]. But, as mentioned earlier, the lack of standards for indoor stability testing prevents a like-for-like comparison between technologies and reports from different groups[8,87,117]. Initial performance increases during conditioning should be distinguished from burn-in degradation. Both condition and degradation can depend on illumination history. It is therefore strongly motivated to develop protocols for stability testing[74,87], and a place to start may be the ISOS tests[88].

When it comes to scaling, DSPV modules are already used commercially[97,98], as are OPV modules[118], while PPVs are beginning to enter into commercialization as IPV modules. In all cases, modules well above 10 $cm^2$ have already been demonstrated[97,98,118,119]. Across all technologies, when translating from laboratory-scale cells to commercially-relevant modules, future progress will depend not only on continued molecular and interface engineering to further suppress voltage losses, but also on scalable fabrication strategies that preserve the requirements of an exceptionally-high shunt resistance and morphological uniformity[80,120,121].

***Future materials.*** There has been particular emphasis in the photovoltaics community on developing light-harvesting semiconductors that can emulate the defect tolerance of halide perovskites, but overcome their toxicity and stability limitations[12,122,123]. Many of these 'perovskite-inspired' materials have wide-bandgaps in the range of 1.5-2.2 eV, which are non-optimal for 1-sun applications, but are well-suited for harvesting artificial lighting[12,122,123]. Materials include Sb-, Sn- and Bi-based semiconductors, such as $Sb_2S_3$, Sn-halide perovskites, and BiOI, as well as chalcogenides, such as elemental crystalline Se (c-Se). Among these materials, the classes of materials exhibiting the greatest combination of high performance, stability and scalability thus far are the chalcogenides[12,123–127]. However, the challenge with chalcogenide materials is the prevalent use of CdS as a buffer layer to achieve the highest efficiencies[85,123,128]. This should be substituted with wide-bandgap charge transport layers to overcome the toxicity of Cd (which is more toxic than Pb), as well as the parasitic optical losses from CdS, which cause a reduction in IPV performance under colder white light sources, as discussed earlier[67,85]. Equally, it is important to focus on materials that have a wide phase-stable processing space, low energetic disorder, as well as band-like transport. For example, kesterites have bandgaps that can be tuned to the 1.4–2.2 eV range, are stable, and are composed of earth-abundant elements, but are held back currently by low shunt resistances ($10^2$–$10^4$ $\Omega$ $cm^2$)[129], due to the narrow phase-stable region for $Cu_2ZnSnS_4$, with the prevalence of conducting phase impurities that severely limit the consistency in performance under low illuminances[130]. For $Sb_2S_3$ and c-Se, low-light performance likewise depends on recombination, collection and parasitic conduction (Supplementary Note 5). Both materials have reached $PCE_i$ ~20%[126,131], and have high theoretical limits of 45–50% $PCE_i$ (1.7–1.9 eV bandgap)[123,125,132]. Promisingly, both materials have now demonstrated high performances without requiring CdS as a buffer layer, such that high $PCE_i$ at high CCTs could still be achieved[126,133], along with high stability in ambient air without encapsulation[123,125]. Furthermore, $Sb_2S_3$ is deposited by hydrothermal, chemical bath or spray deposition methods[123,134,135], while c-Se is grown by thermal evaporation[126], all of which are scalable methods.

For molecular systems, future materials development should translate chemical tunability into stable performance across indoor spectra and intensities. OPV priorities include reducing energetic disorder and voltage losses while maintaining efficient charge collection[49]. For DSPV, reducing regeneration losses must preserve efficient electron injection and dye regeneration[91,92]. Across both technologies, these molecular and interfacial improvements should be retained during scale-up and assessed at the intended operating illumination. Distinguishing apparent shunt behaviour from conducting leakage pathways will help identify which materials interventions are needed (Supplementary Note 5).

## [H1] Energy Storage for Autonomous Edge Computing Devices

The highly variable nature of lighting available indoors throughout the day demands energy storage devices that combine high energy storage density (ESD), *i.e.*, large amount of energy stored per unit mass or per unit volume/area, with high charge-discharge efficiency to minimize energy losses during intermittent operation. This ESD should be maintained over multiple cycles throughout the 10- to 20-year lifetime of an edge computing device, while also accommodating rapid fluctuations in energy input and output. Among the available energy storage approaches, electrostatic energy storage is uniquely positioned to meet these requirements, as it relies exclusively on field-driven charge displacement[136,137]. As a result, electrostatic capacitors offer ultrafast response times, high power delivery, and exceptional endurance, making them especially attractive for applications requiring rapid and reliable energy delivery. In contrast, electrochemical energy systems, such as batteries and electrochemical supercapacitors, typically provide higher energy densities but suffer from intrinsic limitations associated with diffusion-controlled processes and interfacial chemical reactions. These limitations lead to reduced power capability, low charging rates, finite cycle life, and safety concerns, which restrict their suitability for reliable operation at high-speed. Moreover, the stringent requirements of IoT and integrated electronic platforms demand energy storage solutions that are fully compatible with on-chip integration, further motivating the exploitation of electrostatic alternatives[138,139]. There are a wide range of materials options for the dielectrics for electrostatic capacitors, ranging from linear and paraelectric ceramics and thin films to polymers, each offering distinct tradeoffs in permittivity, recoverable energy storage density, breakdown voltage, thermal stability, and endurance. But herein, we specifically focus on (anti)ferroelectric materials because their giant, field-induced switchable polarization enables orders-of-magnitude higher energy densities and CMOS compatibility, which are essential for miniaturized energy storage systems[140–142]. Therefore, in this section, we first elaborate on the challenges with conventional electrochemical energy storage devices, before delving into the opportunities electrostatic capacitors offer to address these challenges, followed by examining the potential of lead-free (anti)ferroelectric hafnium zirconium oxide ($Hf_{1-x}Zr_xO_2$, hereafter HZO) for electrostatic capacitors.

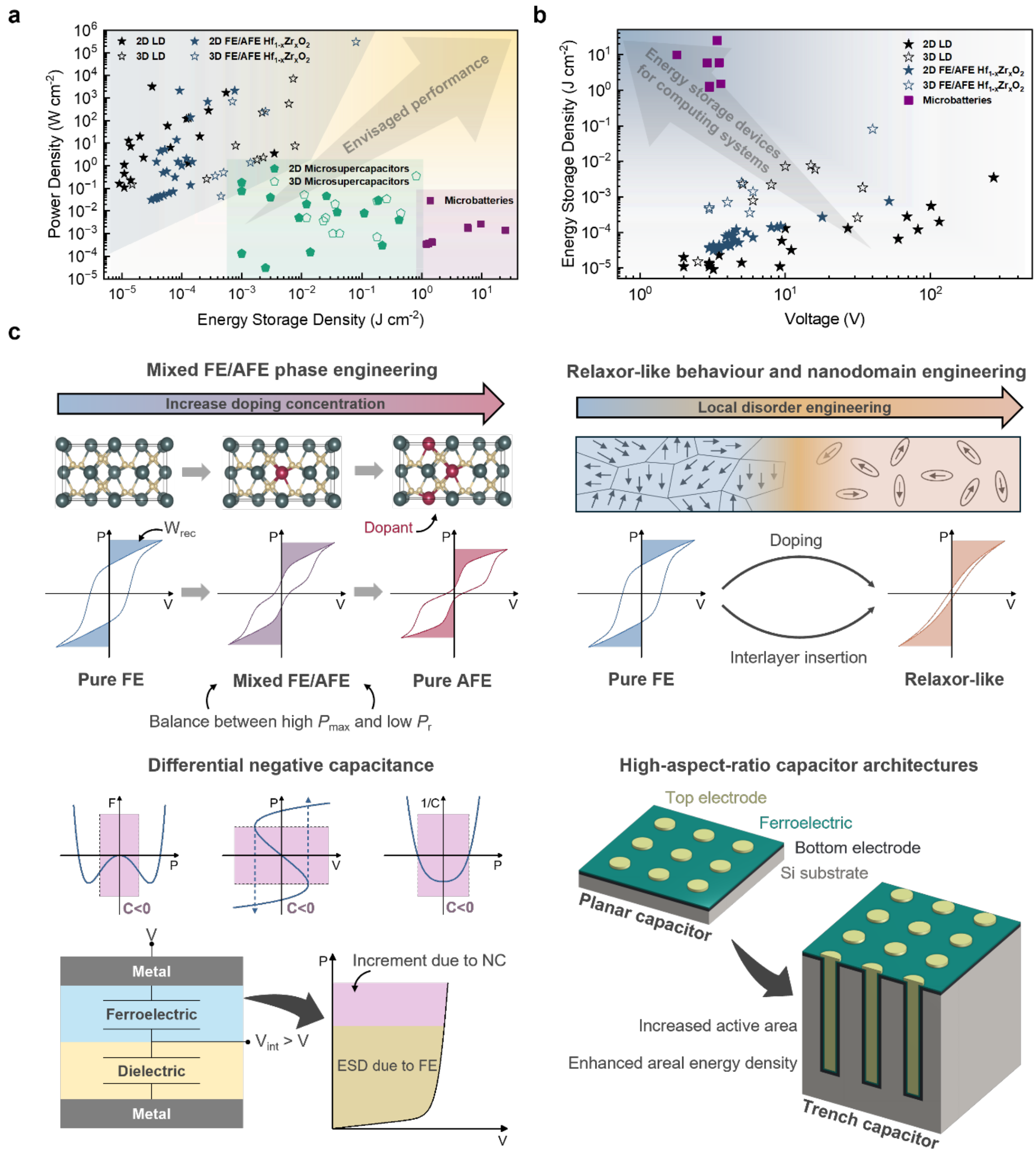

**Fig. 2 | Energy storage devices for edge computing systems. a,** Power density against energy areal storage density, and **b,** areal energy storage density as a function of operating voltage for electrostatic capacitors and microbatteries. The semi-transparent arrow highlights the target regime of low-voltage, high energy storage density operation for edge computing systems. All data provided in Supplementary Tables 4–7. **c,** Key strategies for enhancing the energy storage performance of ferroelectric $Hf_{1-x}Zr_xO_2$-based thin film capacitors, including the engineering of a mixed ferroelectric/antiferroelectric phase, inducing relaxor-like states and nanodomain engineering, exploiting the negative capacitance effect, and incorporation of the optimized films into three-dimensional capacitor architectures.

## [H2] Challenges with Electrochemical Energy Storage

Currently, electrochemical methods are by far the most commonly used for energy storage, with Li-ion batteries the most widely-used commercial electrochemical energy storage device[143,144]. Li-ion batteries support continuous discharge rates of 1C to 2C (Supplementary Table 7). C is the rate at which a battery is discharged or charged relative to its maximum capacity per hour, and can be calculated by dividing the discharge or charge current by the battery's rated capacity. In this context, 1C signifies the battery will be fully charged or discharged in one hour, while 5C signifies in 1/5 of an hour. The low continuous discharge rate, but high ESD, makes Li-ion batteries suitable for low-power edge devices, such as wireless sensor nodes, industrial sensors, and Internet of Things[145,146]. While many low-power IoT sensing nodes are designed for periodic data acquisition, and can operate for several years on a single battery[146], the significant power requirements of edge computing tasks introduce several specific limitations. For instance, intensive computation generates heat, which can reduce battery efficiency and compel processors to reduce speed (throttle) to avoid damage[147,148]. In addition, AI workloads with sudden spikes in current demand may cause temporary voltage drops if the battery is nearly empty or cold, potentially leading to device reboot or crash during a task execution[147]. For example, DI devices (*e.g.*, NVIDIA Jetson series of GPU modules, AI accelerators, compact edge servers) often require 2C to 5C because current demand triples instantly during inference tasks. In high-performance DI devices (*e.g.*, rugged servers with dedicated GPUs (RTX/Tesla) in remote locations), where large, sharp energy spikes last only milliseconds (>5C), batteries become inefficient and often fail to maintain the required voltage for operation. It is therefore essential to develop novel materials solutions capable of dealing with power spikes in computation loads, and which have high recoverable ESD and power density.

In this context, electrochemical capacitors (or supercapacitors), with a discharge rate in the range of 100-1000C, offer the high-power density required for the rapid burst nature of edge AI tasks, making them an attractive solution. Their very low equivalent series resistance allows them to deliver significant power instantly, with minimal transient voltage sag during sudden power demand, avoiding a common limitation with batteries[149]. Moreover, electrochemical capacitors also prevent the system from rebooting during intense computational tasks. However, achieving high ESD and power density, with high efficiency, often motivates the use of hybrid energy storage systems (HESS). The most typical HESS relies on a battery for baseline energy supply and a supercapacitor to handle transient power demands[150,151]. This hybrid approach offers several advantages, including improved battery lifetime, enhanced energy recovery, reduced overall system size, and optimized energy delivery under dynamic load conditions, while providing high energy density, fast response, and the ability to smooth complex power fluctuations[150,152,153]. While this approach improves performance, it increases the system complexity, introduces additional challenges in energy flow management, and increases system cost from approximately US$30-50 $kWh^{-1}$ (batteries) to US$250-400 $kWh^{-1}$ (supercapacitors)[151]. Nonetheless, miniaturizing these technologies for on-chip integration remains challenging due to materials, fabrication, and safety issues[139,154]. In addition, they suffer from having a high self-discharge rate, which requires a regular energy harvesting source (*e.g.*, IPV). Their gradually decreasing output voltage requires additional power management chips to maintain stable operation of the processing unit. Furthermore, the initial hardware cost for a high-capacity supercapacitor bank is 10-20 times higher than a simple Li-ion battery[151].

Another potential strategy is the use of photocapacitors that combine energy harvesting and storage functionalities into a single device[155]. This integration reduces the areal and mass footprint of the device, as well as reducing circuit complexity, and is well-suited to small sensors and wearables. For instance, polyviologen-based supercapacitors can power a three-layer IoT network for 72 hours using only ambient light at 500 lx[155]. However, the efficiency in low-light conditions, long-term stability, as well as the high complexity for manufacturing are current challenges that need to be addressed for this early-stage energy storage technology.

**Box 2 | (Anti)Ferroelectric Materials**

(Anti)ferroelectric materials are a class of dielectric compounds characterized by the presence of electric dipoles that can be reoriented by an external electric field[34,156]. In ferroelectric materials, these dipoles align spontaneously in the same direction, creating a permanent polarization that can be flipped. Instead, in antiferroelectric materials, adjacent dipoles align in opposite directions, resulting in zero net polarization until a sufficiently high field forces them into a parallel state. $Hf_{1-x}Zr_xO_2$ has emerged as a promising lead-free (anti)ferroelectric material. When stabilized in the non-centrosymmetric orthorhombic polymorph (space group $Pca2_1$, bottom panel in the figure below), $Hf_{1-x}Zr_xO_2$ exhibits ferroelectric behaviour. In contrast, its antiferroelectric behaviour stems from an electric-field-induced phase transition from the centrosymmetric tetragonal phase (space group $P4_2/nmc$, top panel) to a polar orthorhombic phase. The representative polarization–voltage curves (hysteresis loops) of these two polymorphs are sketched on the right-hand side. The blue shaded regions indicate the recoverable energy storage density (ESD), while the yellow shaded regions represent the energy loss ($W_{loss}$) during a charge-discharge cycle.

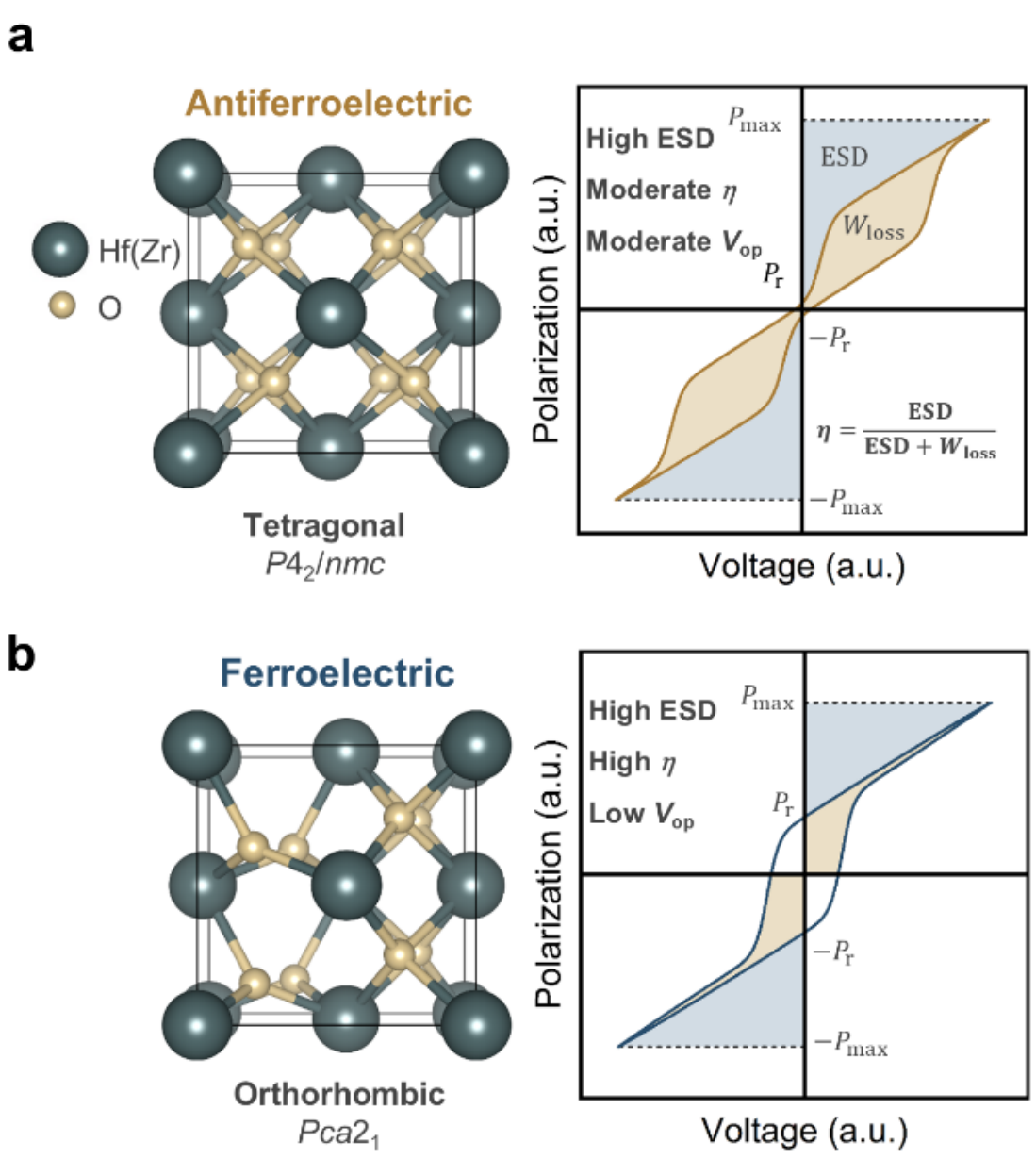

These materials store energy electrostatically, rather than chemically[30]. Inside a capacitor, an insulating material (the dielectric) sits between two conductive metal plates. When a voltage is applied, electric charges accumulate directly on the surfaces of these plates, creating an electric field across the dielectric. Given that the energy is stored without the need for chemical bonds to break or reform, this energy can be released almost instantaneously, while millions of charge-discharge cycles can be achieved with minimal degradation. This fundamental physics advantage allows capacitors to deliver the large, rapid bursts of power required for DI.

In the context of in-memory computing, (anti)ferroelectric materials can be used to mimic the behaviour of the human brain, where multi-state programmability emulates the learning process of strengthening or weakening the connection between two neurons, a property known as synaptic weight[2,157,158], where strengthening (potentiation) and weakening (depression) as for a biological synapse, occurs. Application of controlled voltage pulses of suitable amplitude and time width leads to a partial switching, enabling multi-state programmability for (anti)ferroelectric materials. By varying the strength or number of these electrical pulses, one can incrementally tune the material's electrical conductance.

## [H2] Strong Potential of Electrostatic Capacitors

Electrostatic capacitors, with a discharge rate >10,000 C ($10^{-6}$–$10^{-3}$ s discharge times)[27,28], are intrinsically compatible with high-power, short-duration workloads at the chip level. Current commercial on-chip silicon capacitors already provide well-established reliability, with device operating lifetimes on the order of 10 years, and operating temperatures ranging from –55 °C to 125 °C[159–162]. Despite these advantages, micro- and nano-fabricated electrostatic capacitors that simultaneously deliver high power density and robustness against extensive cycling and temperature for on-chip integration have not yet been systematically demonstrated. Moreover, commercially available electrostatic capacitors typically exhibit much lower ESD than batteries or supercapacitors. These limitations restrict the execution of energy-intensive computational bursts, despite their superior power density. Consequently, enhancing the ESD of electrostatic capacitors has become an active area of research.

In this context, recent advances in ferroelectric (FE) and antiferroelectric (AFE) materials (Box 2) have opened up new opportunities for overcoming these limitations. In particular, the discovery of FE[163,164] and AFE[165,166] behaviour in eco-friendly HZO thin films has generated significant interest due to their compatibility with complementary metal oxide semiconductor (CMOS) devices, owing to their low thermal budget requirements (growth at temperatures below 400 ºC) and thickness scalability (<10 nm). The FE behaviour in HZO thin films is governed by the stabilization of a non-centrosymmetric orthorhombic phase (space group $Pca2_1$; Box 2), which gives rise to reversible spontaneous polarization, whereas the AFE behaviour arises from an electric-field-induced phase transition from the centrosymmetric tetragonal phase (space group $P4_2/nmc$) to the polar orthorhombic phase (Box 2). These structural characteristics result in distinct polarization-voltage responses and energy storage properties (Fig. 2a). A particularly promising materials route to address these constraints lies in 3D AFE HZO-based capacitors. State of the art demonstration of 3D AFE HZO-based capacitors has shown that performance improvements compared to other works arise from: i) the increased intrinsic energy storage performance that was engineered through HZO composition to enhance volumetric ESD (115 J cm$^{-3}$ at ≈6 MV cm$^{-1}$); ii) the scaling of the energy storage performance in superlattices that overcome the conventional thickness limitations of AFE HZO; and iii) the integration of these optimized superlattices into 3D capacitors resulting in ESD per unit area nearly an order of magnitude higher and power densities over two orders of magnitude higher than state-of-the-art electrostatic capacitors (≈80 mJ cm$^{-2}$ and 300 kW cm$^{-2}$, respectively), with discharge times of ≈250 ns and >$10^9$ cycles[167]. These results suggest the potential of AFE HZO-based capacitors to achieve high energy storage capacity (as shown in Fig. 2c, and Supplementary Tables 4 and 5), while clearly outperforming the power values reported for electrochemical micro-supercapacitors and microbatteries (Fig. 2b, Supplementary Tables 6 and 7, respectively). However, the reported energy storage performance of 3D HZO-based capacitors was achieved under relatively high operating voltages (>30 V), as depicted in Fig. 2b [167]. While high applied voltages are inherently favourable for maximizing ESD, practical integration with ultra-low-power systems and energy harvesters, such as IPVs, requires operation at substantially lower voltages, as mentioned at the start of the energy harvesting section. Recent studies on 3D AFE HZO-based capacitors have demonstrated promising strategies for low-voltage operation, including dopant engineering of AFE $HfO_2$ films and the use of intermediate dielectric layers, achieving planar ESD on the order of mJ $cm^{-2}$ at operating voltages below 6 V[168,169]. On the other hand, as will be discussed in the following section, FEs based on HZO systems shows even greater promise to reduce the operation voltage.

**[H2] Ferroelectric Hafnium Zirconium Oxide-Based 3D Capacitors**

From a systems-level perspective, CMOS-compatible FE HZO-based 3D capacitors are a promising route towards simultaneously achieving high power and energy density in commercial edge computing nodes. CMOS compatibility has been increasingly desired because it enables the integration of energy storage directly with on-chip processing and control circuitry, eliminating bulky off-chip components and supporting compact, efficient IoT devices. Yet, the potential of FE HZO-based capacitors in 3D architectures remains largely unexplored, making this discussion both novel and timely.

FE HZO thin films have been shown to exhibit exceptionally high polarization (>45 μC $cm^{-2}$) even for films thinner than 15 nm, and operating voltages below 6 V[170–173]. FE switching of HZO-based capacitors has been reported at voltages as low as ~1 V[174], with a coercive field voltage of ~0.55 V ($E_c$ = 0.65 MV $cm^{-1}$). Furthermore, HZO capacitors have demonstrated energy losses of only ≈5%. This combination suggests that high ESD can be attained in ultra-thin films with low operating voltages, opening a route toward compact and low-voltage devices (Fig. 2b).

Despite these advantages, FE HZO-based thin films present intrinsic challenges that must be addressed for practical energy storage. In particular, the presence of hysteresis reduces the recoverable energy density and efficiency, as part of the stored charge cannot be released due to intrinsic memory effects. A possible strategy to overcome this issue, as shown in Fig. 2c, could involve the stabilization of mixed FE/AFE behaviour, as demonstrated in Si-doped $ZrO_2$ thin films[175]. This approach could combine the high polarizability of FEs with the low remanent polarization of AFEs, enabling higher ESD together with improved efficiency[175].

Another promising route to reduce hysteresis loss in FE HZO is the promotion of relaxor-like (or superparaelectric-like) behaviour. In such states, long-range FE order is suppressed, and polarization arises from nanoscale polar regions, leading to reduced remanent polarization. While relaxor behaviour is well established in perovskite oxides[30,176,177], it remains scarcely explored in HZO-based systems. Nevertheless, recent reports on AFE HZO-based capacitors provide valuable insights. For instance, the insertion of an ultrathin $Al_2O_3$ layer in 15 nm-thick $Hf_{0.2}Zr_{0.8}O_2/Al_2O_3/Hf_{0.2}Zr_{0.8}O_2$ capacitors promoted the formation of dispersed nanopolar regions and induced relaxor-like behaviour, achieving an ESD of 87.7 J $cm^{-3}$ and reduced energy losses as heat in charge-discharge process (efficiency of 68.6%)[178]. Similarly, $Al^{3+}$ doping in 15 nm-thick $Hf_{0.5}Zr_{0.5}O_2$ films were shown to promote the formation of nanodomains and consequently relaxor-like AFE behaviour, yielding ESD above 100 J $cm^{-3}$ and efficiencies over 80%[179]. These results demonstrate that nanodomain engineering in AFE HZO can suppress hysteresis and enhance efficiency. Extending this strategy to FE HZO thin films, where large polarization values are already achievable at low operating voltages, could provide a highly promising route to capacitors combining high ESD and efficiency. However, the thermal stability of polar nanoregions remains limited; therefore, particular attention must be given to the robustness against thermal and electrical cycling, which has only been demonstrated through careful control of coexisting phases[180]. Moreover, by taking advantage of the complementary properties of relaxor AFE and relaxor FE materials in electrostatic capacitors, we may be able to address both energy density and efficiency challenges.

Negative capacitance (NC) behaviour in FE HZO capacitors[167,181,182] has also attracted attention as a strategy to enhance energy storage performance. On a local scale, the ferroelectric free energy landscape allows a region where the differential capacitance is effectively negative. By coupling the ferroelectric with a dielectric in series, this locally unstable state can be stabilized: the ferroelectric operates locally in the NC state while the overall stack retains positive capacitance. This amplifies the voltage at the dielectric/ferroelectric interface, boosting total capacitance and enhancing ESD[182,183]. Through careful materials engineering, electrostatic capacitors can thus be designed with reduced hysteresis and high efficiency. For instance, bilayer structures combining FE $Hf_{0.5}Zr_{0.5}O_2$ with dielectric $Ta_2O_5$ seem to show the presence of transient NC, achieving ultrahigh ESD vales exceeding 100 J $cm^{-3}$, efficiencies above 95%, and robust endurance over $10^8$ cycles[183]. Nevertheless, the required operating voltages (>15 V) remain above the levels desired for integration with on-chip energy harvesters and low-power edge electronics. Therefore, while dielectric-ferroelectric architectures offer a promising pathway to exploit NC effects, substantial device optimization is still required to translate these concepts into practical low-voltage systems. Another promising result was achieved in $Hf_{0.5}Zr_{0.5}O_2$ thin film capacitors when the ferroelectricity disappears suddenly after polarization fatigue[184]. The authors claimed an energy density of 584 J $cm^{-3}$ with close to 100% efficiency, meaning that nearly all of the stored energy could be recovered during discharge, with negligible losses due to hysteresis or leakage, which clearly surpass the performance of 2D AFE capacitors [29,184,185]. This performance was attributed to a distorted orthorhombic phase with ordered oxygen vacancies forming a new $Pca2_1$ phase, giving rise to multiple polar regions and an ultrahigh dielectric constant exceeding 900. While the precise mechanisms and stability of such effects remain under discussion, these findings point to an intriguing and promising route toward further enhancing ESD in hafnia-based systems.

Overall, further research into dielectric material selection, layer thickness control, and multilayer design are required to optimize low-voltage and high-efficiency electrostatic capacitors. Exploring new architectures and mixed-phases, including nanocomposite configurations, may offer additional pathways to tailor performance and stability. In addition, other CMOS-compatible FE materials, such as (Al,Sc,B)N, have been theoretically predicted to achieve ESD values beyond 308 J $cm^{-3}$ with efficiencies close to 100%[186]. Exploring III-nitride compounds and their combination with FE HZO in a single heterostructure therefore represents an unexplored yet promising strategy for ES [187]. Finally, the incorporation of these optimized stacks into 3D or flexible capacitor architectures could enable compact, high-performance ES solutions suitable for edge computing devices operating within distributed intelligence systems. We additionally anticipate that dielectric capacitors can be coupled with batteries to create a HESS where the battery provides high ESD (long-term power), while the dielectric capacitor provides high power density (instantaneous response), thus addressing all requirements of edge computing.

## [H1] Low-Power In-Memory Computing

Improving the efficiency of IMC is paramount to meet the increasing functionality requirements of edge computing, while working within the limitations of the energy supply available from harvesting ambient energy sources. Box 3 discusses some aspects of efficiency quantification. Here, we discuss how emerging ferroelectric technologies can reach performance and efficiencies compatible with IPV energy generation. We focus on the potential of emerging (anti)ferroelectric materials (Box 2) to reduce the power requirements of IMC, evaluating their advantages and current drawbacks against other emerging platforms. Whilst other approaches to IMC are under investigation (*e.g.*, resistive and magnetic switching or phase-change materials), (anti)ferroelectric hafnium zirconium oxide is especially promising because it can address challenges with both IMC and energy storage, while most other IMC approaches cannot, while being fully compatible with CMOS and large-scale manufacturing.

**[H2] Emerging Memristor Technologies for Edge Computing**

In DI, memory access constitutes a substantial fraction of the total energy costs. Even without external communication, in many systems, memory access accounts for up to 90% of total energy consumption[20]. The emerging paradigm of memristor-based in-memory computing (MIMC) can offer significant advantages over conventional CMOS-based systems for enabling high DI computational performance at low power supply, by obviating the need for external memory access[17]. Memristors are a new class of fundamental electronic memory devices, wherein information can be stored as the electrical resistance of the device, rather than its charge state (like in CMOS[188]). Several memristor concepts are already commercially available, although mostly as direct replacement for conventional embedded memory, such as flash and static random-access memory (SRAM)[189]. Beyond embedded memory replacement, memristors are very appealing for neural network (NN) implementations, which have shown great promise for in-memory data analysis[190], combining memory and computing physically in the same place, and hence drastically reducing energy consumption by avoiding data movement.

In a NN, layers of neurons are linked through weighted "synaptic" connections (Fig. 3b), and the propagation of data along these constitutes matrix-vector multiplications[191]. The compute effort thereof scales quadratically with the number of neurons $N$, i.e. as $O(N^2)$, and makes up the major fraction of the overall NN activity[192]. MIMC provides a means to map such NN architectures directly onto computing hardware, reducing the scaling to O(1).

If we take the example of an autonomous drone, to provide it with edge intelligence functionality such as navigation optimization and on-the-fly learning, a system with >1 TOPS performance has been estimated to be required[193]. For more focused tasks, such as face recognition (*e.g.*, ArcFace-lite), anomaly detection (1D CNN), or voice command processing (DeepSpeech-lite) ≤1 TOPS can be sufficient[194].

There are numerous IMC systems based on SRAM or DRAM (dynamic random-access memory), where the former is more prevalent and has achieved energy efficiencies and performance densities of 10-1000 TOPS $W^{-1}$ and 0.1-1000 TOPS $mm^{-2}$, respectively[195]. The corresponding total throughput (in TOPS) is near to or sufficient for the required 1 TOPS (Fig. 3a), but at power consumptions much too high (typically in the mW to W range) to work fully autonomously by utilizing IPV.

Here, MIMC is starting to emerge as a strong alternative. It can be seen in Fig. 3a that several MIMC implementations are approaching power levels that IPV can provide, while at the same time approaching compute efficiencies required for edge intelligence. In MIMC, the synaptic weights of NN can be implemented by non-volatile memristors, obviating the need for additional external memory and leading to energy savings from not having to move data constantly. With memristive crossbar arrays, computation of synaptic operations can then be carried out with complexity scaling of O(1) instead of O($N^2$), as numerous individual digital computations and memory access operations are replaced by a single parallel analogue calculation.

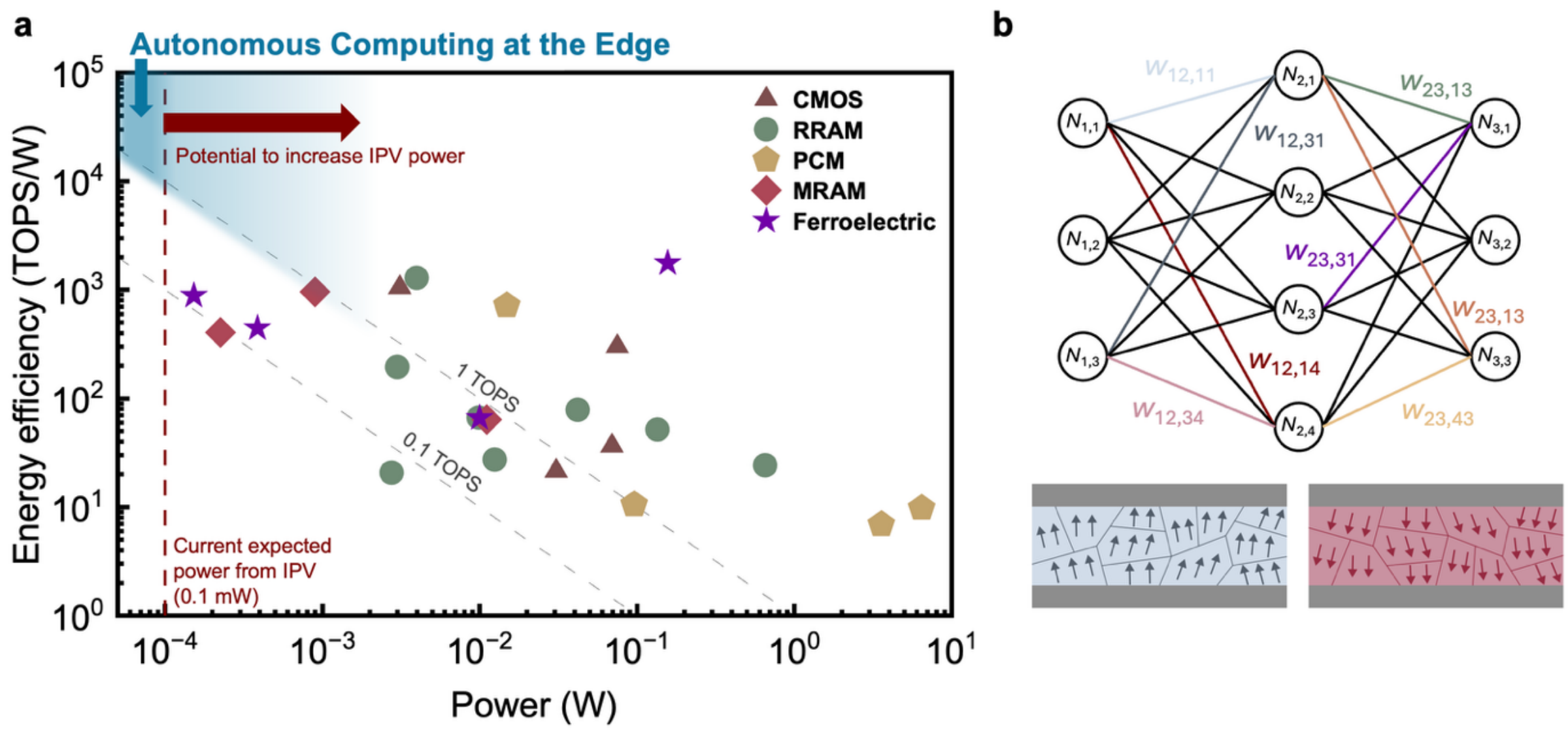

**Fig. 3 | Comparison of emerging memory technologies for in-memory computing. a,** Energy efficiency (TOPS $W^{-1}$) as a function of power consumption for representative in-memory computing implementations based on standard complementary metal-oxide-semiconductor (CMOS) components, resistive random access memory (RRAM), magnetoresistive random access memory (MRAM), phase-change memory (PCM) and ferroelectric technologies. The blue shaded region represents the ideal region, where the energy efficiency of IMC is adequate for 1 TOPS computational performance to be driven by 100 μW power provided by an IPV. Following through with the materials strategies detailed earlier could lead to IPV reliably producing >100 μW, which then relaxes the energy efficiency requirements of IMC, and this is indicated by the fading shading in the blue region sketched. All data are provided in Supplementary Table 8. **b,** Schematic of a neural network (NN) and a ferroelectric memristor with two opposite polarizations. In a NN, the different nodes $N_{i,j}$ in each layer are connected to the nodes in the next layer through 'weights' $w_{k,l}$. In a software implementation, these weights would be stored in conventional memory as model parameters, whereas in IMC, they could be stored directly as the resistance of a device, implemented for example as the controllable polarization of a ferroelectric device.

The MIMC systems in Fig. 3a provide performance efficiencies of up to a few hundred TOPS $W^{-1}$, very competitive with SRAM IMC[195], but at lower overall power. This is remarkable because these emerging technologies still generally face challenges at the materials and individual device level, while SRAM is an extremely mature base building block.

Purely at the level of TOPS $W^{-1}$ benchmarking, MIMC is getting close to the required operational area for compatibility with IPV, but need improvements by about another order of magnitude. With details at the device and materials levels discussed below, there is potential to achieve this by lowering the current levels of individual devices, hence increasing power efficiency. In addition, MIMC has the potential benefit of a smaller cell footprint than the 6T minimum of SRAM, which is critical for DI, given the limited space available on each node. However, the absence of reported area efficiency in most publications suggests that this is another area where progress is still needed, *i.e.*, further scaling of FE devices is necessary. Finally, whilst a more stable power supply through the combination of IPV and energy storage could possibly accommodate volatile DRAM/SRAM IMC designs, the non-volatile capabilities of MIMC can provide an additional advantage over CMOS.[129] Although several memristor technologies seem to be capable of similar performance, in this perspective we discuss the foundations and some outstanding challenges of FE/AFE in particular, given their advantage of being capable of both energy storage and IMC.

**Box 3 | Quantifying Computing Performance**

While the overview in Fig. 3 compares different in-memory computing systems in terms of tera operations per second per watt (TOPS $W^{-1}$) for inference, the overall performance of IMC systems can typically not be reduced to a single metric. For the same reasons, the TOPS $W^{-1}$ figure reported for any one IMC system will typically not be suitable for a general purpose comparison. Rather, reported values are often for highly specific operating and model conditions. There are several underlying conditions that may strongly relativize the reported numbers, such as hardware utilization, bit precision, or sparsity.

Hardware utilization, for example, can be strongly influenced by batching. If, for example, a single input does not utilize the entire IMC provisions, processing single inputs at a time can activate the entire hardware pipeline, but utilize only a fraction of it. With efficient batching, multiple inputs can be processed at the same time, leading to a much higher hardware utilization. For some applications, however, such as real-time visual data analysis, batching may not be feasible, as it increases the latency (time from input to result) for each individual input, as multiple inputs are effectively delayed while waiting to be accumulated into each batch. Hence, a reported TOPS $W^{-1}$ for optimal batching in this particular system may not provide the same TOPS $W^{-1}$ for a different use case. Similarly, if a particular model does not fit into one IMC array, and each input needs to be shuttled through several memory blocks, a system will be more efficient when continuously utilizing the full on-chip bandwidth, rather than just small parts of it with repeated wait times in between.

In terms of precision (number of bits used to represent data), IMC TOPS $W^{-1}$ are often reported for 8-bit precision or lower, where lower precision leads to faster processing, and hence higher TOPS $W^{-1}$, but may compromise result accuracy – again depending on application. For sparse models or inputs (containing many data with zero values), processing the sparse elements will still count as operations, increasing TOPS $W^{-1}$, while they are much more efficiently calculated than dense 64-bit components, and so sparsity can be another important aspect to relativise the reported performance.

Some reported TOPS $W^{-1}$ values are limited to the memory array itself, excluding peripheral overhead such as power-hungry analogue-to-digital converters (ADCs) to improve the reported value. This is a questionable practice, as some systems cannot function without ADCs. Such considerations should at least be clearly highlighted.

Nonetheless, reported TOPS $W^{-1}$ can provide a valuable demonstration of what is possible for optimised systems, and hence for the progress of a technology, the presentation of which is the purpose of Fig. 3.

It is worth highlighting that for centralised AI chips in data centres, in addition to the power consumption determined by the TOPS $W^{-1}$ efficiency, a large overhead may be required for constant centralized communication[8]. For DI, which ideally only needs rare communication events, there is an additional gain by obviating this required communication power.

Finally, it is worth noting that especially in hardware accelerators with external memory, the overall system performance can be limited by the memory bandwidth rather than the computational TOPS $W^{-1}$ which the system is capable of. This is illustrated by the roofline model, and IMC architectures can shift these bounds dramatically.

Although not covered in this article, when diverging from architectures based on vector-matrix multiplications, and moving to fully event-based functionalities, such as spiking neural networks, TOPS $W^{-1}$ may not be meaningful anymore. Other metrics, such as operations per synaptic event, will become more meaningful in such cases.

**[H2] Potential of Hafnium Zirconium Oxide Devices for Edge Computing**

The fundamental building blocks of FE MIMC systems include FE tunnel junctions (FTJs), field-effect transistors (FeFETs), and capacitors (FeCaps and FeRAM). At the device level, these ferroelectric-based devices show some of the best-known memory performance in terms of energy consumption on the order of fJ $bit^{-1}$, combined with ultra-fast read/write times on the order of 1-10 ns, and excellent retention of tens of years,[189] more than fulfilling the ten-year requirement for non-volatile storage. Other performance metrics depend on the device type: whilst the switching endurance of FeCaps and FeRAM (1T-1C arrays) is excellent (>$10^{15}$ cycles, matching DRAM and SRAM[189]), FeFETs and FTJs still suffer from cyclability issues[185].

The main reason is the formation of a “dead layer” (DL), by chemical reactions or intermixing between the electrodes and the ferroelectric[196], and the presence of oxygen vacancies[185]. The choice of electrode and synthesis conditions are key factors to optimize cycling endurance, but they cannot fully eliminate DLs, particularly when CMOS compatibility is a constraint[197,198]. Instead, the insertion of an interlayer such as $Al_2O_3$ can prevent the formation of a DL and resulting depolarization of the device [199], but at the cost of increasing the operating voltage and consequently the energy per bit.

Another current challenge for FTJs is the restriction to low on-currents, which impede rapid readout, and hence demonstrations of efficient array/system integration. This could be addressed, for example, by stabilizing ferroelectricity in tunnel barriers with thicknesses <5 nm by controlling the oxygen vacancy concentration and strain in the FE thin films[200]. This can simultaneously address the challenge of dielectric breakdown as one of the critical failure mechanisms, and reduce the coercive (“switching”) voltage.

In terms of systems implementations for edge computing[179], FE-based demonstrations have achieved among the best inference performance in terms of TOPS $W^{-1}$ at low power, as summarized in Fig. 3[201]. With more examples provided in Supplementary Table 8, systems implementations based on FeFETs are at the most mature stage for FE edge computing. A crossbar macro utilizing a multi-level FeFET cell demonstrated an inference efficiency of 885.4 TOPS $W^{-1}$ at a power of 0.15 mW, close to the requirement for edge intelligence. Given that high-resolution analogue-to-digital (ADC)/digital-to-analogue converters (DACs) consume as much as 80–88 % of the total energy consumption in IMC[202], the power needed for edge computing could be significantly reduced by the implementation of ADC-free arrays[202]. Initial ADC-free demonstrations have achieved efficient image edge detection at a low power consumption of ~10 fJ per operation without accuracy loss[57,203].

Most systems demonstrations are for inference tasks. But for online learning, meaning autonomous weight updates, programming considerations become important. The main challenge for the programming step is reducing the coercive ("switching") field $E_C$ as much as possible. In epitaxial HZO systems, it has been shown that grain boundaries can cause domain pinning and hence increase $E_C$[204]. Another critical materials property is the concentration of oxygen vacancies. As discussed before, these are responsible for DL formation, but oxygen vacancies can also contribute to reducing $E_C$ if a controlled amount of them can intercalate from an adjacent layer into the HZO[205]. Optimizing the oxygen vacancy concentration in the FE and adjacent layers will be critical to stabilise the FE phase and reduce $E_C$, whilst not causing domain pinning, leakage, or state degradation. The control of oxygen vacancies by La doping has, for example, led to a coercive field <1 MV cm$^{-1}$ [206], which would enable device programming voltages <1 V, highly promising to lower the power consumption of future systems integrations, which so far typically require operating voltages ≥2 V (refer to Supplementary Table 1).

Finally, a large gain may be possible by exploiting not only the variable resistive, but also capacitive properties of FE devices ('memcapacitors'), where much of the stored energy can be recovered instead of being lost to Joule heating (charge is recovered upon discharging)[207–209], and energy efficiencies of $10^3$–$10^4$ TOPS W$^{-1}$ have been predicted[207,210,211]. Outstanding challenges include expanding the memory window to support multiple programmable impedance levels per device for higher-density weight storage, and improving materials uniformity of the FE thin films[212]. This can be achieved by improving the remanent polarization, as described in the previous section, and optimizing interfacial properties to minimize charge trapping or depolarization fields.

Besides HZO, there are other FE materials that can be integrated with CMOS, including AlScN[208,210], $Bi_2SeO_5$ [211], $In_2Se_3$ and $MoS_2$ [213]. These materials possess properties very well suited to IMC applications as well, but also require further optimization to be seen as a promising solution for edge computing. Most of their studies focus on investigating their physical properties at this point, and not yet on devices that can be implemented in edge computing.

In addition to FE materials, AFE materials can also attain sizable polarization at zero external field, with read/write speeds of 2 ns and excellent endurance of $10^{12}$ cycles at 85 °C[214]. Furthermore, the multiple peaks of the coercive field in an AFE can facilitate multi-level states[215,216] and enable logic-in-memory functions[217]. This can be exploited in capacitors, FETs, or AFE FTJs, which have demonstrated, for example, an endurance of $\approx 10^8$ cycles and tunnel electroresistance exceeding 100 [218]. On the other hand, volatile polarization in AFE can be an advantage for the development of spiking neural networks. Neuromorphic devices based on Zr-rich AFE HZO transistors[219] and capacitors[220] feature good endurance ($>10^9$ switching cycles)[221] and power consumption as low as 1.36 fJ $spike^{-1}$ [220]. In addition to neuronal functionality, synaptic behaviour can be added in the same device by eliminating trapped charges in an AFE transistor[219], or by combining the functionality of AFE and FE transistors[221]. However, the energy efficiencies and performance densities of these devices have not been investigated at chip levels.

Overall, initial FE-based MIMC systems are among the closest to fulfilling the requirements for high-performance autonomous edge AI, despite having by far not been studied as long as the mature DRAM and SRAM platforms. The key outstanding challenge to improving FE performance further is controlling the oxygen vacancy distributions in the FE and adjacent films, affecting all of phase stabilisation, leakage, coercive field, and switching endurance.

## [H1] Conclusions and Outlook

In conclusion, recent advances with IPV and (anti)ferroelectric materials have now made autonomous edge computing a distinct possibility as an alternative to centralized AI. Historically, there was a large gulf in performance between the power that could be delivered by IPVs and the power required to carry out edge computing beyond basic functionalities. Recent developments with absorber materials and interface engineering for IPV, coupled with the rise of neuromorphic computing offer the opportunity to bridge this gulf, where 1 TOPS of computational performance is accessible with 0.1–1 mW of power, which can now be delivered by IPV. As of now, DSPV, OPV and PPV are the main technologies capable of producing 0.1–1 mW power with reasonable footprints of 10 $cm^2$ and common indoor illuminances of 200–1000 lx, whilst also being cost-effective to produce. This is achievable with >35% $PCE_i$, and was accomplished by tailoring the spectral response of the light-harvesting materials to the indoor artificial light source, whilst minimising bulk absorber and interface defect densities to maximise shunt resistances and minimise dark current densities to ensure high $V_{OC}$ at low illuminances. All technologies have entered into commercial production, passing industry-standard outdoor stability tests (IEC 61215), and demonstrating the capability of scaling to >10 $cm^2$. There is therefore a strong likelihood that these PV materials could fulfil the scaling and stability requirements of IPVs for edge computing. On the computing side, orders of magnitude improvements in efficiency were realized by switching from conventional GPUs ($10^{-1}$ to $10^{1}$ TOPS $W^{-1}$) to IMC architectures, where the significant energy costs of memory access are obviated, along with ADC-free peripheral circuits. Ferroelectric-based in-memory computing is particularly promising because the FE materials can be integrated with transistors (FeFETs), leading to multi-level storage at low current levels (~10 nA) which have enabled computational efficiencies reaching close to $10^3$ TOPS $W^{-1}$ at powers reaching values almost as low as 100 μW. Furthermore, ferroelectric materials, such as HZO, are capable of energy storage as electrostatic capacitors. These can deliver discharge rates >$10^4$C, far surpassing electrochemical capacitors ($10^2$–$10^3$C) and Li-ion batteries (1–2C), which are essential because of the spiking nature of energy requirements for edge computing. Whilst electrostatic capacitors have lower energy storage densities than Li-ion batteries, these can be improved through doping and engineering negative capacitance. The materials technologies are therefore in place to enable edge computing. Moving forward, co-development of the IPV, IMC and energy storage devices together, rather than in isolation, will be critical.

For IPVs, the key outlook is to move beyond simple $PCE_i$ metrics under a single artificial light source to quantify progress. Standardization of IPV testing is currently under development, with the field now converging towards a standard light source for quantifying performance that can be compared between laboratories[13,222]. The next step will be to determine and define standardized conditions for real-world indoor testing, which will be the basis for energy rating standards that enable the quantification of the power and voltage produced from IPVs under realistic conditions, *e.g.*, environments with a mixture of artificial lighting and diffuse sunlight. This will ensure absorber optimization is not biased towards one light source, and quantify the specific range of conditions that each IPV technology can be used for edge computing. Further, efforts will be required to develop accelerated degradation testing of IPV, and reveal underlying materials bottlenecks for stability. Although indoor testing conditions are generally expected to be less strenuous than outdoor 1-sun testing, the absence of trap filling from high irradiance illumination may affect degradation pathways and repeatability. Further improvements in the performance of DSPV, OPV and PPV will require engineering of the absorber layers to improve shunt resistance and mobility-lifetime product, while minimizing interface recombination. This requires, for example, improving the phase stability of WBG perovskites, reducing energetic disorder and sub-gap states in OPVs, and closing the ~0.9 V gap between the dye excited-state energy and photovoltaic in DSPV. Furthermore, there are a wide range of alternative inorganic and molecular light harvesters have theoretical efficiency limits exceeding 40% $PCE_i$, and offer benefits in terms of low toxicity, low synthetic complexity and high stability. Fundamental materials discovery and development will remain essential to realise this potential. There is also scope to explore the use of multi-junctions for IPV. Whilst these have increased complexity than single-junction devices, with less of a gain in efficiency compared to outdoor applications because of the narrower light source spectra, they enable higher open-circuit voltages. This may be beneficial in reducing the overall footprint of the IPV by requiring fewer cells to be connected in series to produce the voltages needed for operating the energy storage, IMC and peripheral circuits, but careful investigations into how current matching is affected by changes in light source will be necessary.

The discovery of ferroelectricity in $HfO_2$ and $ZrO_2$-based thin films approximately 15 years ago has transformed hafnia into one of the most promising CMOS-compatible materials for electrostatic energy storage. While remarkable progress has been made in increasing recoverable energy density, reducing operating voltage, and demonstrating three-dimensional capacitor architectures, further improvements are still required. Future work should focus on simultaneously increasing recoverable energy density and charge-discharge efficiency, while reducing coercive voltage and remanent polarization. Achieving ultra-high charge-discharge efficiency can be particularly challenging due to the hysteresis loss from ferroelectric switching. However, materials engineering strategies, including relaxor-like behaviour, mixed ferroelectric–antiferroelectric phases and negative capacitance heterostructures, could enable these advances. Whether HZO can ultimately satisfy the requirements of next-generation edge devices or whether alternative CMOS-compatible ferroelectrics, such as AlScN, will be required remains an open question. For in-memory computing, HZO-based devices have already demonstrated competitive energy efficiencies and low-power operation, yet large-scale deployment still requires improved switching endurance, lower programming voltages, greater device uniformity, and better control of oxygen vacancies and interfacial effects. Emerging ferroelectric materials, including III-nitrides and two-dimensional ferroelectrics, provide additional opportunities to overcome these limitations. Ultimately, the greatest challenge is not only improving individual devices, but also demonstrating fully integrated platforms combining IPVs, ferroelectric energy storage, and ferroelectric in-memory computing using scalable CMOS-compatible manufacturing.

Similar arguments apply to the IMC application of AFE materials. For purely inference-based systems, state retention is the most important materials and device metric, whereas endurance and programming power consumption becomes important for online learning. For both, the suppression of dead layers and control of oxygen vacancies is paramount, and for systems-level implementations, this has to be achieved reliably on the wafer scale. While the accurate control of oxygen vacancy distributions can be exploited to control phase stability, endurance, and retention, arguably, a step change could be achieved by controlling these parameters in a different way, effectively nullifying the effects of oxygen vacancies. Ultimately, AFE and FE IMC systems implementations need to be more efficient than silicon CMOS to warrant large-scale development. Research so far suggests that this is possible, and initial FE IMC demonstrations compete strongly with CMOS, achieving compute efficiencies close to $10^3$ TOPS $W^{-1}$, which can enable certain edge AI functionalities. For fully autonomous real-time and online learning capabilities, $10^4$ TOPS $W^{-1}$ will be required, where further strong improvements in the aforementioned materials properties will be critical. However, these are not the only limiting factor, as peripheral (CMOS) electronics often make up the majority of power consumption, necessitating strong interdisciplinary collaboration between materials science, electrical engineering, and computer science.

A core motivation for the development of IPV-powered edge computing is to improve the sustainability of the AI ecosystem by lowering energy and water usage, but it is equally important to address the rapid increase in critical raw minerals used and electronic waste that comes with the proliferation of data centres [4]. The latter challenge will still be encountered by DI if edge computing becomes successful. It is therefore essential to consider the future life cycle of the materials used in DI, particularly addressing the questions of how the materials required for fabricating IPV and (anti)ferroelectric materials will be securely and sustainably sourced, as well as how waste will be handled. These challenges call for a shift towards circular design principles for the IPV, energy storage device and IMC, where repairability, modularity, material transparency and recyclability are considered from the earliest stages of technology development.

These challenges call for a shift towards 'safe and sustainable by design' principles for the IPV, energy storage and IMC devices, where repairability, modularity, material transparency and recyclability are considered from the earliest stages of technology development. In this context, recent European policies, including the Circular Economy Action Plan, the Ecodesign for Sustainable Products Regulation and the Critical Raw Materials Act, provide an important framework to accelerate the transition towards more sustainable electronics. In parallel, the European Union is pursuing a closely interconnected strategy for the clean and digital transitions, with technological sovereignty, supply-chain resilience, and industrial competitiveness becoming increasingly important policy objectives with particular attention devoted to strategic sectors such as photovoltaics. Similar policy frameworks are in place globally, for example in the the United States (Chips and Science Act, National AI Initiative,

DOE Solar Energy Technologies Office), China (Made in China 2025, New Generation AI Development Plan), India (National AI Strategy) and South Korea. As DI develops, the policies to encourage and accelerate the translation of edge computing technologies should prioritize: a) targeted public procurement for real-world validation, b) enhanced merged financing mechanisms that bridge the gap from laboratory innovation to industrial deployment, with increased partnership between public research funding and private funding, c) the set up of pilot infrastructures where the manufacturing processes can be scaled and optimized.

Overall, the challenges of realizing IPV-powered DI commercially are both daunting and exhilarating: the need to simultaneously optimize performance, stability, scalability, and sustainability pushes the boundaries of what is possible in thin-film semiconductor research toward the next frontier. At the same time, there is an unprecedented opportunity to shape a technology that will directly touch billions of lives, will define the future of digital infrastructure, and will contribute to overall global sustainability. If the scientific community, industry, investors, and policymakers can align their efforts, IPV-powered DI may well become a cornerstone technology of the 21st century.

## References

1. Mehonic, A. & Kenyon, A. J. Brain-inspired computing needs a master plan. *Nature* vol. 604 255–260 Preprint at https://doi.org/10.1038/s41586-021-04362-w (2022).
2. Liu, Y. *et al.* Ferroelectric-based neuromorphic memory devices for bio-inspired computing. *Nature Reviews Electrical Engineering* **2**, 773–787 (2025).
3. *Energy and AI*. https://www.iea.org/reports/energy-and-ai (2025).
4. AI has an environmental problem. Here’s what the world can do about that. *UN Environment Programme* https://www.unep.org/news-and-stories/story/ai-has-environmental-problem-heres-what-world-can-do-about (2024).
5. Li, P., Yang, J., Islam, M. A. & Ren, S. Making AI Less ‘Thirsty’. *Commun. ACM* **68**, 54–61 (2025).

6. Shirvanimoghaddam, M. *et al.* Towards a Green and Self-Powered Internet of Things Using Piezoelectric Energy Harvesting. *IEEE Access* **7**, 94533–94556 (2019).

7. Sek Chai. How Edge Computing Can Solve AI's Energy Crisis. *builtin* https://builtin.com/artificial-intelligence/edge-ai-energy-solution (2025).

8. Mathews, I., Kantareddy, S. N., Buonassisi, T. & Peters, I. M. Technology and Market Perspective for Indoor Photovoltaic Cells. *Joule* **3**, 1415–1426 (2019).

9. Grey, C. P. & Hall, D. S. Prospects for lithium-ion batteries and beyond—a 2030 vision. *Nature Communications* vol. 11 Preprint at https://doi.org/10.1038/s41467-020-19991-4 (2020).

10. Harrop, P. *Battery Elimination in Electronics and Electrical Engineering 2018-2028*. (2017).

11. Pecunia, V., Occhipinti, L. G. & Hoye, R. L. Z. Emerging Indoor Photovoltaic Technologies for Sustainable Internet of Things. *Adv. Energy Mater.* **11**, 2100698 (2021).

12. Grandhi, G. K. *et al.* Promises and challenges of indoor photovoltaics. *Nature Reviews Clean Technology* **1**, 132–147 (2025).

13. Hoye, R. L. Z. *et al.* Reaching a consensus on indoor photovoltaics testing. *Joule* **9**, 102127 (2025).

14. Chakraborty, A. *et al.* Photovoltaics for indoor energy harvesting. *Nano Energy* vol. 128 Preprint at https://doi.org/10.1016/j.nanoen.2024.109932 (2024).

15. Müller, D. *et al.* Indoor Photovoltaics for the Internet-of-Things - A Comparison of State-of-the-Art Devices from Different Photovoltaic Technologies. *ACS Appl. Energy Mater.* **6**, 10404–10414 (2023).

16. Michaels, H. *et al.* Emerging indoor photovoltaics for self-powered and self-aware IoT towards sustainable energy management. *Chem. Sci.* **14**, 5350–5360 (2023).

17. Kim, T. H. *et al.* Record indoor performance of organic photovoltaics with long-term stability enabled by self-assembled monolayer-based interface management. *Nano Energy* **112**, 108429 (2023).

18. Ma, Q. *et al.* One-step dual-additive passivated wide-bandgap perovskites to realize 44.72%-efficient indoor photovoltaics. *Energy Environ. Sci.* **17**, 1637–1644 (2024).

19. NVIDIA Expands its AI Edge Capabilities. *Edge AI + Vision Alliance* (2026).

20. Mutlu, O., Olgun, A. & Yüksel, I. E. Memory-Centric Computing: Solving Computing's Memory Problem. in *2025 IEEE International Memory Workshop, IMW 2025 - Proceedings* (Institute of Electrical and Electronics Engineers Inc., 2025). doi:10.1109/IMW61990.2025.11026935.

21. Sangwan, V. K. & Hersam, M. C. Neuromorphic nanoelectronic materials. *Nature Nanotechnology* vol. 15 517–528 Preprint at https://doi.org/10.1038/s41565-020-0647-z (2020).

22. Yang, J. J., Strukov, D. B. & Stewart, D. R. Memristive devices for computing. *Nature Nanotechnology* vol. 8 13–24 Preprint at https://doi.org/10.1038/nnano.2012.240 (2013).

23. Waser, R., Dittmann, R., Staikov, G. & Szot, K. Redox-Based Resistive Switching Memories – Nanoionic Mechanisms, Prospects, and Challenges. *Advanced Materials* **21**, 2632–2663 (2009).

24. Chanthbouala, A. *et al.* Solid-state memories based on ferroelectric tunnel junctions. *Nat. Nanotechnol.* **7**, 101–104 (2012).

25. Chanthbouala, A. *et al.* A ferroelectric memristor. *Nat. Mater.* **11**, 860–864 (2012).

26. Boyn, S. *et al.* Learning through ferroelectric domain dynamics in solid-state synapses. *Nat. Commun.* **8**, 14736 (2017).

27. Diao, C. *et al.* Overviews of dielectric energy storage materials and methods to improve energy storage density. *Journal of Materials Science: Materials in Electronics* **33**, 21199–21222 (2022).

28. Palneedi, H., Peddigari, M., Hwang, G.-T., Jeong, D.-Y. & Ryu, J. High-Performance Dielectric Ceramic Films for Energy Storage Capacitors: Progress and Outlook. *Adv. Funct. Mater.* **28**, 1803665 (2018).

29. Silva, J. P. B., Sekhar, K. C., Pan, H., Macmanus-Driscoll, J. L. & Pereira, M. Advances in Dielectric Thin Films for Energy Storage Applications, Revealing the Promise of Group IV Binary Oxides. *ACS Energy Letters* vol. 6 2208–2217 Preprint at https://doi.org/10.1021/acsenergylett.1c00313 (2021).

30. Jayakrishnan, A. R. *et al.* Are lead-free relaxor ferroelectric materials the most promising candidates for energy storage capacitors? *Progress in Materials Science* vol. 132 Preprint at https://doi.org/10.1016/j.pmatsci.2022.101046 (2023).

31. Cheng, Y. *et al.* Overcoming Energy Storage-Loss Trade-Offs in Polymer Dielectrics Through the Synergistic Tuning of Electronic Effects in π-Conjugated Polystyrenes. *Advanced Science* **12**, 2415738 (2025).

32. Xie, B. *et al.* Constructing superrelaxor critical state towards giant energy storage in lead-free dielectric ceramics. *Nature Communications* **17**, (2026).

33. Xiong, X. *et al.* Giant Energy-Storage in Pb-Free Relaxor Ferroelectrics via Atomic-level Design. *Advanced Materials* **38**, e17815 (2026).

34. Catalan, G., Gruverman, A., Íñiguez-González, J., Meier, D. & Trassin, M. A modern perspective on antiferroelectrics. *Nature Materials* Preprint at https://doi.org/10.1038/s41563-026-02483-z (2026).

35. Feng, Y.-P. *et al.* Direct Observation of Electric-Field-Driven Phase Transitions Associated with Energy Storage in Antiferroelectric Films. *Advanced Science* **13**, e17897 (2026).

36. Dai, K. *et al.* Tailoring antiferroelectric phase transitions in ultrahigh energy storage ceramics. *Phys. Rev. B* **113**, 14104 (2026).

37. Filho, C. P. *et al.* A Systematic Literature Review on Distributed Machine Learning in Edge Computing. *Sensors* vol. 22 Preprint at https://doi.org/10.3390/s22072665 (2022).

38. Rosendo, D., Costan, A., Valduriez, P. & Antoniu, G. Distributed intelligence on the Edge-to-Cloud Continuum: A systematic literature review. *J. Parallel Distrib. Comput.* **166**, 71–94 (2022).

39. Shi, W., Cao, J., Zhang, Q., Li, Y. & Xu, L. Edge Computing: Vision and Challenges. *IEEE Internet Things J.* **3**, 637–646 (2016).

40. Pioli, L., De Macedo, D. D. J., Costa, D. G. & Dantas, M. A. R. Intelligent Edge-powered Data Reduction: A Systematic Literature Review. *ACM Computing Surveys* vol. 56 Preprint at https://doi.org/10.1145/3656338 (2024).

41. Ruan, R. AI-Based intelligent edge autonomous control technology for energy conservation and carbon reduction in smart public buildings. in *Proceedings of 2025 8th International Conference on Computer Information Science and Artificial Intelligence, CISAI 2025* 784–790 (Association for Computing Machinery, Inc, 2025). doi:10.1145/3773365.3773489.

42. Png, E. *et al.* An internet of things upgrade for smart and scalable heating, ventilation and air-conditioning control in commercial buildings. *Appl. Energy* **239**, 408–424 (2019).

43. Fang Quanmin, Q., Feng, Z. & Editors, Q. *Advances in Intelligent Systems and Computing*. vol. 890 (2018).

44. Jayakrishnan, A. R. *et al.* Hard way or hardware? Taking the heat out of AI. *APL Machine Learning* **3**, (2025).

45. *Executive Summary*.

46. Rengaraj, S. AI Is Moving to MCUs, Here's What's Driving It. *multicoreware* (2026).

47. Pazmiño-Ortiz, L., Cuenca-Sánchez, A. & Loarte-Cajamarca, B. Distributed Intelligence in the Artificial Intelligence of Things: A Review of Artificial Intelligence Workload Placement Across the Device-Edge-Fog-Cloud Continuum. *Future Internet* **18**, 296 (2026).

48. Pecunia, V., Occhipinti, L. G. & Hoye, R. L. Z. Emerging Indoor Photovoltaic Technologies for Sustainable Internet of Things. *Adv. Energy Mater.* **11**, 2100698 (2021).

49. Kay, A. M. *et al.* The Thermodynamic Limit of Indoor Photovoltaics Based on Energetically-Disordered Molecular Semiconductors. *Solar RRL* **7**, (2023).

50. Seunarine, K. *et al.* Light power resource availability for energy harvesting photovoltaics for self-powered IoT. *JPhys Energy* **6**, 015018 (2024).

51. Michaels, H., Benesperi, I. & Freitag, M. Challenges and prospects of ambient hybrid solar cell applications. *Chem. Sci.* **12**, 5002–5015 (2021).

52. David Garrett. NDP200 tinyML Vision Processing. in *2022 tinyML Summit* (2022).

53. Qin, D. *et al.* MobileNetV4 -- Universal Models for the Mobile Ecosystem. http://arxiv.org/abs/2404.10518 (2024).

54. Utlralytics YOLO26. (2026).

55. Soliman, T. *et al.* First demonstration of in-memory computing crossbar using multi-level Cell FeFET. *Nature Communications* **14**, (2023).

56. Ghosh, A. *et al. Approximate ADCs for In-Memory Computing*. https://arxiv.org/abs/2408.06390 (2026).

57. Chen, J. *et al.* Low-power edge detection based on ferroelectric field-effect transistor. *Nature Communications* **16**, (2025).

58. Russ, B., Glaudell, A., Urban, J. J., Chabinyc, M. L. & Segalman, R. A. Organic thermoelectric materials for energy harvesting and temperature control. *Nature Reviews Materials* vol. 1 Preprint at https://doi.org/10.1038/natrevmats.2016.50 (2016).

59. Mahapatra, S. Das *et al.* Piezoelectric Materials for Energy Harvesting and Sensing Applications: Roadmap for Future Smart Materials. *Advanced Science* vol. 8 Preprint at https://doi.org/10.1002/advs.202100864 (2021).

60. Costa, C. M. *et al.* Smart and Multifunctional Materials Based on Electroactive Poly(vinylidene fluoride): Recent Advances and Opportunities in Sensors, Actuators, Energy, Environmental, and Biomedical Applications. *Chemical Reviews* vol. 123 11392–11487 Preprint at https://doi.org/10.1021/acs.chemrev.3c00196 (2023).

61. Cheng, T., Shao, J. & Wang, Z. L. Triboelectric nanogenerators. *Nature Reviews Methods Primers* **3**, (2023).

62. Hu, F. *et al.* Ultrabroadband nonlinear Hall rectifier using SnTe. *Nat. Nanotechnol.* **20**, 1588–1595 (2025).

63. Ho, J. K. W., Yin, H. & So, S. K. From 33% to 57%-An elevated potential of efficiency limit for indoor photovoltaics. *J. Mater. Chem. A Mater.* **8**, 1717–1723 (2020).

64. Dong, C. *et al.* Lycopene-Based Bionic Membrane for Stable Perovskite Photovoltaics. *Adv. Funct. Mater.* **31**, 2011242 (2021).

65. Gong, O. Y. *et al.* Van der Waals Force-Assisted Heat-Transfer Engineering for Overcoming Limited Efficiency of Flexible Perovskite Solar Cells. *ACS Energy Lett.* **7**, 2893–2903 (2022).

66. Xu, J. *et al.* Improved humidity resistance and bending stability of flexible perovskite photovoltaic module by Incorporating polymerized networks. *Chemical Engineering Journal* **510**, 161624 (2025).

67. Gao, H. *et al.* Band Gap Adjustable Antimony Selenosulfide Indoor Photovoltaics with 20% Efficiency. *Solar RRL* **8**, 2400389 (2024).

68. An, X. *et al.* Photovoltaic Absorber “Glues” for Efficient Bifacial Selenium Photovoltaics. *Angew. Chem. Int. Ed.* **64**, e202505297 (2025).

69. Jebin, P. R., George, A. S., Mishra, R. K., John, J. & Soman, S. Enhanced indoor photovoltaic efficiency of 40% in dye-sensitized solar cells using cocktail starburst triphenylamine dyes and dual-species copper electrolyte. *J. Mater. Chem. A Mater.* **12**, 32721–32734 (2024).

70. Bi, P. *et al.* A High-Performance Nonfused Wide-Bandgap Acceptor for Versatile Photovoltaic Applications. *Advanced Materials* **34**, (2022).

71. Liu, C. Y. *et al.* Flexible multijunction solar cells embedded inside smart dust modules for outdoor applications to Smart Grids. *Appl. Energy* **306**, (2022).

72. *IEC TS 62607-7-2:2023 Nanomanufacturing ; Key Control Characteristics. Part 7-2, Nano-Enabled Photovoltaics -- Device Evaluation Method for Indoor Light*. (2023).

73. Ma, Q. *et al.* One-step dual-additive passivated wide-bandgap perovskites to realize 44.72%-efficient indoor photovoltaics. *Energy Environ. Sci.* **17**, 1637–1644 (2024).

74. Best practices for measuring the performance and stability of indoor photovoltaic devices. *Nat. Energy* (2026).

75. Schulte, K. L. *et al.* A Path to III-V Photovoltaic Cost Reduction Combining GaAs (211) Spalling with Halide Vapor Phase Epitaxy. in *2025 IEEE 53rd Photovoltaic Specialists Conference (PVSC)* 0362–0362 (Institute of Electrical and Electronics Engineers (IEEE), Montreal, 2025). doi:10.1109/pvsc59419.2025.11132574.

76. Kay, A. M., Riley, D. B., Burwell, G. & Meredith, P. Thermodynamic principles for optimizing multi-junction photovoltaics—Exemplified for perovskite-based indoor photovoltaics. *APL Energy* **3**, (2025).

77. Li, Y. *et al.* In Situ Formation of 2D Perovskite Seeding for Record-Efficiency Indoor Perovskite Photovoltaic Devices. *Advanced Materials* **36**, (2024).

78. Wang, W. *et al.* High-performance organic photovoltaic cells under indoor lighting enabled by suppressing energetic disorders. *Joule* **7**, 1067–1079 (2023).

79. Michaels, H. *et al.* Dye-sensitized solar cells under ambient light powering machine learning: towards autonomous smart sensors for the internet of things. *Chem. Sci.* **11**, 2895–2906 (2020).

80. Burwell, G. *et al.* Scaling Considerations for Organic Photovoltaics for Indoor Applications. *Solar RRL* **6**, (2022).

81. Park, S. Y. *et al.* Alkoxybenzothiadiazole-Based Fullerene and Nonfullerene Polymer Solar Cells with High Shunt Resistance for Indoor Photovoltaic Applications. *ACS Appl. Mater. Interfaces* **10**, 3885–3894 (2018).

82. Müller, M. F., Freunek, M. & Reindl, L. M. Maximum efficiencies of indoor photovoltaic devices. *IEEE J. Photovolt.* **3**, 59–64 (2013).

83. Lübke, D., Hartnagel, P., Angona, J. & Kirchartz, T. Comparing and Quantifying Indoor Performance of Organic Solar Cells. *Adv. Energy Mater.* **11**, 2101474 (2021).

84. Park, S. Y., Labanti, C., Luke, J., Chin, Y. C. & Kim, J. S. Organic Bilayer Photovoltaics for Efficient Indoor Light Harvesting. *Adv. Energy Mater.* **12**, 2103237 (2022).

85. Placidi, M. *et al.* Benchmarking Inorganic Thin-Film Photovoltaics Technologies for Indoor Applications. *Solar RRL* **9**, (2025).

86. Mattia, G. P. & Beraldi, R. Lifespan and energy-oriented load balancing algorithms across sets of nodes in Green Edge Computing. in *Proceedings - 2023 IEEE Cloud Summit, Cloud Summit 2023* 41–48 (Institute of Electrical and Electronics Engineers Inc., 2023). doi:10.1109/CloudSummit57601.2023.00013.

87. Asuo, I. M. *et al.* Stability of Perovskite Indoor Photovoltaics: A Focused Review and a Call for Standardized Stability Reporting. *Advanced Energy Materials* Preprint at https://doi.org/10.1002/aenm.202506091 (2026).

88. Khenkin, M. V. *et al.* Consensus statement for stability assessment and reporting for perovskite photovoltaics based on ISOS procedures. *Nat. Energy* **5**, 35–49 (2020).

89. Kay, A. M. *et al.* Optimising photovoltaic modules for indoor energy-harvesting systems. *JPhys Energy* **7**, (2025).

90. O'Sullivan, J. *et al.* Transparent conducting electrodes for perovskite–silicon tandem solar cells. *Nature Reviews Physics* Preprint at https://doi.org/10.1038/s42254-026-00938-5 (2026).

91. Freitag, M. *et al.* Dye-sensitized solar cells for efficient power generation under ambient lighting. *Nat. Photonics* **11**, 372–378 (2017).

92. Zhang, D. *et al.* A molecular photosensitizer achieves a V oc of 1.24 V enabling highly efficient and stable dye-sensitized solar cells with copper(II/I)-based electrolyte. *Nature Communications* **12**, (2021).

93. Ren, Y. *et al.* Hydroxamic acid pre-adsorption raises the efficiency of cosensitized solar cells. *Nature* **613**, 60–65 (2023).

94. Meethal, S. M. *et al.* Asymmetric dual species copper(ii/i) electrolyte dye-sensitized solar cells with 35.6% efficiency under indoor light. *J. Mater. Chem. A Mater.* **12**, 1081–1093 (2023).

95. Wong, X. W. *et al.* Quasi-Solid-State Dye-Sensitized Solar Cells with over 36% Efficiency Achieved by a Direct-Contact Cell Structure and In Situ Gelation of Copper-Based Electrolytes. *ACS Sustain. Chem. Eng.* **13**, 3432–3440 (2025).

96. Cao, Y. *et al.* 11% efficiency solid-state dye-sensitized solar cells with copper(II/I) hole transport materials. *Nature Communications* **8**, (2017).

97. Solid-State Dye-Sensitized Solar Cell. *Ricoh* https://www.ricoh.com/technology/tech/066_dssc.

98. POWERFOYLE. *EXEGER* https://www.exeger.com/powerfoyle/.

99. Cui, Y. *et al.* Wide-gap non-fullerene acceptor enabling high-performance organic photovoltaic cells for indoor applications. *Nat. Energy* **4**, 768–775 (2019).

100. Ma, L. K. *et al.* High-Efficiency Indoor Organic Photovoltaics with a Band-Aligned Interlayer. *Joule* **4**, 1486–1500 (2020).

101. Zarrabi, N. *et al.* Charge-generating mid-gap trap states define the thermodynamic limit of organic photovoltaic devices. *Nature Communications* **11**, (2020).

102. Yan, J., Rezasoltani, E., Azzouzi, M., Eisner, F. & Nelson, J. Influence of static disorder of charge transfer state on voltage loss in organic photovoltaics. *Nature Communications* **12**, (2021).

103. Chen, C.-Y. *et al.* Perovskite Photovoltaics for Dim-Light Applications. *Adv. Funct. Mater.* **25**, 7064–7070 (2015).

104. Wu, M.-J. *et al.* Bandgap Engineering Enhances the Performance of Mixed-Cation Perovskite Materials for Indoor Photovoltaic Applications. *Adv. Energy Mater.* **9**, 1901863 (2019).

105. Xu, J. *et al.* Key Parameters and Thresholds Values for Obtaining High Performance Perovskite Solar Cells Indoors from Full Br Compositional and Bandgap Engineering. *ACS Applied Energy Materials* vol. 6 10215–10224 Preprint at https://doi.org/10.1021/acsaem.2c03394 (2023).

106. Cheng, R. *et al.* Tailoring Triple-Anion Perovskite Material for Indoor Light Harvesting with Restrained Halide Segregation and Record High Efficiency Beyond 36%. *Adv. Energy Mater.* **9**, 1901980 (2019).

107. Degani, M. *et al.* Compositional Gradient of Mixed Halide 2D Perovskite Interface Boosts Outdoor Stability of Highly Efficient Perovskite Solar Cells. *Adv. Energy Mater.* **15**, (2025).

108. Shen, C., Ye, T., Yang, P. & Chen, G. All–Inorganic Perovskite Solar Cells: Defect Regulation and Emerging Applications in Extreme Environments. *Advanced Materials* vol. 36 Preprint at https://doi.org/10.1002/adma.202401498 (2024).

109. Mei, J. & Yan, F. Recent Advances in Wide-Bandgap Perovskite Solar Cells. *Advanced Materials* vol. 37 Preprint at https://doi.org/10.1002/adma.202418622 (2025).

110. Jiang, B. H. *et al.* Enhancing the Efficiency of Indoor Perovskite Solar Cells through Surface Defect Passivation with Coplanar Heteroacene Cored A–D–A-type Molecules. *Adv. Funct. Mater.* **34**, (2024).

111. Gnanasekaran, P. *et al.* Interfacial Engineering Using C-3 Alkyl Linker-Based Carbazole-Derived SAM Layers to Achieve 41.77% Indoor Efficiency in Wide-Bandgap Perovskite Solar Cells. *Small* **21**, (2025).

112. Pallotta, R. *et al.* Reducing the MAPbI3 microstrain by fast crystallization. *Joule* **9**, (2025).

113. Zanetta, A. *et al.* Vertically oriented low-dimensional perovskites for high-efficiency wide band gap perovskite solar cells. *Nature Communications* **15**, (2024).

114. Scardina, A. *et al.* Spacer cation design: promoting vertical orientation in layered perovskites. *EES Solar* **2**, 402–412 (2026).

115. Li, B. *et al.* Closed-loop manufacturing for sustainable perovskite photovoltaics. *Nature Reviews Materials* vol. 11 10–25 Preprint at https://doi.org/10.1038/s41578-025-00872-5 (2026).

116. Chen, C.-Y. *et al.* Vacuum-deposited perovskite photovoltaics for highly efficient environmental light energy harvesting. *J. Mater. Chem. A Mater.* **7**, 3612–3617 (2019).

117. Hou, X. *et al.* Indoor application of emerging photovoltaics - Progress, challenges and perspectives. *Journal of Materials Chemistry A* vol. 8 21503–21525 Preprint at https://doi.org/10.1039/d0ta06950g (2020).

118. *Epishine Onecell Indoor Solar Cells Data Sheet*. (2026).

119. Bellini, E. GCL claims 19.04% efficiency for perovskite solar panel. *PV Magazine* https://www.pv-magazine.com/2025/04/03/gcl-claims-19-04-efficiency-for-perovskite-solar-panel/ (2025).

120. Jiang, E. *et al.* Side Connection for High-Efficiency Organic Photovoltaic Modules for Indoor Applications: Shunt Reduction and Performance Improvement. *Solar RRL* **8**, (2024).

121. Beket, G. *et al.* Understanding and Addressing the Performance Asymmetry Issue in Semitransparent Laminated Organic Photovoltaic Devices. *Adv. Funct. Mater.* **35**, (2025).

122. Mosquera-Lois, I. *et al.* Multifaceted nature of defect tolerance in halide perovskites and emerging semiconductors. *Nat. Rev. Chem.* **9**, 287–304 (2025).

123. Chen, X. *et al.* Additive engineering for Sb2S3 indoor photovoltaics with efficiency exceeding 17%. *Light Sci. Appl.* **13**, 281 (2024).

124. Xu, J. *et al.* Progress in perovskite indoor photovoltaics. *EES Solar* Preprint at https://doi.org/10.1039/d5el00210a (2026).

125. Wen, X. *et al.* Illumination-assisted annealing enables selenium solar cells with open-circuit voltage over 1 V and efficiency exceeding 10%. *Nat. Energy* https://doi.org/10.1038/s41560-025-01939-x (2026) doi:10.1038/s41560-025-01939-x.

126. Lu, W. *et al.* Ordering one-dimensional chains enables efficient selenium photovoltaics. *Joule* **8**, 1430–1442 (2024).

127. Lamminen, N. *et al.* Surpassing the 10% Efficiency Threshold in Perovskite-Inspired Indoor Photovoltaics. *ACS Energy Lett.* **10**, 3415–3418 (2025).

128. Nasyori, A. *et al.* In ambient air processed Cu2ZnSnS4 absorber layers from DMSO-based precursors: enhanced efficiency via device post-annealing. *J. Mater. Chem. A Mater.* **13**, 30167–30179 (2025).

129. Jimenez-Arguijo, A. *et al.* Guidelines for the design of efficient and injection-resilient thin film photovoltaic converters for indoor applications. *Journal of Physics: Energy* **7**, 045008 (2025).

130. Gong, Y. *et al.* Attaining 15.1% Efficiency in Cu2ZnSnS4 Solar Cells Under Indoor Conditions Through Sodium and Lithium Codoping. *Solar RRL* **9**, 2400756 (2025).

131. Yan, B. *et al.* Indoor photovoltaics awaken the world's first solar cells. **8**, eadc9923 (2022).

132. Kavanagh, S. R. *et al.* Intrinsic point defect tolerance in selenium for indoor and tandem photovoltaics. *Energy Environ. Sci.* **18**, 4431–4446 (2025).

133. Hussien, H. *et al.* Interface engineering approach of in-air-processed Sb2S3solar cells enabling 7.5% AM 1.5G device efficiency and an 18% indoor milestone performance. *J. Mater. Chem. A Mater.* **13**, 37215–37231 (2025).

134. Tang, R. *et al.* Hydrothermal deposition of antimony selenosulfide thin films enables solar cells with 10% efficiency. *Nat. Energy* **5**, 587–595 (2020).

135. Hussien, H. *et al.* Interface engineering approach of in-air-processed Sb2S3solar cells enabling 7.5% AM 1.5G device efficiency and an 18% indoor milestone performance. *J. Mater. Chem. A Mater.* **13**, 37215–37231 (2025).

136. Yao, Z. *et al.* Homogeneous/Inhomogeneous-Structured Dielectrics and their Energy-Storage Performances. *Advanced Materials* **29**, 1601727 (2017).

137. Dunn, B., Kamath, H. & Tarascon, J.-M. Electrical Energy Storage for the Grid: A Battery of Choices. *Science (1979).* **334**, 928–935 (2011).

138. Kyeremateng, N. A., Brousse, T. & Pech, D. Microsupercapacitors as miniaturized energy-storage components for on-chip electronics. *Nature Nanotechnology* vol. 12 7–15 Preprint at https://doi.org/10.1038/nnano.2016.196 (2017).

139. Dinh, K. H., Roussel, P. & Lethien, C. Advances on Microsupercapacitors: Real Fast Miniaturized Devices toward Technological Dreams for Powering Embedded Electronics? *ACS Omega* vol. 8 8977–8990 Preprint at https://doi.org/10.1021/acsomega.2c07549 (2023).

140. Huang, W. *et al.* Highly Reliable Polymer Dielectric Materials for High-Temperature Energy Storage Capacitors. *Adv. Funct. Mater.* **36**, e75109 (2026).

141. Yuan, Q. *et al.* Ceramic-based dielectrics for electrostatic energy storage applications: Fundamental aspects, recent progress, and remaining challenges. *Chemical Engineering Journal* vol. 446 Preprint at https://doi.org/10.1016/j.cej.2022.136315 (2022).

142. Zhang, T. *et al.* Recent progress in polymer dielectric energy storage: From film fabrication and modification to capacitor performance and application. *Progress in Materials Science* vol. 140 Preprint at https://doi.org/10.1016/j.pmatsci.2023.101207 (2023).

143. *Battery Energy Storage Systems Market (2026 - 2033)*. (2026).

144. Thakur, J., Phogat, P., Shreya, Jha, R. & Singh, S. Non-rechargeable batteries: a review of primary battery technology and future trends. *Physical Chemistry Chemical Physics* vol. 27 4045–4077 Preprint at https://doi.org/10.1039/d4cp04614e (2025).

145. Hasan, K., Tom, N. & Yuce, M. R. Navigating Battery Choices in IoT: An Extensive Survey of Technologies and Their Applications. *Batteries* vol. 9 Preprint at https://doi.org/10.3390/batteries9120580 (2023).

146. Nabavi, S. M. H., Hajforoosh, S. & Nabavi, S. M. A. Battery Selection for Energy-Efficient IoT Devices: A Comparative Study of Longevity Across Environmental Conditions. *International Journal of Electrochemistry* **2025**, (2025).

147. Kharabati, S. & Saedodin, S. A systematic review of thermal management techniques for electric vehicle batteries. *Journal of Energy Storage* vol. 75 Preprint at https://doi.org/10.1016/j.est.2023.109586 (2024).

148. Zhou, Q. *et al.* On-Device Learning Systems for Edge Intelligence: A Software and Hardware Synergy Perspective. *IEEE Internet of Things Journal* vol. 8 11916–11934 Preprint at https://doi.org/10.1109/JIOT.2021.3063147 (2021).

149. Jacob, A. *et al.* Operational Limitations, Degradation Mechanisms, and Safety Challenges in Lithium-Ion Batteries for Electric Vehicle Applications. in *2025 10th International Conference on Communication and Electronics Systems (ICCES)* 103–109 (2025). doi:10.1109/ICCES67310.2025.11337143.

150. Zuo, W. *et al.* Battery-Supercapacitor Hybrid Devices: Recent Progress and Future Prospects. *Advanced Science* vol. 4 Preprint at https://doi.org/10.1002/advs.201600539 (2017).

151. Liu, L., Zhang, X., Liu, Y. & Gong, X. Electrochemical Energy Storage Devices–Batteries, Supercapacitors, and Battery-Supercapacitor Hybrid Devices. *ACS Applied Electronic Materials* vol. 7 2233–2270 Preprint at https://doi.org/10.1021/acsaelm.5c00069 (2025).

152. Hu, Q., Xie, S. & Zhang, J. Data-based power management control for battery supercapacitor hybrid energy storage system in solar DC-microgrid. *Sci. Rep.* **14**, (2024).

153. Rizoug, N., Mesbahi, T., Sadoun, R., Bartholomeüs, P. & Le Moigne, P. Development of new improved energy management strategies for electric vehicle battery/supercapacitor hybrid energy storage system. *Energy Effic.* **11**, 823–843 (2018).

154. Xu, Y. *et al.* Recent progress in electrode materials for micro-supercapacitors. *iScience* vol. 27 Preprint at https://doi.org/10.1016/j.isci.2024.108786 (2024).

155. Flores-Diaz, N. *et al.* Unlocking high-performance photocapacitors for edge computing in low-light environments. *Energy Environ. Sci.* **18**, 4704–4716 (2025).

156. Zubairi, H., Lu, Z., Zhu, Y., Reaney, I. M. & Wang, G. Current development, optimisation strategies and future perspectives for lead-free dielectric ceramics in high field and high energy density capacitors. *Chemical Society Reviews* vol. 53 10761–10790 Preprint at https://doi.org/10.1039/d4cs00536h (2024).

157. Fang, H. *et al.* Ferroelectric Devices for In-Memory and In-Sensor Computing. *Advanced Science* Preprint at https://doi.org/10.1002/advs.202600007 (2026).

158. Yuan, M., Zhang, Z., Li, C., Wang, L. & Sun, L. 2D Ferroelectrics and Antiferroelectrics for Neuromorphic Computing Devices. *Advanced Functional Materials* Preprint at https://doi.org/10.1002/adfm.76227 (2026).

159. Multilayer Ceramic Chip Capacitors: C3216X8L0G156M160AC. *TDK* (2026).

160. Multilayer Ceramic Chip Capacitors: C0510X7S0E105M030BC. *TDK* (2026).

161. Silicon Capacitors: EMSC Series. *Murata Manufacturing Co., Ltd*.

162. Silicon Capacitors: LPSC Series. *Murata Manufacturing Co., Ltd*.

163. Schroeder, U., Park, M. H., Mikolajick, T. & Hwang, C. S. The fundamentals and applications of ferroelectric HfO2. *Nature Reviews Materials* vol. 7 653–669 Preprint at https://doi.org/10.1038/s41578-022-00431-2 (2022).

164. Noheda, B., Nukala, P. & Acuautla, M. Lessons from hafnium dioxide-based ferroelectrics. *Nat. Mater.* **22**, 562–569 (2023).

165. Ganser, R. *et al.* Mechanism of Antiferroelectricity in Polycrystalline ZrO2. *Adv. Funct. Mater.* **34**, (2024).

166. Hoffmann, M. *et al.* Antiferroelectric negative capacitance from a structural phase transition in zirconia. *Nat. Commun.* **13**, (2022).

167. Cheema, S. S. *et al.* Giant energy storage and power density negative capacitance superlattices. *Nature* **629**, 803–809 (2024).

168. Zheng, G. *et al.* Plasma-Enhanced Atomic Layer-Deposited Ti,Si-Doped ZrO2 Antiferroelectric Films for Energy Storage Capacitors. *ACS Appl. Electron. Mater.* **5**, 5907–5915 (2023).

169. Böscke, T. S., Müller, J., Bräuhaus, D., Schröder, U. & Böttger, U. Ferroelectricity in hafnium oxide thin films. *Appl. Phys. Lett.* **99**, (2011).

170. Liu, L. *et al.* Large ferroelectric polarization and high dielectric constant in HfO2-based thin films via Hf0.5Zr0.5O2/ZrO2 nanobilayer engineering. *Journal of Materiomics* **11**, (2025).

171. Yun, Y. *et al.* Intrinsic ferroelectricity in Y-doped HfO2 thin films. *Nat. Mater.* **21**, 903–909 (2022).

172. *2023 IEEE Symposium on VLSI Technology and Circuits : 11-16 June 2023, Kyōto, Japan*. (IEEE, 2023).

173. Guo, J. *et al.* Rhombohedral R3 Phase of Mn-Doped Hf0.5Zr0.5O2 Epitaxial Films with Robust Ferroelectricity. *Advanced Materials* **36**, (2024).

174. Wang, Y. *et al.* A stable rhombohedral phase in ferroelectric Hf(Zr)1+xO2 capacitor with ultralow coercive field. *Science (1979).* **381**, 558–563 (2023).

175. Xu, B., Lomenzo, P. D., Kersch, A., Mikolajick, T. & Schroeder, U. Influence of Si-Doping on 45 nm Thick Ferroelectric ZrO2Films. *ACS Appl. Electron. Mater.* **4**, 3648–3654 (2022).

176. Pan, H. *et al.* Ultrahigh energy storage in superparaelectric relaxor ferroelectrics. *Science (1979).* **374**, 100–104 (2021).

177. Zhao, P. *et al.* High-Performance Relaxor Ferroelectric Materials for Energy Storage Applications. *Adv. Energy Mater.* **9**, (2019).

178. Chen, H. *et al.* Ultrahigh Energy Storage Density in Superparaelectric-Like Hf0.2Zr0.8O2 Electrostatic Supercapacitors. *Advanced Science* **10**, (2023).

179. Shuai, W. *et al.* Superior and ultrafast energy storage performance of relaxorantiferroelectric HfO2-based supercapacitors. *Energy Storage Mater.* **62**, (2023).

180. Yang, B. *et al.* High-entropy enhanced capacitive energy storage. *Nat. Mater.* **21**, 1074–1080 (2022).

181. Jo, S. *et al.* Negative differential capacitance in ultrathin ferroelectric hafnia. *Nat. Electron.* **6**, 390–397 (2023).

182. Íñiguez, J., Zubko, P., Luk'yanchuk, I. & Cano, A. Ferroelectric negative capacitance. *Nature Reviews Materials* vol. 4 243–256 Preprint at https://doi.org/10.1038/s41578-019-0089-0 (2019).

183. Hoffmann, M. *et al.* Negative Capacitance for Electrostatic Supercapacitors. *Adv. Energy Mater.* **9**, 1901154 (2019).

184. Zhang, W. Di *et al.* Ultrahigh dielectric permittivity in Hf0.5Zr0.5O2 thin-film capacitors. *Nature Communications* **16**, (2025).

185. Silva, J. P. B. *et al.* Roadmap on ferroelectric hafnia- and zirconia-based materials and devices. *APL Mater.* **11**, 089201 (2023).

186. Liu, Z. *et al.* Giant energy density nitride dielectrics enabled by a paraelectric-metaparaelectric phase transition. *Nature Communications* **16**, (2025).

187. Kim, K. Do, Yeom, M. K., Park, H. S., Jeon, G. & Hwang, C. S. Proximity-Induced Ferroelectric Switching in Wurtzite/Fluorite Bilayers for High-Performance Ferroelectric Field-Effect Transistor. *Advanced Materials* **37**, (2025).

188. Chua, L. Memristor-The missing circuit element. *IEEE Transactions on Circuit Theory* **18**, 507–519 (1971).

189. Hellenbrand, M., Teck, I. & MacManus-Driscoll, J. L. Progress of emerging non-volatile memory technologies in industry. *MRS Commun.* **14**, 1099–1112 (2024).

190. Moitra, A., Bhattacharjee, A., Li, Y., Kim, Y. & Panda, P. When in-memory computing meets spiking neural networks—A perspective on device-circuit-system-and-algorithm co-design. *Appl. Phys. Rev.* **11**, 031325 (2024).

191. Gokmen, T. & Vlasov, Y. Acceleration of Deep Neural Network Training with Resistive Cross-Point Devices: Design Considerations. *Front. Neurosci.* **Volume 10-2016**, (2016).

192. Jain, S. *et al.* A Heterogeneous and Programmable Compute-In-Memory Accelerator Architecture for Analog-AI Using Dense 2-D Mesh. *IEEE Trans. Very Large Scale Integr. VLSI Syst.* **31**, 114–127 (2023).

193. Keshavarzi, A., Ni, K., Van Den Hoek, W., Datta, S. & Raychowdhury, A. FerroElectronics for Edge Intelligence. *IEEE Micro* **40**, 33–48 (2020).

194. Do You Really Need 100 TOPS? Right-Sizing AI Performance for Real-World Edge Applications. *Peila* (2025).

195. Sun, J., Houshmand, P. & Verhelst, M. Analog or Digital In-Memory Computing? Benchmarking Through Quantitative Modeling. in *IEEE/ACM International Conference on Computer-Aided Design, Digest of Technical Papers, ICCAD* (Institute of Electrical and Electronics Engineers Inc., 2023). doi:10.1109/ICCAD57390.2023.10323763.

196. Li, S. *et al.* Involvement of Unsaturated Switching in the Endurance Cycling of Si-doped HfO2 Ferroelectric Thin Films. *Adv. Electron. Mater.* **6**, 2000264 (2020).

197. Pantel, D. & Alexe, M. Electroresistance effects in ferroelectric tunnel barriers. *Phys. Rev. B* **82**, 134105 (2010).

198. Wen, Z., Li, C., Wu, D., Li, A. & Ming, N. Ferroelectric-field-effect-enhanced electroresistance in metal/ferroelectric/semiconductor tunnel junctions. *Nat. Mater.* **12**, 617–621 (2013).

199. Max, B., Hoffmann, M., Mulaosmanovic, H., Slesazeck, S. & Mikolajick, T. Hafnia-based double-layer ferroelectric tunnel junctions as artificial synapses for neuromorphic computing. *ACS Appl. Electron. Mater.* **2**, 4023–4033 (2020).

200. Hellenbrand, M. *et al.* Achieving High ON State Current through Ferroelectric Polarization-Dependent Interfacial Resistance Switching in Undoped Orthorhombic HfO2 Films. *Adv. Funct. Mater.* **n/a**, e74667 (2026).

201. Majumdar, S. Can ferroelectric tunnel junction be a game changer as eNVM and in neuromorphic hardware? *APL Machine Learning* **3**, 020902 (2025).

202. Aguirre, F. *et al.* Hardware implementation of memristor-based artificial neural networks. *Nature Communications* vol. 15 Preprint at https://doi.org/10.1038/s41467-024-45670-9 (2024).

203. Kuzum, D., Yu, S. & Philip Wong, H.-S. Synaptic electronics: materials, devices and applications. *Nanotechnology* **24**, 382001 (2013).

204. Kim, J. S. *et al.* Coercive Field Control in Epitaxial Ferroelectric Hf0.5Zr0.5O2 Thin Films by Nanostructure Engineering. *ACS Appl. Mater. Interfaces* **17**, 25442–25450 (2025).

205. Kim, J. S. *et al.* Reducing Coercive Field and Improving Endurance in Ferroelectric Epitaxial Hf0.5Zr0.5O2 Thin Films via Novel Interface Layer Approach. *Advanced Science* **13**, e17314 (2026).

206. Shi, W. *et al.* Record-Low Coercive Field in ALD-Grown HZO Ferroelectric Films: Enabling Ultra-low Operation Voltage in World's First 2-Layer 3D Stacked Oxide Semiconductor 1T1C FeRAM. in *2024 IEEE International Electron Devices Meeting (IEDM)* 1–4 (2024). doi:10.1109/IEDM50854.2024.10873565.

207. Demasius, K. U., Kirschen, A. & Parkin, S. Energy-efficient memcapacitor devices for neuromorphic computing. *Nat. Electron.* **4**, 748–756 (2021).

208. Mikolajick, T. *et al.* Next generation ferroelectric materials for semiconductor process integration and their applications. *J. Appl. Phys.* **129**, (2021).

209. Chen, B. *et al.* Enhanced Ferroelectricity of Hf-Based Memcapacitors by Adopting Ti Insert-Layer and C–V Measurement for Constructing Energy-Efficient Reservoir Computing Network. *Adv. Electron. Mater.* **11**, (2025).

210. Schönweger, G. *et al.* In-Grain Ferroelectric Switching in Sub-5 nm Thin Al0.74Sc0.26N Films at 1 V. *Advanced Science* **10**, (2023).

211. Wu, Q. *et al.* Wafer-scale ultrathin and uniform van der Waals ferroelectric oxide. *Science (1979).* **391**, eadz1655 (2026).

212. Wang, X. *et al.* Hafnium oxide-based nonvolatile ferroelectric memcapacitor array for high energy-efficiency neuromorphic computing. *Nano Energy* **140**, 111011 (2025).

213. Lipatov, A. *et al.* Direct observation of ferroelectricity in two-dimensional MoS2. *NPJ 2D Mater. Appl.* **6**, (2022).

214. Chang, S. .-C. *et al.* FeRAM using Anti-ferroelectric Capacitors for High-speed and High-density Embedded Memory. in *2021 IEEE International Electron Devices Meeting (IEDM)* 33.2.1-33.2.4 (2021). doi:10.1109/IEDM19574.2021.9720510.

215. Liao, C.-Y. *et al.* Multipeak Coercive Electric-Field-Based Multilevel Cell Nonvolatile Memory With Antiferroelectric-Ferroelectric Field-Effect Transistors (FETs). *IEEE Trans. Ultrason. Ferroelectr. Freq. Control* **69**, 2214–2221 (2022).

216. Zeng, B. *et al.* Multiple Polarization States in Hf1−xZrxO2 Thin Films by Ferroelectric and Antiferroelectric Coupling. *Advanced Materials* **37**, 2411463 (2025).

217. Tian, G. *et al.* Highly Reliable Logic-in-Memory by Bidirectional Built-in Electric-Field-Modulated Multistate IGZO/AFE Nonvolatile Memory. *ACS Appl. Electron. Mater.* **5**, 1041–1049 (2023).

218. Hsiang, K. .-Y. *et al.* Dielectric Layer Design of Bilayer Ferroelectric and Antiferroelectric Tunneling Junctions Toward 3D NAND-Compatible Architecture. *IEEE Electron Device Letters* **43**, 1850–1853 (2022).

219. Gao, J. *et al.* Reconfigurable neuromorphic functions in antiferroelectric transistors through coupled polarization switching and charge trapping dynamics. *Nature Communications* **16**, (2025).

220. Xu, K. *et al.* Novel Two-Terminal Synapse/Neuron Based on an Antiferroelectric Hafnium Zirconium Oxide Device for Neuromorphic Computing. *Nano Lett.* **24**, 11170–11178 (2024).

221. Sun, C. *et al.* Novel a-IGZO Anti-Ferroelectric FET LIF Neuron with Co-Integrated Ferroelectric FET Synapse for Spiking Neural Networks. in *Technical Digest - International Electron Devices Meeting, IEDM* vols 2022-December 211–214 (Institute of Electrical and Electronics Engineers Inc., 2022).

222. Chakraborty, A. *et al.* Best practices for measuring the performance and stability of indoor photovoltaic devices. *Nat. Energy* https://doi.org/10.1038/s41560-026-02101-x (2026) doi:10.1038/s41560-026-02101-x.

## Acknowledgements

The authors acknowledge support from the European Union's Horizon 2020 Research and Innovation Programme under the Marie Skłowdowska-Curie grant no. 101168161 (MASAUTO) and 101169056 (MENTOR). In addition, the authors thank UK Research and Innovation (UKRI) for support through the UK Innovation & Knowledge Centre in Neuromorphic Computing Hardware (no. UKRI2784). R.L.Z.H. also thanks UKRI for funding through a Frontier Grant (no. EP/X029900/1), awarded through the 2021 ERC Starting Grant scheme, as well as support from the Royal Academy of Engineering and Science & Technology Facilities Council through the Research Chair scheme (no. RCSRF2324-18-68). G.B. was supported by the EPSRC through the Switch to Net Zero Buildings Place Based Impact Acceleration Account (no. EP/Y024060/1). A.M.K. acknowledges support from the EPSRC through Project MANTISS *'Molecular Absorbers for Novel Transducing Imaging and Sensing Semiconductor Photodetectors'* (UKRI4483). This work was supported by the Portuguese Foundation for Science and Technology (FCT) in the framework of the (i) Strategic Funding UID/04650/2025 (DOI: https://doi.org/10.54499/UID/04650/2025), (ii) project 2024.12826.MIT(https://doi.org/10.54499/2024.12826.MIT) and (iii) project 2024.15998.PEX (https://doi.org/10.54499/2024.15998.PEX). This work has also received funding from the European Union's Horizon 2020 research and innovation programme under the Marie Skłodowska Curie grant agreement no: 101168161 (MASAUTO, doi: 10.3030/101168161). J. P. B. S. also thanks FCT for the contract under the FCT Tenure program 1st edition. N. Estrócio acknowledges his Ph.D. grant from FCT, with reference 2025.00731.BD (https://doi.org/10.54499/2025.00731.BD). G.G. and A.M. acknowledge funding from the European Research Council (ERC) under the European Union's Horizon Europe research and innovation programme ERC Cog 2024 ELOW-DI (Grant Agreement No. 101171012). Funded by the European Union. Views and opinions expressed are, however, those of the authors only and do not necessarily reflect those of the European Union or ERCEA. Neither the European Union nor the granting authority can be held responsible for them. G.G. and A.S. acknowledge the GOPV project (CSEAA_00011), which received funds from Bando Ricerca di Sistema—CSEA—TIPO A Piano triennale 2019–2021 Decreto direttoriale 27 Ottobre 2021 del Ministero della Transizione Ecologica. J.L.M.-D.

acknowledges support from the Royal Academy of Engineering Chair in Emerging Technologies (Grant No. CiET1819\24), and the European Research Council ERC Advanced (Grant Agreement Nos. EU-H2020-ERC-ADG and 882929), EROS. R.L.Z.H., Q.J. and J.L.M.-D. thank the National Science Foundation (NSF) and EPSRC for support through the ECCS-EPOSRC collaborative grant (EPSRC no. EP/Y032942/1, NSF no. ECCS-2313755). J.Í.-G. acknowledges support from the Luxembourg National Research Fund (FNR) through project INTER/NSF/24/18804122/PIEZOHAFNIA. I.F. and F.S. acknowledge financial support from the Spanish Ministry of Science, Innovation and Universities (MCIN/AEI/10.13039/501100011033), through PID2023-147211OB-C21 and Severo Ochoa MATRANS42 (CEX2023-001263-S).

## Author Contributions

R.L.Z.H., J.P.B.S. and G.G. conceived of this review. R.L.Z.H., M.H., J.P.B.S. and M.F. wrote the introduction and energy requirements sections together. The IPV section and IPV outlook were written by R.L.Z.H., A.M.K., G.B., A.S., M.F., A.M., G.K. and G.G. A.M.K., A.S. and A.M. prepared Fig. 1, with input from the others, and A.M.K. and G.B. wrote the detailed supporting notes on IPV in the supplementary information. N.E., J.P.B.S., L.M., J.Í-G., I.F. wrote the energy storage section and outlook, while N.E. and J.P.B.S. prepared Boxes 1 and 2, along with Fig. 3 and Supplementary Table 1, 4–7. M.F. also contributed to Box 1. M.H., N. E. and J.P.B.S. prepared Figure 3 and Supplementary Table 8. M.H., B.N., F.S., B.O., J.L.M.D. and Q.J. wrote the IMC section and outlook, while M.H. prepared Box 3. All authors contributed to discussion of the content and reviewed and/or edited the manuscript before submission.

## Competing interests

R.L.Z.H. consults for *NanoPrint Innovations Ltd*, which is commercialising spatial atomic layer deposition for growing metal oxides for photovoltaics.

## Related links

MASAUTO Marie Curie Doctoral training network: https://masauto-net.eu/ and DOI: 10.3030/101168161

MENTOR Marie Curie Doctoral training network: https://projects.tuni.fi/mentor/

**Supplementary Information for**

# Computing at the Edge Enabled by Indoor Photovoltaics and Ferroelectrics

Robert L. Z. Hoye,[1,*] Markus Hellenbrand,[2,*] Nuno Estrócio,[3,4] Austin M. Kay,[5] Alice Scardina,[6] Alessandro Mezzetti,[6] Ignasi Fina,[7] Jorge Íñiguez-González,[8,9] Luís S. Marques,[3,4] Francesco Matteucci,[10,11] Quanxi Jia,[12] Judith L. MacManus-Driscoll,[2] Bert Offrein,[13] Florencio Sánchez,[7] Beatriz Noheda,[14,15] George Koutsourakis,[16] Marina Freitag,[17] Gregory Burwell,[5] Giulia Grancini,[6,*] Jose P. B. Silva[3,4,*]

[1.] Inorganic Chemistry Laboratory, University of Oxford, South Parks Road, Oxford OX1 3QR, United Kingdom

[2.] Department of Materials Science and Metallurgy, 27 Charles Babbage Road, Cambridge CB3 0FS, United Kingdom

[3.] Physics Center of Minho and Porto Universities (CF-UM-UP), University of Minho, Campus de Gualtar, 4710-057 Braga, Portugal

[4.] Laboratory of Physics for Materials and Emergent Technologies, LapMET, University of Minho, 4710-057 Braga, Portugal

[5.] Sustainable Advanced Materials (Sêr-SAM), Centre for Integrative Semiconductor Materials (CISM), Department of Physics, Swansea University Bay Campus, Swansea SA1 8EN, UK

[6.] Department of Chemistry & INSTM, University of Pavia, Via T. Taramelli 14, 27100, Pavia, Italy

[7.] Institut de Ciència de Materials de Barcelona (ICMAB-CSIC), Carrer dels Til.lers, 08193 Cerdanyola del Vallès, Spain

[8.] Smart Materials Unit, Luxembourg Institute of Science and Technology (LIST), Avenue des Hauts-Fourneaux 5, L-4362, Esch/Alzette, Luxembourg

[9.] Department of Physics and Materials Science, University of Luxembourg, Rue du Brill 41, L-4422 Belvaux, Luxembourg

[10.] Nanoshuttle, via. G. Desa 12, Lecce, Italy

[11.] Fondazione Bruno Kessler, via Sommariva 18, Povo, Italy

[12.] Department of Materials Design and Innovation, University at Buffalo – The State University of New York, Buffalo, NY 14260, USA

[13.] IBM Research GmbH – Zürich Research laboratory, CH-8803, Ruschlikon, Switzerland

[14] CongiGron – Groningen Cognitive Systems and Materials Center, University of Groningen, Nijenborgh 4, 9747AG, Groningen, The Netherlands

[15] Zernike Institute for Advanced Materials, University of Groningen, Nijenborgh 4, 9747AG Groningen, The Netherlands

[16] National Physical Laboratory, Hampton Road, Teddington TW11 0LW, UK

[17] School of Natural and Environmental Science, Newcastle University, Newcastle upon Tyne NE1 7RU, UK

* e-mail: robert.hoye@chem.ox.ac.uk (R. L. Z. H.), mkhh2@cam.ac.uk (M. K. H. H.), giulia.grancini@unipv.it (G. G.), josesilva@fisica.uminho.pt (J. P. B. S.)

**Supplementary Table 1 | Hardware performance comparison of publicly-announced AI accelerators and processors.** Reported values may have been obtained under different operating conditions, numerical precisions, workloads and benchmarking methodologies. Consequently, direct quantitative comparisons should be interpreted with caution. Abbreviations: tera operations per second (TOPS), graphics processing unit (GPU) and processing-in-memory (PIM).

| Platform | Power (W) | Peak TOPS | Ref. |
|---|---|---|---|
| Dataflow | | | |
| AML200 | $1.00\times10^{-1}$ | $2.00\times10^{0}$ | 1 |
| Cambricon-X | $9.54\times10^{-1}$ | $5.44\times10^{-1}$ | 2 |
| Chain-NN | $5.68\times10^{-1}$ | $8.06\times10^{-1}$ | 3 |
| DaDianNao | $1.60\times10^{1}$ | $5.58\times10^{0}$ | 4 |
| DianNao | $4.85\times10^{-1}$ | $4.52\times10^{-1}$ | 5 |
| EIE | $6.00\times10^{-1}$ | $1.02\times10^{-1}$ | 6 |
| El Cano | $7.00\times10^{0}$ | $1.60\times10^{1}$ | 7 |
| EyeQ3 | $3.00\times10^{0}$ | $2.50\times10^{-1}$ | 8 |
| EyeQ4 | $6.00\times10^{0}$ | $2.50\times10^{0}$ | 8 |
| EyeQ5 | $1.00\times10^{1}$ | $2.40\times10^{1}$ | 8 |
| EyeQ6 | $3.30\times10^{1}$ | $3.40\times10^{1}$ | 9 |
| Eyeriss | $2.78\times10^{-1}$ | $6.72\times10^{-2}$ | 10 |
| Hailo-10H | $2.50\times10^{0}$ | $4.00\times10^{1}$ | 11 |
| Hailo-15H | $5.00\times10^{0}$ | $2.00\times10^{1}$ | 12 |
| Hailo-15L | $3.00\times10^{0}$ | $7.00\times10^{0}$ | 13 |
| Hailo-8 | $2.50\times10^{0}$ | $2.60\times10^{1}$ | 14 |
| HyperX HX40416 | $8.00\times10^{0}$ | $6.60\times10^{0}$ | 15 |
| Journey 2 | $2.00\times10^{0}$ | $4.00\times10^{0}$ | 16 |
| Journey 3 | $2.50\times10^{0}$ | $5.00\times10^{0}$ | 16 |
| Journey 5 | $3.00\times10^{1}$ | $1.28\times10^{2}$ | 17 |
| MX3 | $2.00\times10^{0}$ | $6.00\times10^{0}$ | 18 |
| NDP200 | $1.00\times10^{-3}$ | $6.40\times10^{-3}$ | 19 |
| NDP250 | $3.00\times10^{-2}$ | $3.00\times10^{-2}$ | 20 |
| SiMa.ai | $1.00\times10^{1}$ | $5.00\times10^{1}$ | 21 |
| Spyre | $7.50\times10^{1}$ | $3.15\times10^{2}$ | 22 |
| Graphics Processing Unit | | | |
| A10 | $1.50\times10^{2}$ | $2.50\times10^{2}$ | 23 |
| A100 | $3.00\times10^{2}$ | $6.24\times10^{2}$ | 24 |
| A2 | $6.00\times10^{1}$ | $3.60\times10^{1}$ | 25 |
| A30 | $1.65\times10^{2}$ | $3.30\times10^{2}$ | 26 |
| A40 | $3.00\times10^{2}$ | $2.99\times10^{2}$ | 27 |
| A800 | $2.40\times10^{2}$ | $1.25\times10^{3}$ | 28 |
| BR100 | $5.50\times10^{2}$ | $2.05\times10^{3}$ | 29 |
| BR104 | $3.00\times10^{2}$ | $1.02\times10^{3}$ | 29 |
| Flex140 | $7.50\times10^{1}$ | $2.10\times10^{2}$ | 30 |
| Flex170 | $1.50\times10^{2}$ | $5.12\times10^{2}$ | 31 |
| H100 NVL | $4.00\times10^{2}$ | $3.34\times10^{3}$ | 32 |
| H100 SXM | $7.00\times10^{2}$ | $3.96\times10^{3}$ | 32 |
| Jetson AGX Orin | $6.00\times10^{1}$ | $2.75\times10^{2}$ | 33 |
| Jetson AGX Xavier | $3.00\times10^{1}$ | $3.20\times10^{1}$ | 34 |
| Jetson NX Orin | $4.00\times10^{1}$ | $1.57\times10^{2}$ | 33 |
| Jetson Orin Nano | $2.50\times10^{1}$ | $6.70\times10^{1}$ | 33 |
| Jetson Xavier NX | $2.00\times10^{1}$ | $2.10\times10^{1}$ | 34 |
| L4 | $7.20\times10^{1}$ | $4.85\times10^{2}$ | 35 |
| L40 | $3.00\times10^{2}$ | $3.62\times10^{2}$ | 36 |
| L40s | $3.50\times10^{2}$ | $7.33\times10^{2}$ | 37 |
| MI210 | $3.00\times10^{2}$ | $1.81\times10^{2}$ | 38 |
| MI250 | $5.00\times10^{2}$ | $3.62\times10^{2}$ | 39 |
| MI250X | $5.00\times10^{2}$ | $3.83\times10^{2}$ | 39 |
| MI300X | $7.50\times10^{2}$ | $2.61\times10^{3}$ | 40 |
| MI325X | $1.00\times10^{3}$ | $2.61\times10^{3}$ | 41 |
| MI350X | $1.00\times10^{3}$ | $4.61\times10^{3}$ | 42 |
| MI355X | $1.40\times10^{3}$ | $1.01\times10^{4}$ | 43 |
| MTT S2000 | $1.50\times10^{2}$ | $4.24\times10^{1}$ | 44 |
| MTT S50 | $8.50\times10^{1}$ | $2.08\times10^{1}$ | 44 |
| T4 | $7.00\times10^{1}$ | $1.30\times10^{2}$ | 45 |
| In-sensor | | | |
| AIStorm | $2.25\times10^{-1}$ | $2.50\times10^{0}$ | 46 |
| ShiDianNao | $3.20\times10^{-1}$ | $1.94\times10^{-1}$ | 47 |
| Manycore | | | |
| Colossus Mk2 | $2.50\times10^{2}$ | $2.50\times10^{2}$ | 48,49 |
| Coolidge 1 | $3.00\times10^{1}$ | $2.00\times10^{1}$ | 50 |
| Coolidge 2 | $3.00\times10^{1}$ | $5.00\times10^{1}$ | 50 |
| GAP8 | $1.00\times10^{-2}$ | $8.00\times10^{-3}$ | 51 |
| GAP9 | $5.00\times10^{-2}$ | $5.00\times10^{-2}$ | 52 |
| MTIA | $3.50\times10^{1}$ | $1.02\times10^{2}$ | 53 |
| MTIA2i | $8.50\times10^{1}$ | $3.64\times10^{2}$ | 54 |
| NorthPole | $7.35\times10^{1}$ | $2.00\times10^{2}$ | 55,56 |
| XMOS | $8.00\times10^{-1}$ | $5.10\times10^{-2}$ | 57 |
| Neuromorphic | | | |
| Tianjic | $9.50\times10^{-1}$ | $1.21\times10^{0}$ | 58 |
| AKD1000 | $3.00\times10^{0}$ | $1.50\times10^{0}$ | 59 |
| AKD1500 | $8.00\times10^{-1}$ | $8.00\times10^{-1}$ | 60 |
| Processing-In-Memory | | | |
| Axelera Thetis | $2.79\times10^{0}$ | $3.93\times10^{1}$ | 61 |
| Lightspeeur 2081S | $3.00\times10^{-1}$ | $2.80\times10^{0}$ | 62 |
| Lightspeeur 2803S | $7.00\times10^{-1}$ | $1.68\times10^{1}$ | 63 |
| Lightspeeur 5801 | $2.24\times10^{-1}$ | $2.80\times10^{0}$ | 64 |
| M1076 | $3.00\times10^{0}$ | $2.50\times10^{1}$ | 65 |
| M1108 | $4.00\times10^{0}$ | $3.50\times10^{1}$ | 65 |
| Tensor | | | |
| A14 Bionic | $6.00\times10^{0}$ | $1.10\times10^{1}$ | 66 |
| A15 Bionic | $6.00\times10^{0}$ | $1.56\times10^{1}$ | 67 |
| A16 Bionic | $1.80\times10^{1}$ | $1.70\times10^{1}$ | 68 |
| AI 100 Pro | $7.50\times10^{1}$ | $4.00\times10^{2}$ | 69 |
| AI 100 Standard | $7.50\times10^{1}$ | $3.50\times10^{2}$ | 69 |
| AI 100 Ultra | $1.50\times10^{2}$ | $8.70\times10^{2}$ | 69 |
| AI 80 Standard | $7.50\times10^{1}$ | $1.90\times10^{2}$ | 69 |
| AI 80 Ultra | $1.50\times10^{2}$ | $6.18\times10^{2}$ | 69 |
| ATOM Max | $3.50\times10^{2}$ | $5.12\times10^{2}$ | 70 |
| Baidu Kunlun | $1.60\times10^{2}$ | $2.30\times10^{2}$ | 71 |
| Ergo | $7.27\times10^{-2}$ | $4.00\times10^{0}$ | 72 |
| Ethos N77 | $8.00\times10^{-1}$ | $4.00\times10^{0}$ | 73 |
| Groq TSP | $3.30\times10^{3}$ | $6.00\times10^{3}$ | 74 |
| Kendryte K210 | $3.00\times10^{-1}$ | $8.00\times10^{-1}$ | 75 |
| KL520 | $5.00\times10^{-1}$ | $3.00\times10^{-1}$ | 76 |
| KL530 | $5.00\times10^{-1}$ | $5.00\times10^{-1}$ | 77 |
| KL720 | $1.56\times10^{0}$ | $1.40\times10^{0}$ | 76 |
| M1 | $1.00\times10^{1}$ | $1.10\times10^{1}$ | 78 |
| M2 | $2.00\times10^{1}$ | $1.58\times10^{1}$ | 79 |
| M3 | $2.00\times10^{1}$ | $1.80\times10^{1}$ | 80 |
| MLU290-M5 | $3.50\times10^{2}$ | $5.12\times10^{2}$ | 81 |
| MLU370-X8 | $2.50\times10^{2}$ | $2.56\times10^{2}$ | 82 |
| Movidius Myriad X VPU | $1.50\times10^{0}$ | $4.00\times10^{0}$ | 83 |
| NovuTensor | $1.50\times10^{1}$ | $1.50\times10^{1}$ | 84 |
| QRB5165 | $7.00\times10^{0}$ | $1.50\times10^{1}$ | 85 |
| Raptor | $5.50\times10^{1}$ | $2.00\times10^{2}$ | 86 |
| RNGD | $1.85\times10^{2}$ | $5.12\times10^{2}$ | 87 |
| STM32N6x7 | $2.00\times10^{-1}$ | $6.00\times10^{-1}$ | 88 |
| TPU Edge | $2.00\times10^{0}$ | $4.00\times10^{0}$ | 89 |
| TPU v1 | $4.00\times10^{1}$ | $9.20\times10^{1}$ | 90 |
| TPU v4 | $1.92\times10^{2}$ | $2.75\times10^{2}$ | 91 |
| TPU v5p | $5.37\times10^{2}$ | $9.18\times10^{2}$ | 92 |
| TPU v6e | $3.83\times10^{2}$ | $1.84\times10^{3}$ | 92 |
| TPU v7p | $9.59\times10^{2}$ | $4.61\times10^{3}$ | 92 |

**Supplementary Note 1 | Simulating Indoor Photovoltaics**

To simulate power generation in photovoltaic (PV) devices illuminated by sources of artificial light, a method recently described by Kay et al. was employed[93]. This technique combines a device's PV external quantum efficiency ($\mathrm{EQE_{PV}}$) spectrum with the spectral photon flux ($\Phi_{\mathrm{source}}$) of the considered light source to determine the photocurrent density ($J_{\mathrm{ph}}$), dark saturation current density ($J_0$), open-circuit voltage ($V_{\mathrm{oc}}$), and ultimately the maximum power point parameters; all while accounting for band-filling effects[94], the chemical potential of radiation[95], non-radiative open-circuit voltage losses, and the failure of the reciprocity principle for states that lie well below the bandgap[96].

To begin, the photocurrent generated in a photovoltaic device under voltage bias $V$ can be evaluated using its $\mathrm{EQE_{PV}}$ spectrum and the spectral flux of the light source via the general expression[97]

$$J_{\mathrm{ph}}(V) = q\int_{E_{\mathrm{lower}}}^{\infty} \mathrm{EQE_{PV}}(E)\cdot\tanh\left(\frac{E-qV}{4k_{\mathrm{B}}T}\right)\cdot\Phi_{\mathrm{source}}(E)\,\mathrm{d}E. \qquad \textbf{(S1.1)}$$

Here, $E$ denotes the photon energy, while $q$ is the elementary charge, $k_{\mathrm{B}}$ is Boltzmann's constant, and $T$ is the temperature of the photovoltaic device. Note that the lower limit of the integral ($E_{\mathrm{lower}}$) accounts for deep trap states (with energies much smaller than the bandgap energy) not obeying the standard reciprocity relation[96], while the hyperbolic factor accounts for the effects of (sub-gap) state-filling[94]. Generally, the largest contribution to the photocurrent is made by above-gap states with energy $E > E_{\mathrm{g}}$ (with $E_{\mathrm{g}}$ being the bandgap energy), meaning that the hyperbolic correction factor tends to one for these states (as $qV < E_{\mathrm{g}} < E$ at typical operating voltages for photovoltaic devices)[93].

The dark saturation current density can likewise be determined using the spectral photon flux of ambient radiation inside the device ($\Phi_{\mathrm{Planck}}$), which may be written in terms of the device's temperature as

$$\Phi_{\mathrm{Planck}}(E)=\frac{2\pi}{h^3c^2}\frac{E^2}{\exp\left(\frac{E}{k_{\mathrm{B}}T}\right)-1}, \tag{S1.2}$$

where $h$ denotes Planck's constant and $c$ is the speed of light in vacuum. Using this expression for ambient radiation, the dark saturation current density can be evaluated in the radiative limit ($J_0 \rightarrow J_0^{\mathrm{rad}}$) using

$$J_0^{\mathrm{rad}}(V)=q\int_{E_{\mathrm{lower}}}^{\infty}\mathrm{EQE}_{\mathrm{PV}}(E)\cdot w(E,V)\cdot\Phi_{\mathrm{Planck}}(E)\,\mathrm{d}E. \tag{S1.3}$$

In this case, the term $w(E,V)$ is a correction factor that accounts for state-filling effects and the chemical potential radiation[97]; it is defined by[93]

$$w(E,V)=\frac{1}{\left[1+\exp\left(\frac{qV-E}{2k_{\mathrm{B}}T}\right)\right]^2}. \tag{S1.4}$$

For states that are well above the gap ($E \gg E_{\mathrm{g}} > qV$), it follows that $w(E,V) \rightarrow 1$. In practice, non-radiative open-circuit voltage losses ($\Delta V_{\mathrm{oc}}^{\mathrm{nr}}$) in the device increase the dark saturation current density, and so in the non-radiative limit, $J_0$ may be calculated from $J_0^{\mathrm{rad}}$ using

$$J_0(V)=\frac{J_0^{\mathrm{rad}}(V)}{\mathrm{EQE}_{\mathrm{EL}}}=J_0^{\mathrm{rad}}(V)\exp\left(\frac{q\Delta V_{\mathrm{oc}}^{\mathrm{nr}}}{k_{\mathrm{B}}T}\right) \tag{S1.5}$$

where $\mathrm{EQE}_{\mathrm{EL}}=\exp\left(-\frac{q\Delta V_{\mathrm{oc}}^{\mathrm{nr}}}{k_{\mathrm{B}}T}\right)$ is the device's electroluminescent external quantum efficiency[98]. In the radiative limit used in **Figure 1c** of the main text, non-radiative open-circuit voltage losses are assumed minimal (i.e., $V_{\mathrm{oc}}^{\mathrm{nr}}=0$), giving $\mathrm{EQE}_{\mathrm{EL}}=1$ and $J_0(V) \rightarrow J_0^{\mathrm{rad}}(V)$.

Using Equation **(S1.1)-(S1.5)**, the total current density ($J$) generated by a photovoltaic device under voltage bias $V$ can be described using the ideal diode equation

$$J(V) = J_0(V)\left[\exp\left(\frac{qV}{k_\mathrm{B}T}\right) - 1\right] - J_\mathrm{ph}(V). \quad \textbf{(S1.6)}$$

By numerically evaluating the photocurrent, the dark saturation current, and the total current produced by the device under different voltage biases, Equation **(S1.6)** can be used to identify the maximum power point voltage ($V_\mathrm{mpp}$), the corresponding current density ($J_\mathrm{mpp}$), and the power conversion efficiency (PCE); which are defined by

$$\left.\frac{\mathrm{d}\,[J(V)\cdot V]}{\mathrm{d}V}\right|_{V=V_\mathrm{mpp}} \equiv 0, \quad \textbf{(S1.7a)}$$

$$J_\mathrm{mpp} \equiv J(V_\mathrm{mpp}), \quad \textbf{(S1.7b)}$$

$$\mathrm{PCE} \equiv \frac{J_\mathrm{mpp}\cdot V_\mathrm{mpp}}{P_\mathrm{source}}. \quad \textbf{(S1.7b)}$$

Here, $P_\mathrm{source} = \int_0^\infty E\cdot\Phi_\mathrm{source}(E)\,\mathrm{d}E$ denotes the integrated irradiance of the light source.

The approach described above was utilized to simulate the power generated by the different materials considered in Figure 1 of the main text. To easily simulate power generation under different light sources, a freely-available computational tool, *PV-Simulator*[99], was constructed by Kay et al.[93] This tool currently holds the $\mathrm{EQE_{PV}}$ spectra of a few hundred PV devices from different material classes, as well as more than one hundred spectrally-unique light sources; enabling a quick comparison of indoor photovoltaic performance under a variety of illumination conditions.

**Supplementary Note 2 | Photovoltaic External Quantum Efficiency and Irradiance Spectra**

To demonstrate the utility of *PV-Simulator*, the photocurrent density, open-circuit voltage, fill factor (FF), and power conversion efficiency have all been determined under indoor illumination conditions in the Shockley-Quessier model, with the result plotted versus the optical gap in **Supplementary Fig. 1a-d**. In this idealized model, all photons above a threshold optical gap ($E_{\mathrm{opt}}$) generate electron-hole pairs at 100% efficiency, whilst photons below this threshold do not. Losses occur only through radiative pathways, and parasitic resistances are assumed negligible[100]. Whilst the original investigation by Shockley and Queisser focussed on solar illumination conditions, **Supplementary Fig. 1** considers two light sources specified in an indoor photovoltaic testing standard from the International Electrotechnical Commission (62607-7-2:2023)[101]. Namely, these light sources are CIE LED-B4 and CIE FL-10, and both spectra were scaled to an irradiance $P_{\mathrm{source}} = \int_0^\infty E \cdot \Phi_{\mathrm{source}}(E)\, dE = 313.2\ \mu\mathrm{Wcm}^{-2}$, which corresponds to an irradiance of 1000 lux under CIE LED-B4 illumination.

In addition to the curves simulated in the Shockley-Queisser limit, **Supplementary Fig. 1** also shows the performance of the four different indoor photovoltaic systems considered in **Figure 1** of the main text, which were simulated using the techniques outlined in **Supplementary Note 1** and the $\mathrm{EQE}_{\mathrm{PV}}$ spectra of each system (**Supplementary Fig. 2a**). Namely, these systems are crystalline silicon (c-Si; square data points), a perovskite photovoltaic (PPV; circle data points), an organic photovoltaic (OPV; triangle data points), and a dye-sensitized photovoltaic device (DSPV; diamond data points). To simulate the performance of these systems, their $\mathrm{EQE}_{\mathrm{PV}}$ spectra were combined with the irradiance ($E \cdot \Phi_{\mathrm{source}}$) spectra shown in **Supplementary Fig. 2b** and **2c**.

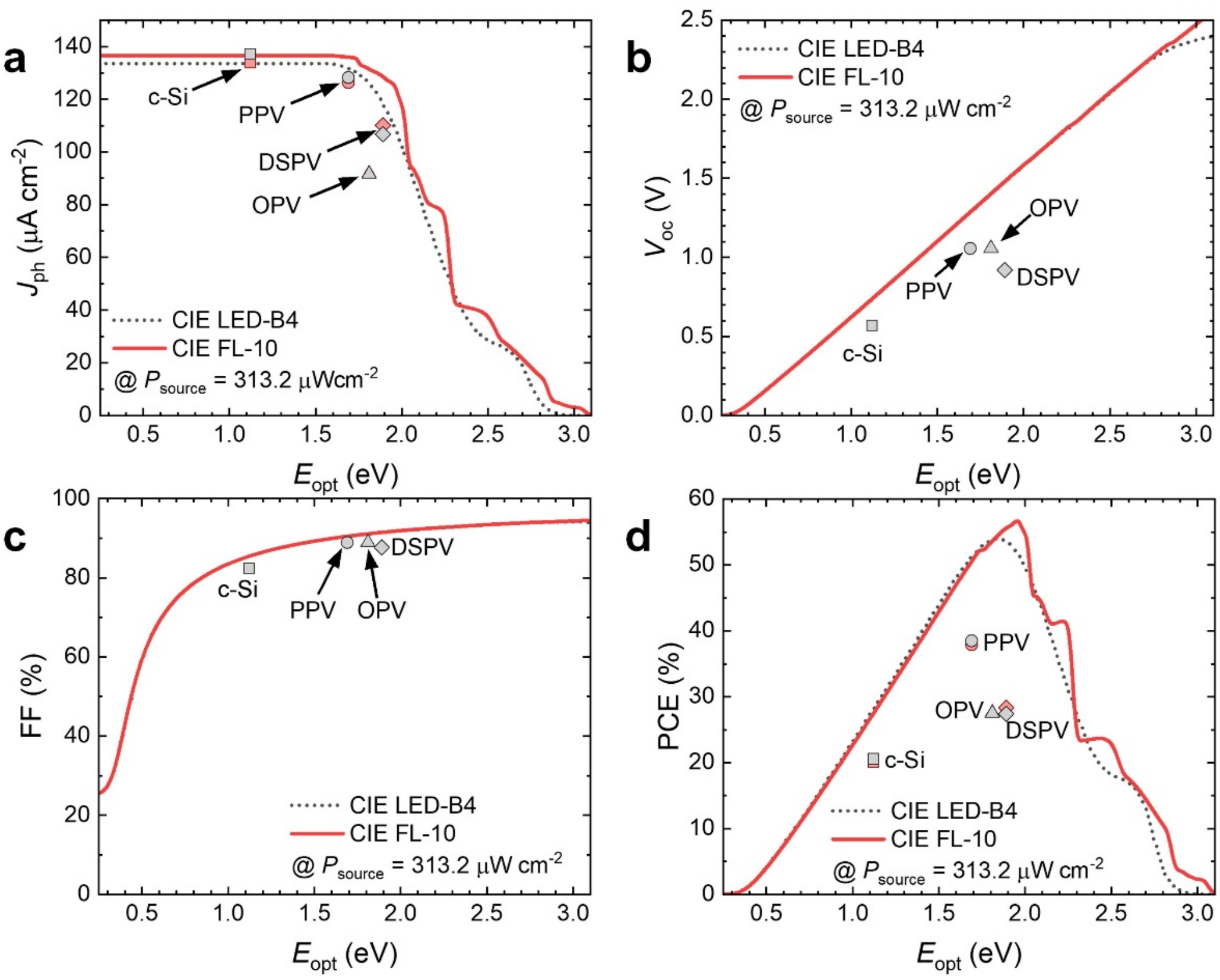

**Supplementary Fig. 1 | Photovoltaic figures-of-merit in the Shockley-Queisser model under indoor illumination conditions.** Under both CIE LED-B4 (dotted black curves) and CIE FL-10 (solid red curves) illumination, the **a,** photocurrent density ($J_{\mathrm{ph}}$), **b,** open-circuit voltage ($V_{\mathrm{oc}}$), **c,** fill factor (FF), and **d,** power conversion efficiency (PCE) have all been determined as a function of the optical gap ($E_{\mathrm{opt}}$). Both irradiance spectra were scaled to an integrated irradiance $P_{\mathrm{source}} = \int_0^{\infty} E \cdot \Phi_{\mathrm{source}}(E)\,\mathrm{d}E = 313.2\ \mu\mathrm{Wcm}^{-2}$. In addition to the curves in the Shockley-Queisser limit, data points illustrate the figures-of-merit of the four systems considered in Figure 1 of the main text, namely these are a crystalline silicon device (c-Si; squares), a perovskite photovoltaic (PPV; circles), an organic photovoltaic (OPV; triangles), and a dye-sensitized photovoltaic (DSPV; diamonds).

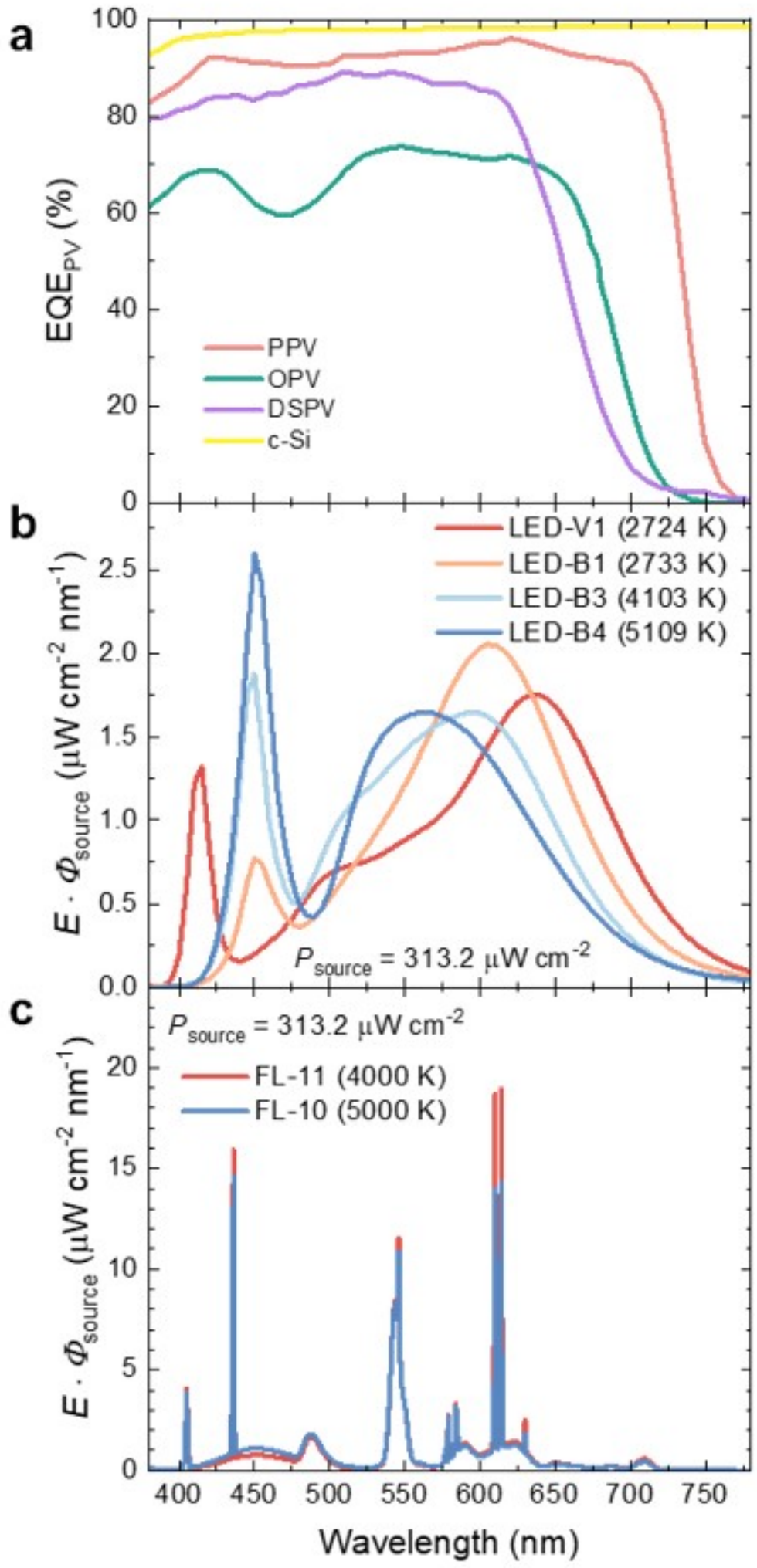

**Supplementary Fig. 2 | Spectral responses of four state-of-the-art indoor photovoltaic systems compared with the spectral irradiances of standard sources of artificial light. a,** Photovoltaic external quantum efficiency ($EQE_{PV}$) spectra plotted versus wavelength ($\lambda$) for the four state-of-the-art indoor photovoltaic systems described in Figure 1 of the main text. International Commission on Illumination (CIE) standard irradiance spectra (given by the product of the photon energy $E$ and the photon flux density $\Phi_{source}$ of the light source) are plotted for light-emitting diodes (LEDs) and fluorescent lamps (FLs) in **b** and **c**, respectively. All irradiance spectra have been scaled to an integrated irradiance $P_{source} = \int_0^\infty E \cdot \Phi_{source}(E)\,dE = 313.2\ \mu Wcm^{-2}$.

**Supplementary Note 3 | Optimizing Photovoltaic Modules for Indoor Energy Harvesting**

While power conversion efficiency (PCE) is a useful figure-of-merit for comparing photovoltaic devices, alternative figures-of-merit can also play pivotal roles for energy harvesting applications, such as the maximum power point voltage ($V_{\mathrm{mpp}}$) and the corresponding current ($I_{\mathrm{mpp}}$)[102]. This is due to the photovoltaic device being just one component of a broader-scale energy-harvesting system, which both captures and stores power in an energy storage device via an energy harvesting board (EHB), or power management integrated circuit (PMIC)[103]. Consequently, the optimal operational voltage and current of the chosen EHB are equally as important as considerations like the chosen photovoltaic material, its associated losses and limitations, and its resulting current-voltage characteristics at the typical illumination conditions of the desired deployment location.

In the following investigations, the performance of photovoltaic modules based on the four materials discussed in **Figure 1** of the main text was determined under different illumination conditions. The $\mathrm{EQE_{PV}}$ spectra of these photovoltaic systems are shown in **Supplementary Fig. 2**. By singulating a monolithic photovoltaic cell into a photovoltaic module consisting of multiple, serially-connected cells, the voltage the module can deliver to an EHB will increase (while the current will decrease proportionately). As illustrated in **Supplementary Fig. 3a**, however, adding more cells to a module while its total area ($A_{\mathrm{module}} = L_x L_y$) remains fixed increases the effective 'dead area' of the device – which is the sum of all the interconnect widths ($\delta$). The associated losses in the geometric fill factor can reduce the PCE of the module, though if $\delta$ is small these losses are almost negligible[102].

Importantly, the number of cells in a photovoltaic module ($N_{\text{cells}}$) can be optimized to achieve the highest energy-harvesting system efficiency ($\eta_{\text{sys}}$) when coupled with a particular EHB. This figure-of-merit quantifies the amount of power that is stored by the whole energy-harvesting system, and it can be written as

$$\eta_{\text{sys}} = \text{PCE} \cdot \eta_{\text{EHB}} \quad \textbf{(S3.1)}$$

Here, the PCE is that of the photovoltaic cell/module, while $\eta_{\text{EHB}}$ is the efficiency of energy-storing by the EHB. Herein, the e-peas AEM13920 is used as an exemplary EHB[104], and its $\eta_{\text{EHB}}$ is plotted as a function of a photovoltaic module's $V_{\text{mpp}}$ and $I_{\text{mpp}}$ in **Supplementary Fig. S3b**. For $V_{\text{mpp}} < 0.5$ V and $V_{\text{mpp}} > 4.0$ V, the efficiency of this EHB tends to nought. Consequently, to achieve optimal energy harvesting for this EHB, a photovoltaic module should aim to deliver between around 2.5 V and 3.5 V across the typical range of light intensities expected at the deployment location. Other EHBs will likely have different optimal voltage ranges.

Using the simulation framework described in **Supplementary Note 1**, the performance of square photovoltaic modules under 1000 lx LED-B4 illumination was evaluated for the four photovoltaic materials discussed in **Figure 1** of the main text, where $N_{\text{cells}}$ was varied while a fixed area $A_{\text{module}} = 25\ \text{cm}^2$ and a reasonable $\delta = 100$ µm were assumed. The resultant figures-of-merit are shown throughout **Supplementary Fig. S3**, including the open-circuit voltage ($V_{\text{oc}}$) and $V_{\text{mpp}}$ in **Supplementary Fig. S3c**, $I_{\text{mpp}}$ in **Supplementary Fig. S3d**, and the PCE in **Supplementary Fig. S3e**. Furthermore, the total system efficiency evaluated using **Equation (S3.1)** is also plotted in **Supplementary Fig. S3e**, with the ratio between $\eta_{\text{sys}}$ and PCE given in **Supplementary Fig. S3f**. From **Supplementary Fig. S3**, it's clear that for all photovoltaic materials $V_{\text{oc}}$ and $V_{\text{mpp}}$ grow linearly with $N_{\text{cells}}$, whilst $I_{\text{mpp}}$ is inversely proportional to $N_{\text{cells}}$. Consequently, as $\delta$ is small relative to the side-length of the module, PCE is mostly independent of $N_{\text{cells}}$. However, as the EHB has an optimal voltage range, then for each photovoltaic material as $N_{\text{cells}}$ increases from one, $\eta_{\text{sys}}$ increases until a maximum is reached. The optimal number of cells for a given photovoltaic material depends on its bandgap and voltage losses, as well as the used EHB. In any case, after the maximum is reached, $\eta_{\text{sys}}$ tends rapidly to zero as $N_{\text{cells}}$ continues to increase – in other words, adding more cells will not improve the efficiency of the energy-harvesting system.

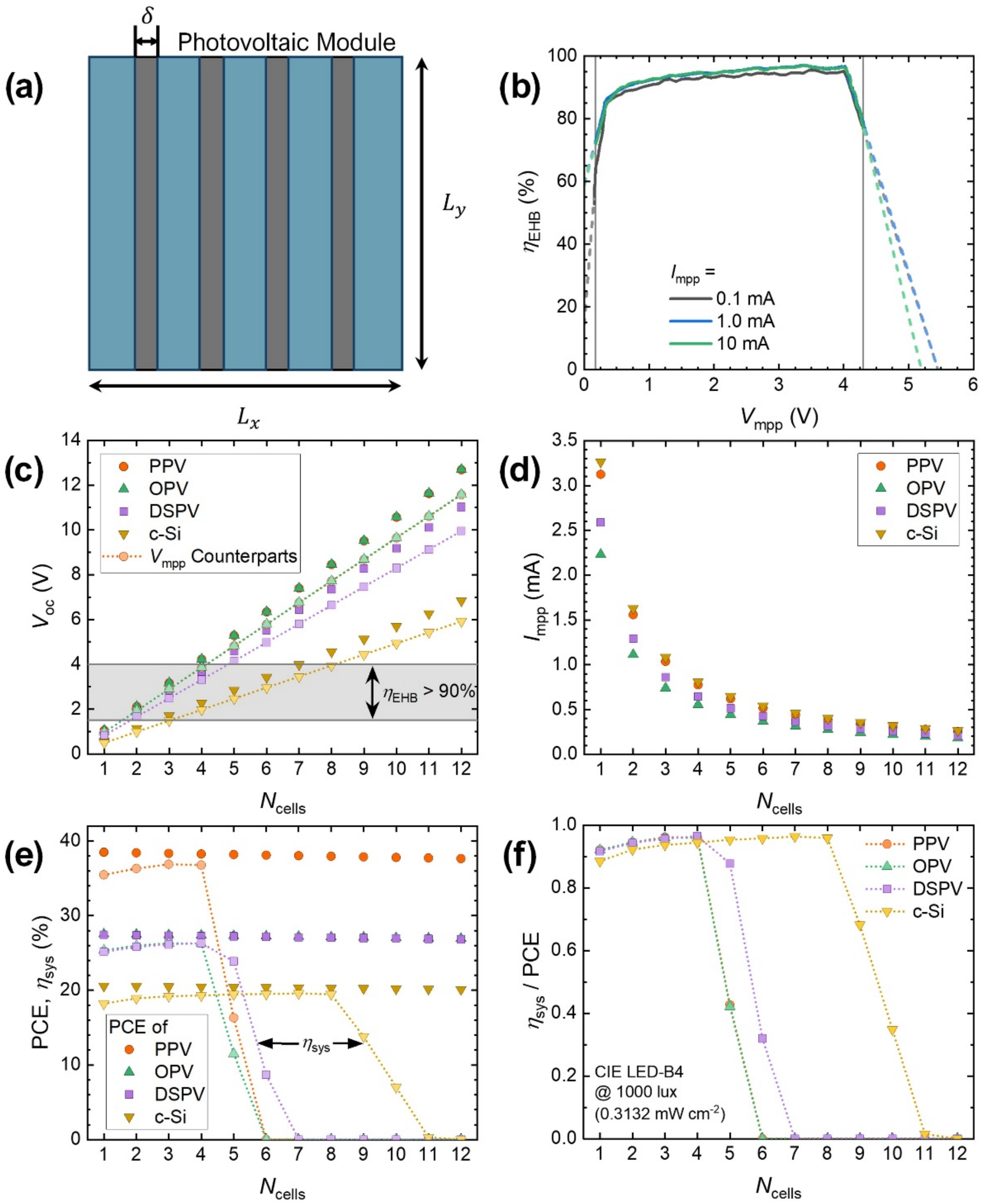

**Supplementary Fig. 3 | Effect of number of cells ($N_{\mathrm{cells}}$) in a photovoltaic module on the total efficiency of the energy-harvesting system it serves. a,** Illustration of a photovoltaic module of fixed area $A_{\mathrm{module}} = L_x L_y = 25\ \mathrm{cm}^2$, where each cell is separated from its neighbours by interconnect width $\delta$. **b,** Efficiency of the e-peas AEM13920 energy-harvesting board ($\eta_{\mathrm{EHB}}$) versus maximum power point voltage ($V_{\mathrm{mpp}}$) of the module[104], where each curve indicates $\eta_{\mathrm{EHB}}$ for different maximum power point currents ($I_{\mathrm{mpp}}$). Note that the dashed lines indicate where $\eta_{\mathrm{EHB}}$ has been extrapolated[102]. For modules of varied $N_{\mathrm{cells}}$ under CIE LED-B4 illumination at an irradiance of $0.3132\ \mathrm{mW\ cm^{-2}}$ (or an illuminance of 1000 lx), the **c**, open-circuit voltage ($V_{\mathrm{oc}}$) and $V_{\mathrm{mpp}}$, **d,** $I_{\mathrm{mpp}}$, and **e,** power-conversion efficiency (PCE) and total system efficiency ($\eta_{\mathrm{sys}}$) are shown for the four photovoltaic systems considered in **Supplementary Fig. S2**. In **c**, the grey shaded region indicates voltages that produce $\eta_{\mathrm{EHB}} > 90\ \%$. Finally, the ratio between $\eta_{\mathrm{sys}}$ and PCE is shown in **f**.

From one indoor location to another – and even from hour to hour – indoor photovoltaics are susceptible to order-of-magnitude variations in light intensity. To explore how such variations affect the efficiency of the energy-harvesting systems discussed in **Supplementary Fig. 3**, irradiance-dependent system efficiencies were determined under CIE LED-B4 illumination, with the results indicated in **Supplementary Fig. 4** for the four photovoltaic systems considered thus far. From this figure, it's clear that for small $N_{\mathrm{cells}}$, $\eta_{\mathrm{sys}}$ follows the expected pattern in PCE by growing logarithmically with the irradiance of the incident light. Alternatively, for large $N_{\mathrm{cells}}$, the voltage generated by the photovoltaics surpasses the optimal voltage for the considered EHB, causing $\eta_{\mathrm{sys}} \rightarrow 0$ at high irradiances. Consequently, the optimal number of cells for a given photovoltaic module depends on its voltage output; this optimal value is therefore highly system-specific. To achieve the highest $\eta_{\mathrm{sys}}$ across typical indoor light intensities (from 50 to 1000 ls; indicated by the vertical dotted lines in **Supplementary Fig. S4**) without risking going over-voltage, the PPV, OPV, and DSPV systems all require $N_{\mathrm{cells}} = 4$. The c-Si device, on the other hand, generates a smaller voltage and therefore requires $N_{\mathrm{cells}} = 7$ to optimize $\eta_{\mathrm{sys}}$. Note that at low irradiances, the PCE can be heavily reduced by parasitic shunt/parallel resistances, which can be particularly prevalent in solution-processible photovoltaics.

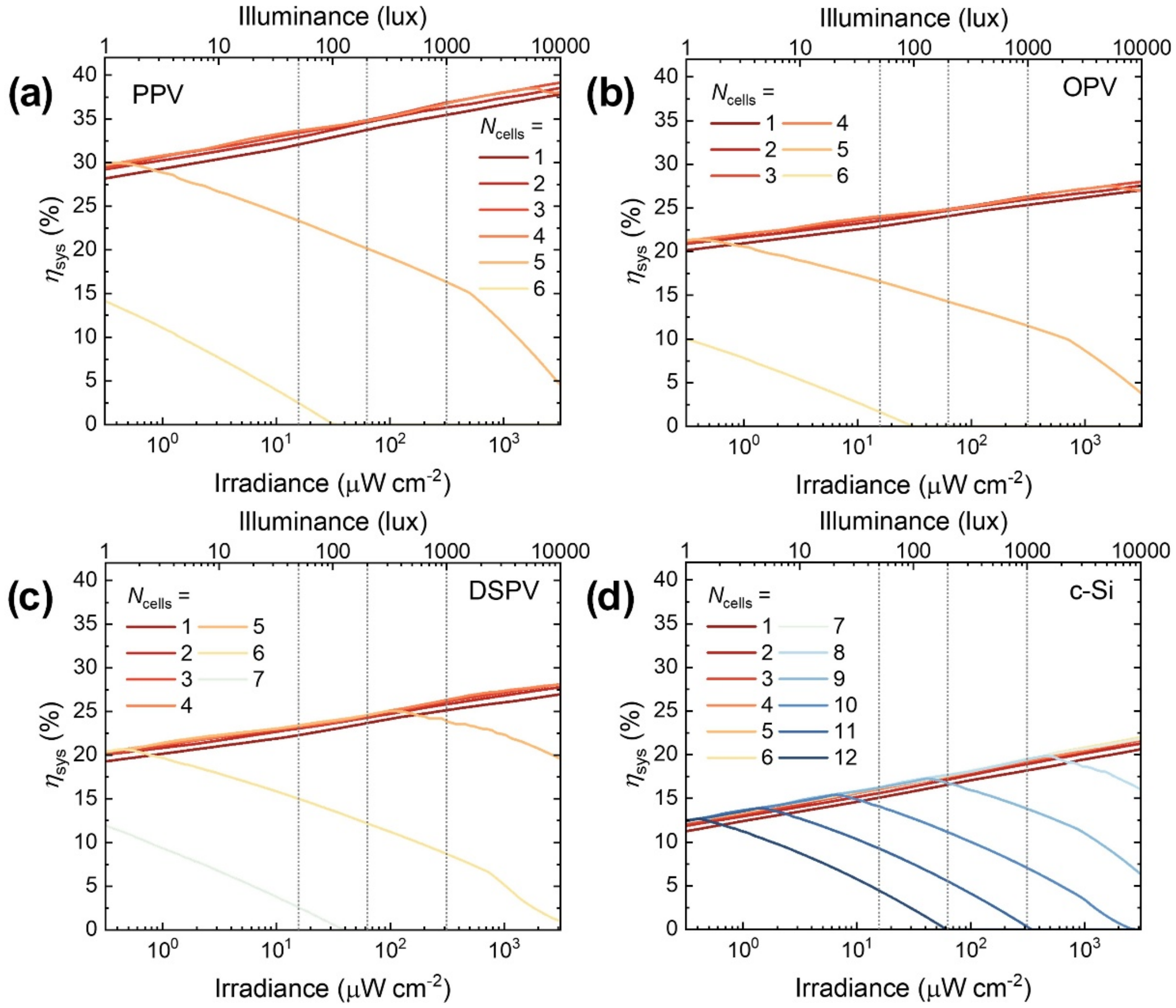

**Supplementary Fig. 4 | Light intensity-dependent system efficiencies of square photovoltaic modules** with area $A_{\mathrm{module}} = 25\ \mathrm{cm}^2$ and a varied $N_{\mathrm{cells}}$ (under CIE LED-B4 illumination). These include **a,** the perovskite photovoltaic (PPV), **b,** the organic photovoltaic, **c,** the dye-sensitized photovoltaic (DSPV), and **d,** the c-Si photovoltaic discussed in **Figure 1** of the main text. In each panel, the radiometric units of irradiance

are used on the lower horizontal scale, while the photometric units of illuminance are used on the upper horizontal scale, with the vertical dotted lines indicating illuminances of 50, 200, and 1000 lx, as specified in indoor photovoltaic testing standard IEC TS 62607-7-2:2023[101].

**Supplementary Note 4 | Energy Harvested in a Typical Deployment Scenario**

With photovoltaic modules based on the four materials systems (PPV, OPV, DSPV, and c-Si) now optimized for use with the e-peas AEM13920 EHB under LED-B4 illumination, the applicability of these modules under realistic illumination conditions was explored using real-world irradiance data measured by Seunarine et al[105]. These irradiance spectra are shown at hourly intervals in **Supplementary Fig. 5**, illustrating how ambient lighting conditions in internal spaces can fluctuate throughout the day; being characterized by daylight (or AM1.5G-like spectral irradiance filtered through a glass window) near noon, and by artificial light from a white LED source in the evening. Note that all of the original spectra have been re-normalized such that the highest-measured light intensity (at 12 noon) would be 1000 lx[102]. Increasing or decreasing this peak light intensity will impact $\eta_{\mathrm{sys}}$, as explored in the previous note.

Using the spectral irradiances plotted in **Supplementary Fig. 5a**, the efficiencies of the energy-harvesting systems based on the four photovoltaic modules were determined using the optimal cell numbers identified in **Supplementary Note 3**, with the resultant $\eta_{\mathrm{sys}}$ plotted versus time of day in **Supplementary Fig. 5b.** The total energy generated by the photovoltaic modules and then stored by the EHB by the end of the day are indicated in **Supplementary Fig. 5c**. It is clear from **Supplementary Fig. 5** that only a fraction of the 83.56 mWhr energy in the light incident on the photovoltaic module will ultimately be stored by the energy-harvesting system, with most power being lost by the photovoltaics. Nevertheless, as the optimal $N_{\mathrm{cells}}$ have been used (for the considered illumination conditions) most of the energy generated by the photovoltaic modules is stored by the EHB, with the PPV providing the most energy and the c-Si photovoltaic providing the least.

To provide some context as to how the energies generated stored by the different systems translate to real-world IoT sensors, a second vertical scale representing the maximum number of daily measurements is provided in **Supplementary Fig. 5c**. As described by Kay et al.[102], this relationship can be derived by assuming a reasonable power consumption per duty cycle of the IoT device, allowing the maximum number of measurements/transmissions per day to be determined from the stored energy (assuming a 100% discharge efficiency from the battery/capacitor to the IoT device). This second scale suggests that an energy-harvesting system using the PPV instead of the c-Si photovoltaic could store enough power to nearly double the number of daily IoT sensor duty cycles from around 2000 to 3500 cycles per day.

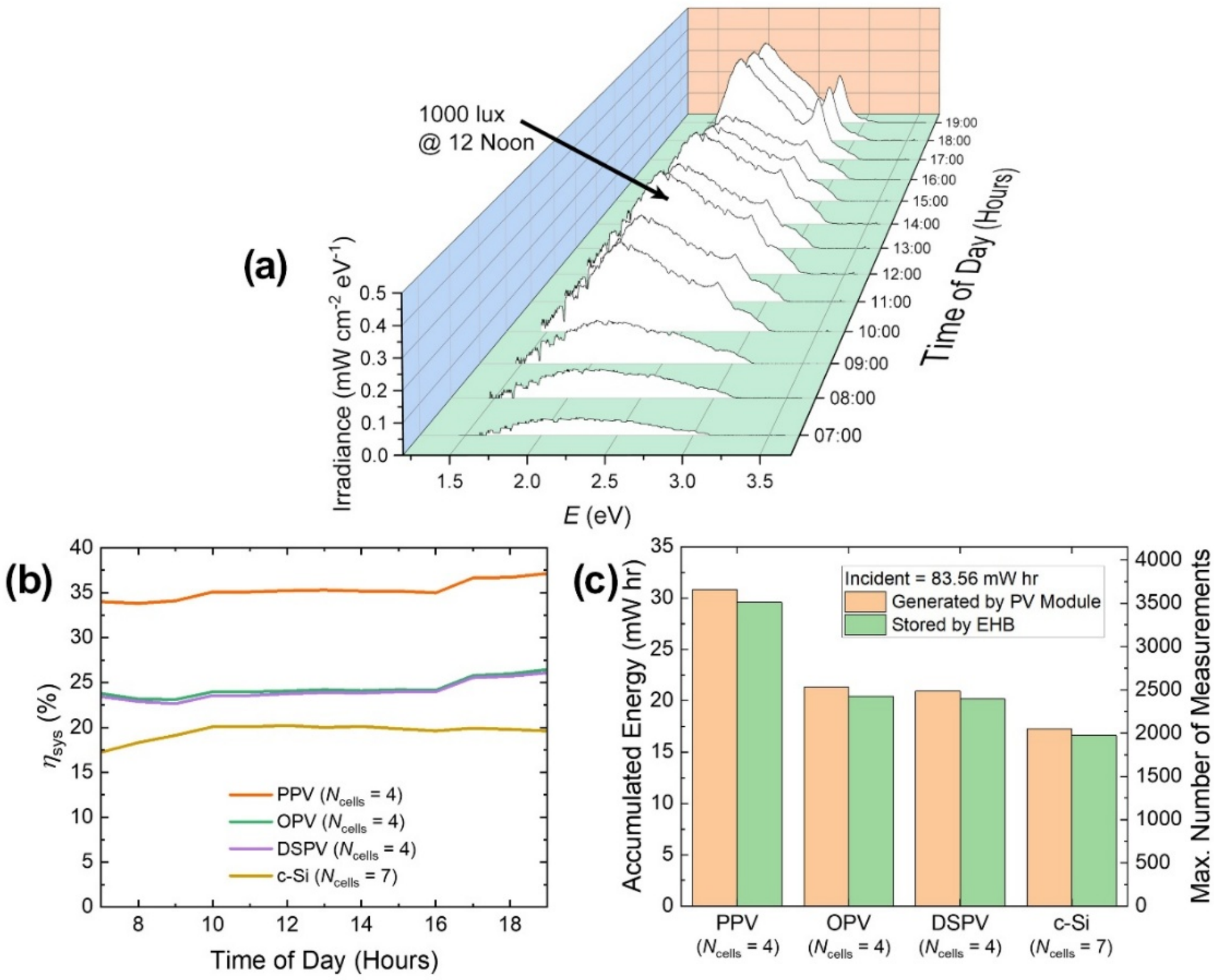

**Supplementary Fig. 5 | Energy generated and stored by energy-harvesting systems based on different photovoltaic modules throughout a 12-hour period in a typical office**, where the hourly spectral irradiances were adapted from measurements by Seunarine et al.[105] **a,** Spectral irradiance throughout a typical day, including an interplay between daylight and artificial light from an LED source. These spectra have been re-normalised such that the highest measured illuminance during the day (at 12 noon) would be 1000 lux. Using these irradiances, the **b** system efficiencies ($\eta_{\text{sys}}$) of the PPV, OPV, DSPV, and c-Si photovoltaic modules were simulated. In turn, the **c** energy generated and stored by the different energy-harvesting systems was evaluated, where a total of 83.56 mWhr of energy is incident on the devices throughout the 12-hour period. Note that the second vertical scale in **c** indicates an estimate for the maximum number of daily IoT sensor measurements (assuming 100% discharge efficiency from the battery/capacitor), as described by Kay et al.[102]

## Supplementary Note 5 | Shunt Resistance and Leakage Losses in Indoor Photovoltaics

Leakage becomes increasingly important as the photocurrent falls from 1-sun illumination to indoor illumination. However, the differential resistance obtained from a current–voltage curve does not, by itself, identify a physical shunt pathway. This note distinguishes these quantities and relates the permissible leakage to a chosen loss in maximum output power.

### Interpreting the Near-Short-Circuit Slope

We retain the current convention of **Supplementary Note 1**: The total current density ($J$) is negative in the power-generating quadrant. For a specified parallel leakage pathway,

$$J(V,\Phi) = J_{\mathrm{PV}}(V_j,\Phi) + J_{\mathrm{leak}}(V_j,\Phi); \qquad V_j = V - r_s J(V,\Phi) \quad \textbf{(S5.1)}$$

Here, $J_{\mathrm{PV}}$ describes the junction without that pathway, $\Phi$ specifies the illumination, and $V_j$ is the internal junction voltage. The series resistance–area product is $r_{\mathrm{s}} = AR_{\mathrm{s}}$, where $R_{\mathrm{s}}$ is the terminal series resistance in Ω and $A$ is the designated device area. For an ohmic pathway, $J_{\mathrm{leak}} = V_j/\rho_{\mathrm{sh}}$, with $\rho_{\mathrm{sh}} = AR_{\mathrm{sh}}$ being the specific shunt resistance in Ω cm², whilst the terminal shunt resistance $R_{\mathrm{sh}}$ is in Ω. A parallel branch is a circuit description; its physical origin must be established separately.

At fixed illumination and in steady state, let the gradient in the total current density with respect to the internal junction voltage be defined by $g = \partial(J_{\mathrm{PV}} + J_{\mathrm{leak}})/\partial V_j$. For a lumped device with constant $r_{\mathrm{s}}$, differentiation gives

$$\frac{dJ}{dV} = \frac{g}{1 + r_{\mathrm{s}} g}; \qquad \rho_{\mathrm{diff}} \equiv \left(\frac{dJ}{dV}\right)^{-1} = r_{\mathrm{s}} + \frac{1}{g} \quad \textbf{(S5.2a)}$$

If the leakage pathway is ohmic, as described above, then $g = g_{\mathrm{PV}} + \frac{1}{\rho_{\mathrm{sh}}}$, where $g_{\mathrm{PV}} = \partial J_{\mathrm{PV}}/\partial V_j$ at the same operating point. Defining $s = dJ/dV$ gives the extraction relation

$$\rho_{\mathrm{sh}} = \left[\frac{s}{1 - r_{\mathrm{s}} s} - g_{\mathrm{PV}}\right]^{-1} \tag{S5.2b}$$

A positive, finite shunt resistance requires the bracketed conductance to be positive. The extraction therefore requires an independently justified junction conductance as well as the measured slope; treating all of that slope as leakage is an additional approximation.

Thus $\rho_{\mathrm{diff}} \approx \rho_{\mathrm{sh}}$ near short circuit only when the leakage is linear, the photocurrent is effectively voltage-independent, the junction differential conductance is negligible relative to $1/\rho_{\mathrm{sh}}$, $r_{\mathrm{s}} \ll \rho_{\mathrm{sh}}$, and transient contributions are absent. At terminal short circuit, $V = 0$ but $V_j = r_{\mathrm{s}} J_{\mathrm{SC}}$, where $J_{\mathrm{SC}} = -J(0) > 0$; the internal voltage drop $V_j$ may be set to zero only if it is negligible on the relevant transport and diode voltage scales.

Voltage-dependent collection and recombination can produce an apparent shunt even without an independent conducting defect. Transport limitations in organic photovoltaics provide a demonstrated example[106]. The photoshunt is an illumination-induced slope contribution, often expressed as $1/\rho_{\mathrm{photo}} = 1/\rho_{\mathrm{p,light}} - 1/\rho_{\mathrm{p,dark}}$. In the applicable regime, $\rho_{\mathrm{photo}}$, rather than its inverse, varies approximately inversely with light intensity.[107] An illuminated linear slope therefore does not prove ohmic leakage. Likewise, a small-signal recombination resistance is not interchangeable with a DC ohmic shunt, and ionic or capacitive transients can invalidate a slope taken from a single voltage sweep.

**A Leakage Criterion at the Operating Point**

For a defined leakage-free reference, let $p_0(V)$ be the delivered power density and let $V_{\mathrm{mpp,0}}$ and $j_{\mathrm{mpp,0}} > 0$ denote its maximum-power voltage and delivered-current magnitude. In the following operating-point analysis, series resistance is neglected, so $V_j = V$. Adding a separable leakage current then gives $p(V) \approx p_0(V) - VJ_{\mathrm{leak}}(V)$, with other junction properties held fixed. To first order in a small leakage perturbation,

$$\frac{\Delta p_{\mathrm{mpp}}}{p_{\mathrm{mpp,0}}} \approx \frac{J_{\mathrm{leak}}(V_{\mathrm{mpp,0}})}{j_{\mathrm{mpp,0}}} \tag{S5.3}$$

Here, $\Delta p_{\mathrm{mpp}} = p_{\mathrm{mpp,0}} - p_{\mathrm{mpp}}$ is the change in delivered power density after adding the leakage current. For a tolerated fractional loss $\varepsilon$, it is required that $J_{\mathrm{leak}}\left(V_{\mathrm{mpp,0}}\right) < \varepsilon\, j_{\mathrm{mpp,0}}$. This result follows from stationarity at the unperturbed maximum, provided the maximum is smooth, interior and non-degenerate. Within the additive approximation, evaluating the perturbed power at $V_{\mathrm{mpp,0}}$ also bounds the optimised loss from above; re-optimising the voltage can only recover power. For finite series resistance, the internal voltage and its feedback on junction and leakage currents must instead be retained; a small terminal voltage drop alone need not justify the approximation for strongly non-linear leakage. A near-zero-bias slope alone does not supply $J_{\mathrm{leak}}(V_{\mathrm{mpp,0}})$.

For leakage demonstrated to be ohmic over the operating range, the corresponding design requirement is

$$\rho_{\mathrm{sh}} > \frac{V_{\mathrm{mpp,0}}}{\varepsilon j_{\mathrm{mpp,0}}} = \frac{{V_{\mathrm{mpp,0}}}^2}{\varepsilon p_{\mathrm{mpp,0}}} \tag{S5.4}$$

The threshold depends on illumination, the reference device, and the tolerated loss. **Supplementary Table 2** gives reproducible examples for a single-diode reference at 300 K with an ideality factor of 1.3 and zero series resistance. The reference is calibrated by $J_{ph}$ = 20 mA cm$^{-2}$ and $V_{OC,0}$ = 1.15 V at 1-sun illumination. For the indoor examples, $J_{ph}$ = 150 µA cm$^{-2}$ at 1000 lx and scales linearly with illuminance at fixed spectral shape. These are illustrative parameters, not fits to the devices compared in the main text. At a 5% tolerance, the thresholds are approximately 1.21×10$^5$, 5.71×10$^5$ and 2.16×10$^6$ Ω cm$^2$ at 1000, 200 and 50 lx; for a 1% tolerance they are five times larger.

The characteristic resistance $\rho_{CH} = V_{OC,0}/J_{SC}$ uses the leakage-free voltage. For this model, $\rho_{CH}$ is approximately 58 Ω cm$^2$ at 1-sun and 6.6×10$^3$, 3.1×10$^4$ and 1.2×10$^5$ Ω cm$^2$ at 1000, 200 and 50 lx. Values of $\rho_{sh}$ near $\rho_{CH}$ indicate a transition; the deeply shunt-limited asymptote requires $\rho_{sh} \ll \rho_{CH}$ and the assumptions below. Equation **(S5.4)** sets the resistance required for a specified small loss. Its 1-sun 5% threshold is 1.07×10$^3$ Ω cm$^2$, illustrating why leakage tolerable outdoors can be consequential indoors.

For a voltage-independent photocurrent and an ohmic shunt, the opposite, deeply shunt-limited regime occurs when diode current and series losses are negligible across the power-generating interval. The device then approaches a photocurrent source loaded by a resistor:

$$V_{OC} \approx J_{SC}\rho_{sh}; \qquad V_{mpp} \approx \frac{J_{SC}\rho_{sh}}{2}; \qquad j_{mpp} \approx \frac{J_{SC}}{2} \tag{S5.5}$$

$$p_{mpp} \approx \frac{J_{SC}{}^2\rho_{sh}}{4}; \qquad \mathrm{FF} \to 0.25 \tag{S5.6}$$

With fixed spectrum and an intensity-independent $\rho_{sh}$, power density then scales quadratically with irradiance and efficiency approximately linearly. This limiting behaviour must not be extrapolated to devices whose leakage, collection or recombination changes with illumination.

**Materials Interpretation and Scale Up**

Suppressing leakage requires identifying the current pathway. Film discontinuities, electrode penetration, poorly isolated edges and defective interconnects can create parasitic paths, but their electrical response need not be ohmic. A local contact between dissimilar transport layers can instead be rectifying. Interface recombination, inefficient charge collection and a separate conducting defect can all reduce measured performance while requiring different materials interventions.

For an intensity-independent ohmic shunt pathway, $j_{\mathrm{mpp},0}$ decreases approximately in proportion with irradiance, while $V_{\mathrm{mpp},0}$ decreases more slowly. Its fractional power penalty therefore grows approximately inversely with irradiance, with a voltage-dependent correction. A strongly rectifying pathway normalized to the same current at a high-light operating point can have a much smaller penalty indoors because the operating voltage falls. Illumination-dependent parasitic conduction requires its own intensity-dependent analysis.

For equipotential ohmic pathways, the terminal conductances of independent defects add in parallel,[108]

$$\frac{1}{R_{\mathrm{sh,tot}}} = \frac{1}{R_{\mathrm{sh,bg}}} + \sum_i \frac{1}{R_{\mathrm{defect},i}}, \quad \textbf{(S5.7)}$$

where $R_{\mathrm{sh,bg}}$ is the background terminal shunt resistance. Normalization by area gives an area-independent $\rho_{\mathrm{sh}}$ only if the leakage conductance scales with area under comparable conditions. Sparse defects, edge leakage, scribes and lateral sheet resistance effects can violate this scaling[108,109]. Device geometry and defect distributions across devices should be reported; a selected small-area result alone does not establish the leakage expected in a module. Segmentation can reduce the impact of some defects, but also introduces interconnect losses and possible new leakage paths[102].

A defensible comparison should report the designated area and perimeter; spectrum and irradiance, with illuminance as supplementary information. Temperature, dark or illuminated conditions, voltage interval, scan rate, direction and preconditioning, and the extraction model are also important parameters to report. Where possible, steady-state dark and illuminated measurements should be combined with intensity, temperature and geometry dependence to distinguish conduction from collection or recombination. For non-linear pathways, $J_{\text{leak}}$ near the intended operating voltage should be reported rather than assigning a single ohmic resistance.

The design thresholds in **Supplementary Table 2** apply to its specified reference model and do not establish an intrinsic ranking of photovoltaic material classes. Likewise, the calculated saturation-current parameters in **Supplementary Note 1** should be distinguished from direct measurements of reverse-bias leakage; comparisons require consistent optical-state cutoffs and treatment of non-radiative losses.

**Supplementary Table 2 | Illustrative shunt-resistance requirements**

| Illumination Condition | $J_{ph}$ (μA cm⁻²) | $V_{mpp,0}$ (V) | $j_{mpp,0}$ (μA cm⁻²) | $\rho_{sh}$ at 5% (Ω cm²) | Optimized loss (%) |
|---|---|---|---|---|---|
| 1-sun | 20 000 | 1.033778 | 19370.282 | 1,067 | 4.984 |
| 1000 lx | 150 | 0.874762 | 144.450 | 121,116 | 4.982 |
| 200 lx | 30 | 0.822658 | 28.823 | 570,844 | 4.981 |
| 50 lx | 7.5 | 0.777873 | 7.189 | 2,163,949 | 4.980 |

Resistance–area products are required to limit the maximum power loss to 5% for the single-diode reference in **Supplementary Note 5**. The current convention is $J_{\text{PV}}$ =

$J_{\text{dark}} - J_{\text{ph}}$, with the total current then being given by Equation **(S5.1)**; the delivered current is $-J$. The ohmic leakage term is $V/\rho_{\text{sh}}$ and series resistance is zero.

$$J_{\text{PV}}(V) = J_0\left[\exp\left(\frac{V}{a}\right) - 1\right] - J_{\text{ph}}; \quad a = nk_{\text{B}}T/q \tag{S5.8a}$$

$$J_0 = \frac{0.020\ \text{Acm}^{-2}}{\exp\left(1.15\frac{V}{a}\right) - 1}, \ n = 1.3, \ T = 300\ \text{K} \tag{S5.8b}$$

For these examples, the dark saturation current density ($J_0$) is held constant. The indoor current–illuminance relation is specified as $J_{\text{ph}}$ = (150 μA cm$^{-2}$)(illuminance/1000 lx) for a fixed spectrum; it is not a universal conversion between lux and photocurrent. For each condition, the leakage-free maximum is obtained by maximizing $-VJ(V)$, and the threshold shunt resistance is then evaluated using Equation **(S5.4)**.

In Supplementary Table 2, the loss column is calculated by re-optimizing voltage after adding the ohmic branch at the unrounded resistance threshold. Displayed resistance values are rounded to the nearest Ω cm². The optimized losses are slightly smaller than 5%, consistent with the conservative operating-point criterion. These examples provide design scales rather than measurements of any photovoltaic material class.

**Estimating Shunt Resistances from Current-Voltage Curves**

To estimate the shunt resistance of the photovoltaic device from its current-voltage curve, the effects of voltage-dependent collection and recombination are assumed minimal, such that the total shunt resistance can be estimated from the rate-of-change of the current density with respect to the voltage bias near short-circuit conditions. After accounting for the parasitic resistances, Equation **(S1.6)** can be rewritten as[93]

$$J(V) = J_0(V_j)\left[\exp\left(\frac{qV_j}{k_\mathrm{B}T}\right) - 1\right] - J_\mathrm{ph}(V_j) + \frac{V_j}{\rho_\mathrm{sh}}, \tag{S5.9}$$

where, as above, the internal voltage drop across the diode ($V_j$) relates to the total voltage drop ($V$) via $V_j = V - r_\mathrm{s}J(V)$. After taking the first derivative of this expression with respect to voltage, one finds that Equation **(S5.2a)** can be rewritten as

$$\frac{\partial J(V)}{\partial V} \approx \frac{\frac{\partial J_0(V_j)}{\partial V}\left[\exp\left(\frac{qV_j}{k_\mathrm{B}T}\right) - 1\right] + \frac{qJ_0(V_j)}{k_\mathrm{B}T}\exp\left(\frac{qV_j}{k_\mathrm{B}T}\right) + \frac{1}{\rho_\mathrm{sh}}}{1 + r_\mathrm{s}\left[\frac{qJ_0(V_j)}{k_\mathrm{B}T}\exp\left(\frac{qV_j}{k_\mathrm{B}T}\right) + \frac{1}{\rho_\mathrm{sh}}\right]} \tag{S5.10}$$

where a weak voltage dependence has been assumed in the photocurrent density due to the reasons stated above. This expression can be simplified at short-circuit conditions, where $V = V_j \approx 0$, giving

$$\left.\frac{\partial J(V)}{\partial V}\right|_{V=0} \approx \frac{\frac{qJ_{00}}{k_\mathrm{B}T} + \frac{1}{\rho_\mathrm{sh}}}{1 + r_\mathrm{s}\left[\frac{qJ_{00}}{k_\mathrm{B}T} + \frac{1}{\rho_\mathrm{sh}}\right]}. \tag{S5.11}$$

Here, the quantity $J_{00}$ denotes the dark saturation current density under negligible voltage bias (where $w(E, V_j) \to 1 \; \forall \; E > V_j = 0$), which gives $J_{00} = q\int_0^\infty \mathrm{EQE}_\mathrm{PV}(E)\Phi_\mathrm{Planck}(E)\,\mathrm{d}E$. Consequently, the device's shunt resistance may be estimated from the gradient of the current density near short-circuit conditions ($J'_{\mathrm{sc}} \equiv \left.\frac{\partial J(V)}{\partial V}\right|_{V=0}$) using

$$\rho_{\mathrm{sh}} \approx \frac{1 - r_s J'_{\mathrm{sc}}}{J'_{\mathrm{sc}}\left[1 + \frac{q r_{\mathrm{s}} J_{00}}{k_{\mathrm{B}} T}\right] - \frac{q J_{00}}{k_{\mathrm{B}} T}}. \qquad \textbf{(S5.12)}$$

While this expression holds for arbitrary values of $r_{\mathrm{s}}$, $J_{00}$, and $J'_{\mathrm{sc}}$, further simplifying approximations can be made**.** For example, if $J_{00} \sim 10^{-12}$ mA cm$^{-2}$, then $r_{\mathrm{s}} J_{00} \ll \frac{k_{\mathrm{B}} T}{q}$ for all reasonable values of $r_{\mathrm{s}}$. In this case, Equation **(S5.12)** simplifies to

$$\rho_{\mathrm{sh}} \approx \frac{1 - r_{\mathrm{s}} J'_{\mathrm{sc}}}{J'_{\mathrm{sc}} - \frac{q J_{00}}{k_{\mathrm{B}} T}} \approx \frac{1}{J'_{\mathrm{sc}} - \frac{q J_{00}}{k_{\mathrm{B}} T}}. \qquad \textbf{(S5.13)}$$

Using this expression, the shunt resistances of the devices considered throughout Figure 1 were estimated (if the values were not already available in the source publications), with the results being summarized in **Supplementary Table 3**.

**Supplementary Table 3 | Summary of the parameters used to estimate shunt resistance** of the indoor photovoltaics from different material classifications (if the values were not already provided), including the devices considered in **Figure 1** of the main text.

| **System** (***PV-Simulator Ref.***) | **Bandgap (eV)** | **Estimated $R_{sh}$. $10^5$ ($\Omega$ cm$^2$)**[a] | **$J_{00}$ (mA cm$^{-2}$)**[b] | **Ref.** |
|---|---|---|---|---|
| $NiO_x$-based Perovskite | 1.69 | 0.07 | $2.66\times10^{-23}$ | 110 |
| PBDB-TF:GS-ISO | 1.81 | 4.3 | $7.66\times10^{-25}$ | 111 |
| XY1:L1 (0.06 M CuII) | 1.89 | 14.7 | $5.36\times10^{-25}$ | 112 |
| c-Si | 1.12 | 7.6 | $1.09\times10^{-13}$ | 113 |
| Se | 1.95 | 0.7 | $1.01\times10^{-26}$ | 114 |

[a.] In the cases where $\rho_{sh}$ was not specified in the source publication, it was calculated using Equation **(S5.13)**.

[b.] Note that the $J_{00}$ values were calculated for states that contribute to the thermodynamic limit, as described by Zarrabi et al.[96] States that are well below the gap may not contribute to the thermodynamic limit.

**Supplementary Table 4 | Benchmark of two-dimensional (2D) and three-dimensional (3D) ceramic linear dielectric-based electrostatic capacitors.** Abbreviations: thickness ($t$), scaffold depth ($d$), areal energy storage density ($U_a$), average areal power density ($P_a$), and maximum applied voltage ($V_{max}$). When only volumetric energy density ($U_v$) was reported, $U_a$ was computed as $U_a = U_v \times t$. $P_a$ was estimated as $P_a = U_a.f$, where the frequency $f$ was determined either as (1) $f = \Delta t_{fall}^{-1}$ with $\Delta t_{fall}$ being the reported fall time of the voltage waveform used in capacitance-frequency or polarization-voltage measurements, or (2) the measurement frequency reported for those same tests. Since different pulse trains may be used, this approach may underestimate $P_a$. In cases where $V_{max}$ was not reported, it was computed as $V_{max} = E_{max}.t$, where $E_{max}$ is the reported maximum applied electric field.

| **Dielectric** | ***t*** **(nm)** | ***d*** **(3D) µm** | $U_a$ **(J cm$^{-2}$)** | ***f*** **Hz** | $P_a$ **(W cm$^{-2}$)** | $V_{max}$ **(V)** | **Ref.** |
|---|---|---|---|---|---|---|---|
| | | | 2D | | | | |
| $Al_2O_3$ | 216 | N/A | $6.5\times10^{-5}$ | $10^5$ | 6.5 | 60.0 | 115 |
| $Al_2O_3$ | 208 | N/A | $1.2\times10^{-4}$ | $10^6$ | $1.2\times10^2$ | 82.0 | 116 |
| $Al_2O_3/HfO_2/Al_2O_3$ | 61 | N/A | $1.1\times10^{-5*}$ | $10^5$ | 1.1 | 9.2*** | 117 |
| $Au:Al_2O_3$ | 211 | N/A | $2.8\times10^{-4}$ | $10^6$ | $2.8\times10^2$ | 69.0 | 116 |
| $(Si, Mg):Al_2O_3$ | 212 | N/A | $2.0\times10^{-4}$ | $10^5$ | $2.0\times10^1$ | 115.0 | 115 |
| $HfO_2$ | 63 | N/A | $1.3\times10^{-4}$ | $10^4$ | 1.3 | 27.0 | 118 |
| $HfO_2$ | 10 | N/A | $1.1\times10^{-5}$ | $4\times10^4$ | $4.4\times10^{-1}$ | 3.0 | 119 |
| $HfO_2$ | 30 | N/A | $5.8\times10^{-5*}$ | $10^6$ | $5.8\times10^1$ | 9.45 | 120 |
| $Hf_{0.5}Zr_{0.5}O_2$ | 9.1 | N/A | $9.1\times10^{-6}$ | $1.7\times10^4$ | $1.6\times10^{-1}$ | 3.2 | 121 |
| $SiO_2$ | 10 | N/A | $3.2\times10^{-5*}$ | $10^5$ | $3.2\times10^3$ | 11.0*** | 122 |
| $SiO_2/TiO_2/SiO_2$ | 20 | N/A | $1.4\times10^{-5*}$ | $5\times10^5$ | 7.2 | 5.0 | 123 |
| $SrTiO_3$ | 46 | N/A | $1.3\times10^{-4*}$ | $10^6$ | $1.3\times10^2$ | 10.0 | 124 |
| $TaO_x$ | 200 | N/A | $5.5\times10^{-4}$ | $3\times10^6$ | $1.7\times10^3$ | 101.0 | 125 |
| $TiO_2$ | 20 | N/A | $1.3\times10^{-5*}$ | $10^6$ | $1.3\times10^1$ | 3.0*** | 126 |
| $Y_2O_3/TiO_2$ | 8.3 | N/A | $2.0\times10^{-5*}$ | $10^6$ | $2.0\times10^1$ | 2.0** | 127 |
| $ZrO_2$ | 8 | N/A | $2.3\times10^{-5*}$ | $10^5$ | 2.3 | 3.5*** | 128 |
| $ZrO_2$ | 8.5 | N/A | $1.4\times10^{-5}$ | $1.7\times10^4$ | $2.3\times10^{-1}$ | 3.0 | 121 |
| $ZrO_2$ | 470 | N/A | $3.5\times10^{-3}$ | $10^3$ | 3.5 | 273.0 | 129 |
| $ZrO_2/TiO_2$ | 11 | N/A | $1.1\times10^{-5*}$ | $10^4$ | $1.1\times10^{-1}$ | 2.0 | 130 |
| | | | 3D | | | | |
| $Al_2O_3$ | 7 | 200 | $2.3\times10^{-3*}$ | $10^3$ | 2.34 | 5.1 | 131 |
| $Al_2O_3$ | 10 | 6 | $7.7\times10^{-3*}$ | $10^2$ | 7.65 | 15.0 | 132 |
| $Al_2O_3$ | 20 | 30 | $7.9\times10^{-4*}$ | $10^4$ | 7.92 | 6.0 | 133 |
| $Al_2O_3$ | 55 | 160 | $1.8\times10^{-3*}$ | $10^3$ | 1.84 | 34.1 | 134 |
| $CuO_2$ | 5-10 | 100 | $7.2\times10^{-3*}$ | $10^6$ | $7.15\times10^3$ | 10.0 | 135 |
| $HfO_2$ | 90 | 50 | $2.6\times10^{-4}$ | $10^3$ | $2.60\times10^{-1}$ | 31.5 | 136 |
| $HfO_2$ | 10 | 1.2 | $1.5\times10^{-5*}$ | $10^4$ | $1.49\times10^{-1}$ | 2.5 | 137 |
| $HfAlO_x$ | 40 | 200 | $6.1\times10^{-3}$ | n/a | $5.66\times10^{2**}$ | 16.0 | 138 |
| $SiO_2/Si_3N_4/SiO_2$ | 15 | 72 | $2.2\times10^{-3*}$ | $10^5$ | $2.24\times10^2$ | 8.0 | 139 |

* $U_a$ values estimated as $U_a = {}^1/_2 C_a \times V_{max}^2$, where $C_a$ is the reported areal capacitance density.

** $P_a$ values were either taken from the reported data or computed as $P_a = U_a/\tau$, where $\tau$ is the reported discharge time from transient charge-discharge measurements.

*** $V_{max}$ values obtained by graphical analysis.

**Supplementary Table 5 | Benchmark of two-dimensional (2D) and three-dimensional (3D) (anti)ferroelectric $Hf_{1-x}Zr_xO_2$-based electrostatic capacitors.** Abbreviations: thickness ($t$), scaffold depth ($d$), volumetric energy storage density (ESD), areal energy storage density ($U_a$), average areal power density ($P_a$), maximum applied voltage ($V_{max}$), linear dielectric (LD), antiferroelectric (AFE), ferroelectric (FE), relaxor ferroelectric (RFE), superlattice (SL), negative capacitance (NC), and multilayer (ML). When only volumetric energy density ($U_v$) was reported, $U_a$ was computed as $U_a = U_v.t$. $P_a$ was estimated as $P_a = U_a.f$, where the frequency $f$ was determined either as (1) $f = \Delta t_{fall}^{-1}$ with $\Delta t_{fall}$ being the reported fall time of the voltage waveform used in capacitance-frequency or polarization-voltage measurements, or (2) the measurement frequency reported for those same tests. Since different pulse trains may be used, this approach may underestimate $P_a$. Since different pulse trains may be used, this approach may underestimate $P_a$. In cases where $V_{max}$ was not reported, it was computed as $V_{max} = E_{max}.t$, where $E_{max}$ is the reported maximum applied electric field.

| Dielectric | $t$ (nm) | $d$ (µm) | Structure | $U_a$ (J cm$^{-2}$) | $f$ Hz | $P_a$ (W cm$^{-2}$) | $V_{max}$ (V) | Ref. |
|---|---|---|---|---|---|---|---|---|
| | | | | **2D** | | | | |
| $Al:HfO_2$ | 10 | n/a | AFE | $4.2\times10^{-5}$ | $10^3$ | $4.2\times10^{-2}$ | 3.9 | 140 |
| $Al:HfO_2$ | 6.5 | n/a | AFE | $4.2\times10^{-5}$ | $10^3$ | $4.2\times10^{-2}$ | 4.2 | 141 |
| $Si:HfO_2$ | 9 | n/a | AFE | $3.6\times10^{-5}$ | $10^3$ | $3.6\times10^{-2}$ | 3.0 | 142 |
| $Si:HfO_2$ | 10 | n/a | AFE | $6.1\times10^{-5}$ | $10^3$ | $6.1\times10^{-2}$ | 4.5 | 143 |
| $Hf_{0.5}Zr_{0.5}O_2$ | 13 | n/a | FE | $1.2\times10^{-4}$ | $1.7\times10^4$ | 2.1 | 4.6 | 121 |
| $Hf_{0.5}Zr_{0.5}O_2$ | 7.1 | n/a | AFE | $3.9\times10^{-5}$ | $10^3$ | $3.9\times10^{-2}$ | 3.5 | 144 |
| $Hf_{0.5}Zr_{0.5}O_2$/ $Hf_{0.25}Zr_{0.75}O_2$ | 10 | n/a | FE/AFE | $7.2\times10^{-5}$ | $10^3$ | $7.2\times10^{-2}$ | 6.0 | 145 |
| $Hf_{0.5}Zr_{0.5}O_2$/ $Ta_2O_5$ | 25 | n/a | NC FE/LD | $2.7\times10^{-4}$ | n/a | $6.8\times10^{2*}$ | $18.0^{**}$ | 146 |
| Si: $Hf_{0.5}Zr_{0.5}O_2$ | 10 | n/a | AFE | $5.0\times10^{-5}$ | $10^3$ | $5.0\times10^{-2}$ | 4.0 | 147 |
| Al: $Hf_{0.5}Zr_{0.5}O_2$ | 15 | n/a | AFE | $1.5\times10^{-4}$ | $10^4$ | 1.5 | 9.3 | 148 |
| Al: $Hf_{0.5}Zr_{0.5}O_2$ | 10 | n/a | AFE | $1.4\times10^{-4}$ | n/a | $1.4\times10^{2*}$ | 8.0 | 149 |
| La: $Hf_{0.5}Zr_{0.5}O_2$ | 10 | n/a | AFE | $4.9\times10^{-5}$ | $10^4$ | $4.9\times10^{-1}$ | 4.0 | 150 |
| $Hf_{0.3}Zr_{0.7}O_2$ | 7.7 | n/a | AFE | $3.5\times10^{-5}$ | $10^3$ | $3.5\times10^{-2}$ | 3.5 | 151 |
| $Hf_{0.3}Zr_{0.7}O_2/Al_2O_3/Hf_{0.3}Zr_{0.7}O_2$ | 10.1 | n/a | AFE/LD/AFE | $5.6\times10^{-5}$ | $10^4$ | $5.6\times10^{-1}$ | 4.5 | 152 |
| $Hf_{0.2}Zr_{0.8}O_2$ | 9 | n/a | NC AFE | $1.0\times10^{-4}$ | $10^4$ | 1.0 | $5.4^{**}$ | 153 |
| $Hf_{0.2}Zr_{0.8}O_2/Al_2O_3$ ML | 94 | n/a | NC AFE/ LD | $7.6\times10^{-4}$ | n/a | $2.1\times10^{3*}$ | $52.0^{**}$ | 153 |
| $Hf_{0.2}Zr_{0.8}O_2/Al_2O_3$/ $Hf_{0.2}Zr_{0.8}O_2$ | 16 | n/a | AFE/LD/AFE | $1.4\times10^{-4}$ | $10^3$ | $1.4\times10^{-1}$ | 8.8 | 154 |
| $ZrO_2$ | 5.3 | n/a | AFE | $3.1\times10^{-5}$ | $10^3$ | $3.1\times10^{-2}$ | $3.2^{**}$ | 155 |
| $ZrO_2$ | 12 | n/a | AFE | $4.4\times10^{-5}$ | $10^5$ | 4.4 | $3.2^{**}$ | 156 |
| $ZrO_2$ | 10 | n/a | AFE | $3.8\times10^{-5}$ | $4\times10^4$ | 1.5 | 3.0 | 119 |
| $ZrO_2$ | 10.3 | n/a | AFE | $8.7\times10^{-5}$ | $1.7\times10^4$ | 1.5 | 4.1 | 121 |
| $ZrO_2/HfO_2$ | 8.8 | n/a | AFE | $4.4\times10^{-5}$ | $10^3$ | $4.4\times10^{-2}$ | 3.5 | 157 |
| $ZrO_2/Hf:Al_2O_3$ | 7 | n/a | FE/ LD | $3.8\times10^{-5}$ | $10^3$ | $3.8\times10^{-2}$ | 3.5 | 158 |
| $TiO_2/ZrO_2/TiO_2$ | 8.7 | n/a | LD/AFE/LD | $8.2\times10^{-5}$ | n/a | $1.4\times10^{1*}$ | 4.4 | 159 |
| $TiO_2/ZrO_2$/TiO2 | 8 | n/a | LD/AFE/LD | $9.2\times10^{-5}$ | n/a | $2.1\times10^{3*}$ | 4.3 | 160 |
| $Al:ZrO_2$ | 7 | n/a | AFE | $7.7\times10^{-5}$ | $4\times10^3$ | $3.1\times10^{-1}$ | 3.9 | 161 |
| $(Si,Ti):ZrO_2$ | 8 | n/a | AFE | $5.2\times10^{-5}$ | $10^5$ | 5.2 | 4.8 | 162 |
| | | | | **3D** | | | | |
| $Al:HfO_2$ | 10 | 8.5 | AFE | $5.0\times10^{-4}$ | $10^3$ | $5.0\times10^{-1}$ | 3.0 | 163 |
| $Si:HfO_2$ | 10 | 8 | AFE | $4.5\times10^{-4}$ | $10^2$ | $4.5\times10^{-2}$ | 3.0 | 164 |
| $Si:HfO_2/Al_2O_3/Si:HfO_2$ | 20 | 8 | AFE | $1.4\times10^{-3}$ | $10^3$ | 1.4 | 6.0 | 165 |
| $Hf_{0.5}Zr_{0.5}O_2$/ $Hf_{0.25}Zr_{0.75}O_2$ | 10 | 70 | FE/AFE | $3.6\times10^{-4}$ | $10^3$ | $3.6\times10^{-1}$ | 5.7 | 145 |
| $Hf_{0.2}Zr_{0.8}O_2/Al_2O_3$ ML | 100 | 100 | NC AFE/LD | $8.0\times10^{-2}$ | $10^6$ | $3.0\times10^5$ | 40.0 | 153 |
| $ZrO_2/Al_2O_3/ZrO_2$ | 7.5 | – | AFE | $7.0\times10^{-4}$ | $10^5$ | $7.0\times10^2$ | $4.0^{**}$ | 156 |
| $(Si,Ti):ZrO_2$ | 8 | 40 | AFE | $2.5\times10^{-3}$ | $10^5$ | $2.5\times10^2$ | 5.0 | 162 |

* $P_a$ values taken from reported data or computed as $P_a = U_a/\tau$, where $\tau$ is the reported discharge time.

** $V_{max}$ values obtained by graphical analysis.

**Supplementary Table 6 | Benchmark of 2D and 3D electrochemical microsupercapacitors.** The values shown in this table are taken from benchmark state-of-the-art microcapacitors review papers[166–168]. Abbreviations: electrode thickness ($t$), areal energy storage density ($U_a$), average areal power density ($P_a$), activated carbon (AC), acetonitrile (AN), carbon-microelectrochemical system (C-MEMS), carbide-derived carbon (CDC), carbon nanotube (CNT), exfoliated graphene (EG), 1-ethyl-3-methylimidazolium tetrafluoroborate (EMI, $BF_4$), layer double hydroxides (LDH), laser-induced graphene (LIG), nanowires (NW), onion-like carbon (OLC), poly(pyrrole) (PPy), propylene carbonate (PC), polyvinyl alcohol (PVA), reduced graphene oxide (rGO), sodium carboxymethylcellulose (CMC).

| **Electrode** | ***t* (µm)** | **Electrolyte** | **$U_a$ (J cm$^{-2}$)** | **$P_a$ (W cm$^{-2}$)** | **$V_{max}$ (V)** | **Ref.** |
|---|---|---|---|---|---|---|
| **2D** | | | | | | |
| AC | 5 | 1M $Et_4NBF_4$ in PC | $1.2\times10^{-2}$ | $2.0\times10^{-2}$ | 3.0 | 169 |
| AC | 2 | 1M $Et_4NBF_4$ in PC | $2.6\times10^{-2}$ | $4.5\times10^{-2}$ | 2.5 | 170 |
| CDC | 4.1 | 2M EMI, $BF_4$ in AN | $1.9\times10^{-1}$ | $3.0\times10^{-2}$ | 3.0 | 171 |
| CNT | 20 | $ZnSO_4$ | $1.1\times10^{-1}$ | $8.0\times10^{-3}$ | 1.8 | 172 |
| CNT | 260 | 0.5M $H_2SO_4$ | $10^{-3}$ | $1.3\times10^{-4}$ | 0.9 | 173 |
| EG | 0.7 | $H_2SO_4$/PVA | $1.0\times10^{-3}$ | $7.6\times10^{-2}$ | 0.8 | 174 |
| Graphene | 25 | $H_2SO_4$/PVA | $9.0\times10^{-3}$ | $5.0\times10^{-3}$ | 1.0 | 175 |
| KOH/AC | 60/70 | $H_3PO_4$/PVA | $2.5\times10^{-3}$ | $3.0\times10^{-5}$ | 1.0 | 176 |
| LIG/$Fe_3O_4$ | 2.5/7.3 | $H_2SO_4$/PVA | $2.2\times10^{-1}$ | $3.0\times10^{-4}$ | 1.5 | 177 |
| OLC | 7 | 1M $Et_4NBF_4$ in PC | $10^{-3}$ | $1.8\times10^{-1}$ | 3.0 | 169 |
| rGO/$MoS_2$ | 6 | $MgSO_4$ | $1.4\times10^{-2}$ | $1.5\times10^{-4}$ | 1.8 | 178 |
| TiC-CDC | 3.2 | 1M $H_2SO_4$ | $3.0\times10^{-3}$ | $4.0\times10^{-2}$ | 0.9 | 179 |
| $Ti_3C_2$/Co–Al–LDH | – | 6M KOH | $3.9\times10^{-2}$ | $8.8\times10^{-3}$ | 1.5 | 180 |
| Zn/AC | 50/80 | $ZnSO_4$/CMC | $4.2\times10^{-1}$ | $4.0\times10^{-3}$ | 1.5 | 181 |
| **3D** | | | | | | |
| AC/etched cavities | 336 | 1M $Et_4NBF_4$ in PC | $2.6\times10^{-1}$ | $3.4\times10^{-2}$ | 2.5 | 182 |
| AC/etched channels | 90 | 1M $naNO_3$ | $1.1\times10^{-2}$ | $5.2\times10^{-2}$ | 0.5 | 183 |
| AC/SU-8 | 250 | PVA-based | $4.3\times10^{-2}$ | $1.0\times10^{-3}$ | 0.8 | 184 |
| $LiMnO_2$/NiSn | 10 | $LiClO_4$ | $2.3\times10^{-2}$ | $3.6\times10^{-3}$ | 3.2 | 185 |
| $MNO_2$/$MnO_2$ | 68 | 0.5 $Na_2SO_4$ | $3.6\times10^{-2}$ | $2.0\times10^{-2}$ | 0.8 | 186 |
| $MnO_2$/Si NW | 50 | Ionic liquid | $3.3\times10^{-2}$ | $1.0\times10^{-3}$ | 2.2 | 187 |
| PPy/C-MEMS | 140 | 0.1M KCl | $2.5\times10^{-2}$ | $5.0\times10^{-3}$ | 0.8 | 188 |
| $Ti_3C_2T_x$ | 75 | $H_2SO_4$/PVA | $1.8\times10^{-1}$ | $7.0\times10^{-4}$ | 0.6 | 189 |
| $RuO_2$/C nanowall | 12 | 1M $Na_2SO_4$ | $1.8\times10^{-1}$ | $2.8\times10^{-2}$ | 0.9 | 190 |
| $RuO_2$/porous Au | 80 | PVA-based | $4.5\times10^{-1}$ | $8.0\times10^{-3}$ | 0.9 | 191 |
| $RuO_2$/Si-$SiO_2$-Au | 10/80 | 0.1M $Na_2SO_4$ | $1.2\times10^{-2}$ | $5.0\times10^{-3}$ | 1.0 | 192 |
| $SnO_x$/Ni | 0.3 | $LiPF_6$ | $8.1\times10^{-1}$ | $3.5\times10^{-1}$ | 3.0 | 193 |

**Supplementary Table 7 | Benchmark of Li-based microbatteries.** Abbreviations: areal capacitance density ($C_a$), discharge areal current density ($J_{dis}$), energy storage density (ESD), areal energy storage density ($U_a$), average areal power density ($P_a$), average discharge voltage ($V_{dis}$), carbonized metal–organic frameworks (c-MOF), dimethyl carbonate (DMC), 1,2-dimethoxyethane (DME), 1,3-dioxolane (DOL), ethylene carbonate (EC), $LiNi_{0.8}Co_{0.15}Al_{0.05}O_2$ (NCA), lithium bis(trifluoromethanesulfonyl)imide (LiTFSI), $Li_4Ti_5O_{12}$ (LTO), $LiFePO_4$ (LFP), $LiMn_2O_4$ (LMO) propylene carbonate (PC).

| Anode/Cathode | Electrolyte | $C_a$ (Ah cm$^{-2}$) | $J_{dis}$ A cm$^{-2}$ | $C$-rate | $U_a$ (J cm$^{-2}$) | $P_a$ (W cm$^{-2}$) | $V_{dis}$ (V) | Ref. |
|---|---|---|---|---|---|---|---|---|
| $LiCoO_2$/Li | LiPON | $1.1\times10^{-4}$ | N/A | 1 | 1.5 | $4.2\times10^{-4}$ | 3.6 | 194 |
| LiMO/Li | Li glassy inorganic thin film | $8.6\times10^{-5}$ | N/A | 1 | 1.2 | $3.4\times10^{-4}$ | 3.0 | 194 |
| $LiCoO_2$/Li | LiPON | $1.1\times10^{-4}$ | N/A | 1 | 1.3 | $3.6\times10^{-4}$ | 3.0 | 194 |
| B:Si/NCA | SU-8 | $5.6\times10^{-4}$ | $6.6\times10^{-4}$ | N/A | 5.8 | $1.9\times10^{-3}$** | 2.9*** | 195 |
| LTO/LFP | 1M $LiClO_4$ in EC:DMC | $1.5\times10^{-3}$ | N/A | 1 | 9.7 | $2.7\times10^{-3}$ | 1.8 | 196 |
| c-MOF/LFP | 1M LiTFSI in DME:DOL | $2.0\times10^{-3}$ | N/A | 0.2 | $2.5\times10^{1}$ | $1.4\times10^{-3}$ | 3.4*** | 197 |
| LMO | 1 M $LiClO_4$ in PC | $4.7\times10^{-4}$ | N/A | 1 | 5.9 | $1.7\times10^{-3}$ | 3.5$^{\delta}$ | 198 |

* $U_a$ values estimated as $U_a = C_a.V_{dis}$

** $P_a$ values estimated as $P_a = J_{dis}.V_{dis}$ or as $P_a = U_a.C$-rate

*** $V_{dis}$ values estimated as $V_{dis} = U_a/C_a$

$^{\delta}$ $V_{dis}$ value obtained by graphical analysis

**Supplementary Table 8 | Performance overview of various in-memory computing technologies**. Note that most of the data cannot be taken at face value, as the details of reporting vary greatly. For example, for normalized values, some reports may normalize with the power or area only of the area of the analogue component of the chip, whereas others normalise with the full chip contribution including complementary metal-oxide-semiconductor (CMOS) components. As the same seems to happen in commercially available CMOS accelerators, the provided data still provides a general overview of available performance metrics. Some reports, commendably, provide differentiated values. Abbreviations: dynamic power consumption (DPC), magnetoresistive random access memory (MRAM), resistive random access memory (RRAM) and tera operations per second (TOPS).

| TOPS $W^{-1}$ | TOPS | Power (mW) | TOPS $mm^{-2}$ | CMOS node | Array size | Cell | Operating voltage*** (V) | Ref. |
|---|---|---|---|---|---|---|---|---|
| | | | | RRAM | | | | |
| 11-195 | 0.418-0.036 | 2-3 * | >0.070-0.006 * | 22 nm | 4 Mb | 1T-1R | 0.8 | [199] |
| 24.1 | 0.157 | 654 | N/A | 90 nm | 128x64 b | 1T-1R | 1.2 | [200] |
| 51.4 | 6.97 | 135 * | 0.284 | 22 nm | 4 Mb | 1T-1R | 0.8 | [201] |
| 78.4 | 3.29 * | 42 ** | 0.15 * | 130 nm | 158.8 Kb | 2T-2R | 4.2-5 | [202] |
| 66.5 | 0.66 | 9.9 | 0.24 * | 40 nm | 4M synapses | 1T-1R | 1.1 | [203] |
| 1286.4 | 5.12 | 3.98 * | 0.28 * | 22 nm | 8 Mb | 1T-1R | 0.8 | [204] |
| 20.7 | 0.0571 | 2.76 * | 0.003 * | 65 nm | 256x64 (64 Kb) | 1T-1R | 1.2 (gate drive) | [205] |
| 27.2 | 0.34 | 12.5 * | 0.056 | 28 nm | 1024x1024 | 1T-1R | 0.8 | [206] |
| | | | | PCM | | | | |
| 9.76 | 63.1 | 6460 | 1.55 | 14 nm | 256x256 | 8T-4R | 0.85 (digital) | [207] |
| 261-718 | 3.9 | 14.9 * | 0.216 * | 40 nm | 2M cells | 1T-1R | 0.75-0.95 | [208] |
| 6.94 | 24.8 * | 3570 | N/A | 14 nm | 512x 2048? | 1T-1R? | 1.8 | [209] |
| 10.5 | 1.008 | 96 * | 1.59 | 14 nm | 256x256 | 8T-4R | 0.85 | [210] |
| 33.3 | N/A | N/A | N/A | 40 nm | 4M | 1T-1R | N/A | [211] |
| | | | | MRAM | | | | |
| 405 | 0.091 | 0.225 | 4.43 | 28 nm | 64x64 | 2T-2MTJ | 0.8 | [212] |
| 30.1-68.0 | N/A | N/A | N/A | 22 nm | 6.6Mb | 1T-1MTJ | N/A | [213] |
| 21.4-63.9 | 0.71 | 11.1 * | 0.018 | 22 nm | 47.25 Mb | ? | 0.8 | [214] |
| 20-59.8 | N/A | N/A | N/A | 28 nm | 1 Mb | 2T-2MTJ | 1-1.8 | [215] |
| 25.1-960.2 | 0.26-0.91 | 10-0.9 * | 0.014-0.051 | 22 nm | 4 Mb | 1T-1MTJ | 0.65-0.8 | [216] |
| | | | | CMOS | | | | |
| 16.02-21.38 | 0.49-0.60 | 30.6-28.1 | 1.19-1.44 | 22 nm | 8x2 KB | SRAM 6T+3T1C | 0.72-0.82 | [217] |
| 36.5 | 1.35 | 12.5-69.4 | 0.2 | 28 nm | N/A | SRAM | 0.6-1.0 | [218] |
| 51-300 | 1.64-9.63 | 0.87-74.9 | 0.36-2.38 | 28 nm | N/A | SRAM | 0.469-0.9 | [219] |
| 1050-1879 | 5.8-23.5 | 3.1-22.4 * | 3-12.1 | 22 nm | 1 Mb | SRAM | 0.6-0.8 | [220] |
| Ferroelectric | | | | | | | | |
| 196 | N/A | N/A | N/A | 28 nm | N/A | 1T-2FeFET | ≤5 | [221] |
| 26.3 | N/A | N/A | N/A | N/A | N/A | 1R-2FeFET | 4 | [222] |
| 224.4 | N/A | N/A | N/A | N/A | 512x512 | 1T-1FTJ | a.u. | [223] |
| 885.4 | 0.136 | 0.153 | N/A | 28 nm | 32x32 | 1FeFET-1R | 2 | [224] |
| 323 | N/A | N/A | N/A | N/A | 16x16 | 1C | 3 | [225] |
| 66 | 0.66 | 10 * | N/A | N/A | 32x1024 | 2T-*n*FeFET | 3-5 | [226] |
| 111.6 | N/A | N/A | N/A | 28 nm | 128x128 | 1FeFET-1C | 2 | [227] |
| 40.65 | N/A | N/A | 77 | 32 nm | 1080x128 | FeFET-based | 1.5-4.5 | [228] |
| 1763.9 | 27.7 | 157* | 3.51 | 180 nm | 520 kb | FeFET-based | 4 | [229] |
| 443 | 0.172 | 0.388 | N/A | 28 nm | | FeFET-based | 0.7 | [230] |

* Estimated

** Obtained by graphical analysis

*** Note that the provided voltage can have different meanings in different publications. In most cases, it is the supply voltage of the chip, but in some cases it is not clearly indicated. In some cases it is denoted in the table.

## Supplementary References

1. Products — AML100 and AML200 analog AI chips. *Aspinity* https://www.aspinity.com/products.html.

2. Zhang, S. *et al.* Cambricon-X: an accelerator for sparse neural networks. in *2016 49th Annual IEEE/ACM International Symposium on Microarchitecture (MICRO)* 1–12 (2016). doi:10.1109/MICRO.2016.7783723.

3. Wang, S., Zhou, D., Han, X. & Yoshimura, T. Chain-NN: an energy-efficient 1D chain architecture for accelerating deep convolutional neural networks. in *Design, Automation & Test in Europe Conference & Exhibition (DATE), 2017* 1032–1037 (2017). doi:10.23919/DATE.2017.7927142.

4. Chen, Y. *et al.* DaDianNao: a machine-learning supercomputer. in *2014 47th Annual IEEE/ACM International Symposium on Microarchitecture* 609–622 (2014). doi:10.1109/MICRO.2014.58.

5. Chen, T. *et al.* DianNao: a small-footprint high-throughput accelerator for ubiquitous machine-learning. *ACM SIGPLAN Not.* **49**, 269–284 (2014).

6. Han, S. *et al.* EIE: efficient inference engine on compressed deep neural network. *ACM SIGARCH Comput. Archit. News* **44**, 243–254 (2016).

7. Peddie, J. Blaize GSP streams AI graphs at the edge. *Jon Peddie Research* https://www.jonpeddie.com/news/blaize-gsp-streams-ai-graphs-at-the-edge/ (2026).

8. Mobileye. The State of AV/ADAS at Mobileye/Intel. (2019).

9. Mobileye. *Wikipedia* (2026).

10. Chen, Y.-H., Krishna, T., Emer, J. S. & Sze, V. Eyeriss: an energy-efficient reconfigurable accelerator for deep convolutional neural networks. *IEEE J. Solid-State Circuits* **52**, 127–138 (2017).

11. Hailo-10H AI Accelerator. *Hailo* https://hailo.ai/products/ai-accelerators/hailo-10h-ai-accelerator/.

12. Hailo-15H AI Vision Processor. *Hailo* https://hailo.ai/products/ai-vision-processors/hailo-15h-ai-vision-processor/.

13. Hailo-15L AI Vision Processor. *Hailo* https://hailo.ai/products/ai-vision-processors/hailo-15-ai-vision-processor/.

14. Hailo-8 AI Accelerator. *Hailo* https://hailo.ai/products/ai-accelerators/hailo-8-ai-accelerator/.

15. Coherent logix configures edge AI. *Edge AI and Vision Alliance* https://www.edge-ai-vision.com/2020/12/coherent-logix-configures-edge-ai/ (2020).

16. Horizon Journey. *Horizon Robotics* https://www.horizon.auto/en/solutions/horizon-journey.

17. Journey™ 5 automotive-grade computing solution obtains ASIL-B certification-horizon robotics. *Horizon Robotics* https://www.horizon.auto/en/news/horizon-news/433.

18. Products - MemryX. *MemryX* https://memryx.com/products/ (2022).

19. Halfacree, G. Syntiant's NDP200 promises 6.4GOP/s of edge AI compute in a tiny 1mW power envelope. *Hackster.io* https://www.hackster.io/news/syntiant-s-ndp200-promises-6-4gop-s-of-edge-ai-compute-in-a-tiny-1mw-power-envelope-96590283ffbc.

20. Syntiant unveils NDP250 neural decision processor with next-gen core 3 architecture. *Syntiant* https://www.syntiant.com/press-release/syntiant-unveils-ndp250-neural-decision-processor-with-next-gen-core-3-architecture/ (2024).

21. MLSoC family. *SiMa AI* https://www.wayofproduct.com/p/133-cutting-through-ai-hype-building/.

22. Bonshor, G. IBM's Spyre AI Accelerator Deep Dive. *More Than Moore* https://morethanmoore.substack.com/p/ibms-spyre-ai-accelerator-deep-dive (2026).

23. NVIDIA A10 Tensor Core GPU. *NVIDIA* https://www.nvidia.com/en-us/data-center/products/a10-gpu/.

24. NVIDIA A100 Tensor Core GPU. *NVIDIA* https://www.nvidia.com/en-us/data-center/a100/.

25. GPU NVIDIA A2 Tensor Core. *NVIDIA* https://www.nvidia.com/pt-br/data-center/products/a2/.

26. NVIDIA A30 Tensor Core GPU. *NVIDIA* https://www.nvidia.com/en-us/data-center/products/a30-gpu/.

27. NVIDIA A40. *NVIDIA* https://www.nvidia.com/en-us/data-center/a40/.

28. NVIDIA A800 40GB Active Graphics Card. *NVIDIA* https://www.nvidia.com/en-eu/products/workstations/a800/.

29. Shilov, A. Chinese Biren's New GPUs Have 77 Billion Transistors, 2 PFLOPS of AI Performance. *Tom's Hardware* https://www.tomshardware.com/news/chinese-biren-rolls-out-new-gpus-with-77-billion-transistors-2-pflops-of-ai-performance (2022).

30. Watts, D. ThinkSystem Intel Flex 140 12GB Gen4 Passive GPU. *Lenovo Press* https://lenovopress.lenovo.com/lp1830-thinksystem-intel-flex-140-12gb-gen4-passive-gpu (2026).

31. Watts, D. ThinkSystem Intel Flex 170 16GB Gen4 Passive GPU. *Lenovo Press* https://lenovopress.lenovo.com/lp1829-thinksystem-intel-flex-170-16gb-gen4-passive-gpu (2026).

32. NVIDIA H100 GPU. *NVIDIA* https://www.nvidia.com/en-us/data-center/h100/.

33. NVIDIA Jetson Orin. *NVIDIA* https://www.nvidia.com/en-us/autonomous-machines/embedded-systems/jetson-orin/.

34. NVIDIA Jetson Xavier. *NVIDIA* https://www.nvidia.com/en-us/autonomous-machines/embedded-systems/jetson-xavier-series/.

35. NVIDIA L4 Tensor Core GPU. *NVIDIA* https://www.nvidia.com/en-us/data-center/l4/.

36. NVIDIA L40 GPU. *NVIDIA* https://www.nvidia.com/en-us/data-center/l40/.

37. NVIDIA L40S. *NVIDIA* https://www.nvidia.com/en-us/data-center/l40s/.

38. AMD Instinct MI210. *TechPowerUp* https://www.techpowerup.com/gpu-specs/instinct-mi210.c3857.

39. AMD Instinct MI250. *TechPowerUp* https://www.techpowerup.com/gpu-specs/instinct-mi250.c3847.

40. AMD Instinct MI300X. *TechPowerUp* https://www.techpowerup.com/gpu-specs/instinct-mi300x.c4179.

41. AMD Instinct MI325X. *TechPowerUp* https://www.techpowerup.com/gpu-specs/instinct-mi325x.c4231.

42. AMD Instinct MI350X. *TechPowerUp* https://www.techpowerup.com/gpu-specs/instinct-mi350x.c4308.

43. AMD Instinct MI355X. *TechPowerUp* https://www.techpowerup.com/gpu-specs/instinct-mi355x.c4309.

44. Klotz, A. New Chinese office GPU can double as a budget 1080p gaming GPU — MTT S50 wields 2,048 MUSA cores, 8GB VRAM, 85W TGP. *Tom's Hardware* https://www.tomshardware.com/pc-components/gpus/new-chinese-office-gpu-can-double-as-a-budget-1080p-gaming-gpu (2024).

45. NVIDIA T4. *NVIDIA* https://www.nvidia.com/en-us/data-center/tesla-t4/.

46. Merritt, R. Startup Accelerates AI at the Sensor. *EE Times* https://www.eetimes.com/startup-accelerates-ai-at-the-sensor/ (2018).

47. Du, Z. *et al.* ShiDianNao: Shifting vision processing closer to the sensor. in 92–104 (2015). doi:10.1145/2749469.2750389.

48. Graphcore Colossus MK2 GC200 (IPU). *Business Systems International* https://www.bsi.uk.com/graphcore-ipu-mk2-gc200.

49. Moor Insights & Strategy. *The Graphcore Second-Generation IPU*. https://www.graphcore.ai/hubfs/MK2-%20The%20Graphcore%202nd%20Generation%20IPU%20Final%20v7.14.2020.pdf (2020).

50. Kalray. *MPPA® DPU Coolidge™*. https://www.kalrayinc.com/wp-content/uploads/2023/10/Kalray_MPPA-Coolidge_flyer_2P_NoNda_EXTERNAL_9.08.pdf (2023).

51. GreenWaves Technologies. *GAP8 IoT Application Processor*. https://gwt-website-files.s3.eu-central-1.amazonaws.com/Product+Brief+GAP8+-+V1_9.pdf (2021).

52. Halfacre, G. GreenWaves Technologies Announces GAP9 IoT Application Processor with Major Improvements Over GAP8. *Hackster.io* https://www.hackster.io/news/greenwaves-technologies-announces-gap9-iot-application-processor-with-major-improvements-over-gap8-399ba4acb602 (2019).

53. Firoozshahian, A. *et al.* MTIA: first generation silicon targeting meta's recommendation systems. in *Proceedings of the 50th Annual International Symposium on Computer*

*Architecture* (Association for Computing Machinery, New York, NY, USA, 2023). doi:10.1145/3579371.3589348.

54. Coburn, J. *et al.* Meta's Second Generation AI Chip: Model-Chip Co-Design and Productionization Experiences. in *Proceedings of the 52nd Annual International Symposium on Computer Architecture* 1689–1702 (Association for Computing Machinery, New York, NY, USA, 2025). doi:10.1145/3695053.3731409.

55. Akopyan, F. *et al.* Breakthrough Edge AI Inference Performance Using Northpole in 3U Vpx Form Factor. in 1–5 (2024). doi:10.1109/HPEC62836.2024.10938474.

56. Cassidy, A. S. *et al.* 11.4 IBM NorthPole: An Architecture for Neural Network Inference with a 12nm Chip. in vol. 67 214–215 (2024).

57. Gwennap, L. XMOS XCORE.AI ADDS VECTOR UNIT: New Iteration Targets Neural Networks on IoT Devices. *Microprocessor Report* (2020).

58. Pei, J. *et al.* Towards artificial general intelligence with hybrid tianjic chip architecture. *Nature* **572**, 106–111 (2019).

59. BrainChip Inc. *AKD1000 PCIe Card Hardware Product Brief*. https://brainchip.com/wp-content/uploads/2025/04/AKD1000-PCIe-Card-Product-Brochure-v1.0.pdf (2025).

60. BrainChip Inc. *AKD1500 Edge AI Co-Processor Product Brief*. https://brainchip.com/wp-content/uploads/2025/10/AKD1500-Product-Brief-V2.4-Oct.25.pdf (2025).

61. Ward-Foxton, S. Axelera Demos AI Test Chip After Taping Out in Four Months. *EE Times* https://www.eetimes.com/axelera-demos-ai-test-chip-after-taping-out-in-four-months/ (2022).

62. Mixos. Gyrfalcon starts shipping AI chip. *Electronics-Lab* https://www.electronics-lab.com/gyrfalcon-starts-shipping-ai-chip/ (2018).

63. Mixos. Gyrfalcon Launches Second-Gen AI Accelerator Chip. *Electronics-Lab* https://www.electronics-lab.com/gyrfalcon-launches-second-gen-ai-accelerator-chip/ (2018).

64. Ward-Foxton, S. Gyrfalcon Unveils Fourth AI Accelerator Chip. *EE Times* https://www.eetimes.com/gyrfalcon-unveils-fourth-ai-accelerator-chip/ (2019).

65. Ward-Foxton, S. Mythic Resizes its AI Chip. *EE Times* https://www.eetimes.com/mythic-resizes-its-analog-ai-chip/ (2021).

66. A14 Bionic: benchmarks and specs. *NanoReview.net* https://nanoreview.net/en/soc/apple-a14-bionic.

67. A15 Bionic: benchmarks and specs. *NanoReview.net* https://nanoreview.net/en/soc/apple-a15-bionic.

68. A16 Bionic: benchmarks and specs. *NanoReview.net* https://nanoreview.net/en/soc/apple-a16-bionic.

69. Cloud AI 100 Ultra. *Qualcomm* https://www.qualcomm.com/data-center/products/cloud-ai-100-ultra.

70. ATOM™-max. *Rebellions* https://rebellions.ai/rebellions-product/atom-max/.

71. J. Ouyang, X. Du, Y. Ma, & J. Liu. 3.3 Kunlun: A 14nm High-Performance AI Processor for Diversified Workloads. in *2021 IEEE International Solid-State Circuits Conference (ISSCC)* vol. 64 50–51 (2021).

72. McGregor, J. Perceive Exits Stealth With Super Efficient Machine Learning Chip For Smarter Devices. *Forbes* https://www.forbes.com/sites/tiriasresearch/2020/04/06/perceive-exits-stealth-with-super-efficient-machine-learning-chip-for-smarter-devices/ (2026).

73. Ian, F. Arm Ethos-N ML Inference Processors: Powering Exciting User Experiences on Edge Devices. *Arm Developer* https://developer.arm.com/community/arm-community-blogs/b/ai-blog/posts/arm-ml-processor-exciting-ux-on-edge-devices (2019).

74. Prickett, N. H. Groq Shares Recipe for TSP Nodes, Systems. *The Next Platform* https://www.nextplatform.com/ai/2020/09/29/groq-shares-recipe-for-tsp-nodes-systems/1636424 (2020).

75. Wu, E. Get Started with K210: Hardware and Programming Environment. *Seeed Studio* https://www.seeedstudio.com/blog/2019/09/12/get-started-with-k210-hardware-and-programming-environment/ (2019).

76. Ward-Foxton, S. Kneron Attracts Strategic Investors. *EE Times* https://www.eetimes.com/kneron-attracts-strategic-investors/ (2021).

77. Wodecki, B. Kneron Unveils the KL530: Its Most Energy-efficient Chip to Date. *AI Business* https://aibusiness.com/verticals/kneron-unveils-the-kl530-its-most-energy-efficient-chip-to-date (2021).

78. Hinum, K. Apple M1. *Notebookcheck* https://www.notebookcheck.net/Apple-M1-Processor-Benchmarks-and-Specs.503613.0.html.

79. Hinum, K. Apple M2. *Notebookcheck* https://www.notebookcheck.net/Apple-M2-Processor-Benchmarks-and-Specs.632312.0.html.

80. Hinum, K. Apple M3. *Notebookcheck* https://www.notebookcheck.net/Apple-M3-Processor-Benchmarks-and-Specs.764609.0.html.

81. Yu, Z., Yang, Y. & Luo, Y. Adopting MLU for geneformer model. in *Proceedings of the 2024 9th International Conference on Information and Education Innovations* 37–43 (Association for Computing Machinery, New York, NY, USA, 2024). doi:10.1145/3664934.3664943.

82. MLU370-X8 Smart Accelerator Card. *Cambricon* https://www.cambricon.com/index.php?m=content&c=index&a=lists&catid=406.

83. Alcorn, P. Intel Unveils Movidius Myriad X Vision Processing Unit. *Tom's Hardware* https://www.tomshardware.com/news/intel-movidius-vpu-ai-inference,35327.html (2017).

84. Hui, Y., Lien, J. & Lu, X. Characterizing and Accelerating End-to-End EdgeAI Inference Systems for Object Detection Applications. in *2021 IEEE/ACM Symposium on Edge Computing (SEC)* 01–12 (2021). doi:10.1145/3453142.3491294.

85. Qualcomm Robotics RB5/RB6 - Why the RB5/RB6? *RidgeRun Developer Wiki* https://developer.ridgerun.com/wiki/index.php/Qualcomm_Robotics_RB5/Introduction/Platform_Comparison.

86. Smith, M. S. CES 2024: Neuchips Demos Low-Power AI Upgrade for PCs: Not the most powerful at the show, but 55 watts is plenty accessible. *IEEE Spectrum* https://spectrum.ieee.org/neuchips-low-power-ai (2024).

87. Furiosa RNGD. *FuriosaAI* https://furiosa.ai/rngd.

88. STM32N6x7. *STMicroelectronics* https://www.st.com/en/microcontrollers-microprocessors/stm32n6x7.html.

89. Type1WV - AI Accelerator Module (featuring the Coral Edge TPU™ from Google). *Murata Manufacturing* https://www.murata.com/en-eu/products/connectivitymodule/edge-ai/overview/lineup/type1wv.

90. Sato, K. & Young, C. An in-depth look at Google's first Tensor Processing Unit (TPU). *Google Cloud Blog* https://cloud.google.com/blog/products/ai-machine-learning/an-in-depth-look-at-googles-first-tensor-processing-unit-tpu (2017).

91. TPU v4. *Google Cloud Documentation* https://docs.cloud.google.com/tpu/docs/v4?hl=pt.

92. Morgan, T. P. With "ironwood" TPU, google pushes the AI accelerator to the floor. *The Next Platform* https://www.nextplatform.com/compute/2025/04/09/with-ironwood-tpu-google-pushes-the-ai-accelerator-to-the-floor/1660461 (2025).

93. Kay, A. M. *et al.* The Thermodynamic Limit of Indoor Photovoltaics Based on Energetically-Disordered Molecular Semiconductors. *Sol. RRL* **7**, 2300277 (2023).

94. Wong, J., Omelchenko, S. T. & Atwater, H. A. Impact of semiconductor band tails and band filling on photovoltaic efficiency limits. *ACS Energy Lett.* **6**, 52–57 (2020).

95. Wurfel, P. The chemical potential of radiation. *J. Phys. C: Solid State Phys.* **15**, 3967 (1982).

96. Zarrabi, N. *et al.* Charge-generating mid-gap trap states define the thermodynamic limit of organic photovoltaic devices. *Nat. Commun.* **11**, 5567 (2020).

97. Kay, A. M., Riley, D. B., Burwell, G. & Meredith, P. Thermodynamic principles for optimizing multi-junction photovoltaics—exemplified for perovskite-based indoor photovoltaics. *APL Energy* **3**, 036102 (2025).

98. Rau, U. Reciprocity relation between photovoltaic quantum efficiency and electroluminescent emission of solar cells. *Phys. Rev. B* **76**, 085303 (2007).

99. Kay, A. M. Austin-M-kay/Photovoltaic_Performance_Simulator. (2026).

100. Shockley, W. & Queisser, H. J. Detailed Balance Limit of Efficiency of p-n Junction Solar Cells. *J. Appl. Phys.* **32**, 510–519 (1961).

101. International Electrotechnical Commission. Nanomanufacturing — Key control characteristics — Part 7-2: Nano-enabled photovoltaics — Device evaluation method for indoor light.

102. Kay, A. M. *et al.* Optimising photovoltaic modules for indoor energy-harvesting systems. *J. Phys.: Energy* **7**, 035019 (2025).

103. Pecunia, V., Occhipinti, L. G. & Hoye, R. L. Z. Emerging Indoor Photovoltaic Technologies for Sustainable Internet of Things. *Advanced Energy Materials* **11**, 2100698 (2021).

104. AEM13920. *E-peas* https://e-peas.com/product/aem13920/.

105. Seunarine, K. *et al.* Light power resource availability for energy harvesting photovoltaics for self-powered IoT. *J. Phys.: Energy* **6**, 015018 (2024).

106. Würfel, U., Neher, D., Spies, A. & Albrecht, S. Impact of charge transport on current–voltage characteristics and power-conversion efficiency of organic solar cells. *Nat. Commun.* **6**, 6951 (2015).

107. Kim, E., Christen, L. & Kirchartz, T. Correlating the photoshunt with charge-collection losses in organic solar cells. *Adv. Energy Mater.* **16**, 2403129 (2026).

108. Meredith, P. & Armin, A. Scaling of next generation solution processed organic and perovskite solar cells. *Nat. Commun.* **9**, 5261 (2018).

109. Burwell, G. *et al.* Scaling considerations for organic photovoltaics for indoor applications. *Sol. RRL* **6**, 2200315 (2022).

110. Ma, Q. *et al.* One-step dual-additive passivated wide-bandgap perovskites to realize 44.72%-efficient indoor photovoltaics. *Energy Environ. Sci.* **17**, 1637–1644 (2024).

111. Bi, P. *et al.* A high-performance nonfused wide-bandgap acceptor for versatile photovoltaic applications. *Adv. Mater.* **34**, 2108090 (2022).

112. Michaels, H. *et al.* Emerging indoor photovoltaics for self-powered and self-aware IoT towards sustainable energy management. *Chem. Sci.* **14**, 5350–5360 (2023).

113. Freunek, M., Freunek, M. & Reindl, L. M. Maximum efficiencies of indoor photovoltaic devices. *IEEE J. Photovolt.* **3**, 59–64 (2013).

114. Lu, W. *et al.* Ordering one-dimensional chains enables efficient selenium photovoltaics. *Joule* **8**, 1430–1442 (2024).

115. Feng, Q., Yao, M., Su, Z. & Yao, X. Significantly enhanced energy density of amorphous alumina thin films via silicon and magnesium co-doping. *Ceramics International* **44**, 11160–11165 (2018).

116. Xie, J., Yao, M., Gao, W., Su, Z. & Yao, X. Significantly enhanced dielectric constant and energy density in Au/Al2O3 nanocomposite thin films. *Journal of Alloys and Compounds* **772**, 324–331 (2019).

117. Ding, S.-J., Zhu, C., Li, M.-F. & Zhang, D. W. Atomic-layer-deposited Al2O3–HfO2–Al2O3 dielectrics for metal-insulator-metal capacitor applications. *Appl. Phys. Lett.* **87**, 053501 (2005).

118. Zhang, L. *et al.* ALD preparation of high-k HfO2 thin films with enhanced energy density and efficient electrostatic energy storage. *RSC Adv.* **7**, 8388–8393 (2017).

119. Kim, S. J. *et al.* Stress-Induced Crystallization of Thin Hf1–XZrXO2 Films: The Origin of Enhanced Energy Density with Minimized Energy Loss for Lead-Free Electrostatic Energy Storage Applications. *ACS Appl. Mater. Interfaces* **11**, 5208–5214 (2019).

120. Yu, X. *et al.* A high-density MIM capacitor (13 fF/μm2) using ALD HfO2 dielectrics. *IEEE Electron Device Letters* **24**, 63–65 (2003).

121. Magagnin, G. *et al.* Comparative performance of fluorite-structured materials for nanosupercapacitor applications. *APL Mater.* **12**, 071124 (2024).

122. Zhu, B. *et al.* High density supercapacitors with atomic layer deposited SiO2 dielectric as solid state electrolyte. *J. Appl. Phys.* **137**, 225303 (2025).

123. Zhang, Q.-X., Zhu, B., Zhang, L.-F. & Ding, S.-J. Atomic-layer-deposited SiO2/TiO2/SiO2 sandwiched dielectrics for metal–insulator–metal capacitor application. *Microelectronic Engineering* **122**, 1–4 (2014).

124. Chiang, K. C. *et al.* High-Temperature Leakage Improvement in Metal–Insulator–Metal Capacitors by Work–Function Tuning. *IEEE Electron Device Letters* **28**, 235–237 (2007).

125. Liu, Q., Xue, Q., Wang, Y., Wei, X. & Hao, J. Bifunctional Device with High-Energy Storage Density and Ultralow Current Analog Resistive Switching. *Advanced Electronic Materials* **7**, 2000902 (2021).

126. Cheng, C. H. *et al.* High Density and Low Leakage Current in TiO2 MIM Capacitors Processed at 300 °C. *IEEE Electron Device Letters* **29**, 845–847 (2008).

127. Wu, Y.-H. *et al.* High-Performance Metal–Insulator-Metal Capacitor Using Stacked TiO2/Y2O3 as Insulator. *IEEE Electron Device Letters* **32**, 1107–1109 (2011).

128. Lin, S. H., Chiang, K. C., Yeh, F. S. & Chin, A. Improved Stress Reliability of Analog Metal–Insulator–Metal Capacitors Using \hboxTiO_2/\hboxZrO_2 Dielectrics. *IEEE Electron Device Letters* **30**, 1287–1289 (2009).

129. Wang, Y., Wang, Y., Zeng, H. & Wei, X. Ultra-high energy storage density of transparent capacitors based on linear dielectric ZrO2 thin films with the thickness scaled up to hundreds nanometers. *Appl. Phys. Lett.* **120**, 023904 (2022).

130. R. Padmanabhan, N. Bhat, Y. Morozumi, S. Mohan, & S. Kaushal. High-performance stacked TiO2-ZrO2 and Si-doped ZrO2 metal-insulator-metal capacitors. in *2014 IEEE International Conference on IC Design & Technology* 1–4 (2014). doi:10.1109/ICICDT.2014.6838596.

131. Sharma, R. *et al.* Large-Area Rolled-Up Nanomembrane Capacitor Arrays for Electrostatic Energy Storage. *Advanced Energy Materials* **4**, 1301631 (2014).

132. Han, F. *et al.* Dielectric capacitors with three-dimensional nanoscale interdigital electrodes for energy storage. *Science Advances* **1**, e1500605 (2015).

133. Klootwijk, J. H. *et al.* Ultrahigh Capacitance Density for Multiple ALD-Grown MIM Capacitor Stacks in 3-D Silicon. *IEEE Electron Device Letters* **29**, 740–742 (2008).

134. Mu, J., Chou, X., Ma, Z., He, J. & Xiong, J. High-Performance MIM Capacitors for a Secondary Power Supply Application. *Micromachines* **9**, 69 (2018).

135. Liu, Z. *et al.* Anomalous high capacitance in a coaxial single nanowire capacitor. *Nat Commun* **3**, 879 (2012).

136. Zhang, Y. *et al.* 3D HfO2 Thin Film MEMS Capacitor with Superior Energy Storage Properties. *Advanced Functional Materials* **33**, 2305733 (2023).

137. Hourdakis, E., Kochylas, I., Botzakaki, M. A., Xanthopoulos, N. J. & Gardelis, S. Si nanowire-based micro-capacitors fabricated with metal assisted chemical etching for integrated energy storage applications. *Solid-State Electronics* **196**, 108408 (2022).

138. Strambini, L. *et al.* Three-dimensional silicon-integrated capacitor with unprecedented areal capacitance for on-chip energy storage. *Nano Energy* **68**, 104281 (2020).

139. Brunet, M. & Kleimann, P. High-Density 3-D Capacitors for Power Systems On-Chip: Evaluation of a Technology Based on Silicon Submicrometer Pore Arrays Formed by Electrochemical Etching. *IEEE Transactions on Power Electronics* **28**, 4440–4448 (2013).

140. Park, M. H. *et al.* Effect of acceptor doping on phase transitions of HfO2 thin films for energy-related applications. *Nano Energy* **36**, 381–389 (2017).

141. Zhou, J. *et al.* A 6.5 nm thick anti-ferroelectric HfAlOx film for energy storage devices with a high density of 63.7 J cm−3. *J. Phys. D: Appl. Phys.* **55**, 014003 (2021).

142. Hoffmann, M. *et al.* Ferroelectric phase transitions in nanoscale HfO2 films enable giant pyroelectric energy conversion and highly efficient supercapacitors. *Nano Energy* **18**, 154–164 (2015).

143. Ali, F. *et al.* Silicon-doped hafnium oxide anti-ferroelectric thin films for energy storage. *J. Appl. Phys.* **122**, 144105 (2017).

144. Kim, K. D. *et al.* Scale-up and optimization of HfO2-ZrO2 solid solution thin films for the electrostatic supercapacitors. *Nano Energy* **39**, 390–399 (2017).

145. He, Y. *et al.* Superhigh energy storage density on-chip capacitors with ferroelectric Hf0.5Zr0.5O2/antiferroelectric Hf0.25Zr0.75O2 bilayer nanofilms fabricated by plasma-enhanced atomic layer deposition. *Nanoscale Adv.* **4**, 4648–4657 (2022).

146. Hoffmann, M. *et al.* Negative Capacitance for Electrostatic Supercapacitors. *Advanced Energy Materials* **9**, 1901154 (2019).

147. Lomenzo, P. D., Chung, C.-C., Zhou, C., Jones, J. L. & Nishida, T. Doped Hf0.5Zr0.5O2 for high efficiency integrated supercapacitors. *Appl. Phys. Lett.* **110**, 232904 (2017).

148. Shuai, W. *et al.* Superior and ultrafast energy storage performance of relaxorantiferroelectric HfO2-based supercapacitors. *Energy Storage Materials* **62**, 102931 (2023).

149. Shin, J. *et al.* Ultra-high energy storage performance of field-induced ferroelectric Al2O3-inserted Hf0.5Zr0.5O2 thin films for electrostatic supercapacitors. *Energy Storage Materials* **79**, 104306 (2025).

150. Kozodaev, M. G. *et al.* La-doped Hf0.5Zr0.5O2 thin films for high-efficiency electrostatic supercapacitors. *Appl. Phys. Lett.* **113**, 123902 (2018).

151. Park, M. H. *et al.* Thin HfxZr1-xO2 Films: A New Lead-Free System for Electrostatic Supercapacitors with Large Energy Storage Density and Robust Thermal Stability. *Advanced Energy Materials* **4**, 1400610 (2014).

152. Das, D., Gaddam, V. & Jeon, S. Insertion of Dielectric Interlayer: A New Approach to Enhance Energy Storage in $Hf_x$Zr1-x$O_2$ Capacitors. *IEEE Electron Device Letters* **42**, 331–334 (2021).

153. Cheema, S. S. *et al.* Giant energy storage and power density negative capacitance superlattices. *Nature* **629**, 803–809 (2024).

154. Chen, H. *et al.* Ultrahigh Energy Storage Density in Superparaelectric-Like Hf0.2Zr0.8O2 Electrostatic Supercapacitors. *Advanced Science* **10**, 2300792 (2023).

155. Lomenzo, P. D. *et al.* Harnessing Phase Transitions in Antiferroelectric ZrO2 Using the Size Effect. *Advanced Electronic Materials* **8**, 2100556 (2022).

156. Pešić, M., Hoffmann, M., Richter, C., Mikolajick, T. & Schroeder, U. Nonvolatile Random Access Memory and Energy Storage Based on Antiferroelectric Like Hysteresis in ZrO2. *Advanced Functional Materials* **26**, 7486–7494 (2016).

157. Yang, K. *et al.* Energy conversion and storage using artificially induced antiferroelectricity in HfO2/ZrO2 nanolaminates. *Composites Part B: Engineering* **236**, 109824 (2022).

158. Silva, J. P. B. *et al.* Energy storage performance of ferroelectric ZrO2 film capacitors: effect of HfO2:Al2O3 dielectric insert layer. *J. Mater. Chem. A* **8**, 14171–14177 (2020).

159. Yi, S.-H., Lin, H.-C. & Chen, M.-J. Ultra-high energy storage density and scale-up of antiferroelectric TiO2/ZrO2/TiO2 stacks for supercapacitors. *J. Mater. Chem. A* **9**, 9081–9091 (2021).

160. Yi, S.-H., Chan, Y.-C., Mo, C.-L., Lin, H.-C. & Chen, M.-J. Enhancement of energy storage for electrostatic supercapacitors through built-in electric field engineering. *Nano Energy* **99**, 107342 (2022).

161. Lee, S. W., Oh, Y., Kim, C. J., Kang, B. S. & Ahn, J.-H. High Energy Storage Performance of Nanometer-Thick Al-Doped ZrO2 Films with Antiferroelectricity for Electrostatic Capacitors. *ACS Appl. Nano Mater.* **8**, 18080–18087 (2025).

162. Zheng, G. *et al.* Plasma-Enhanced Atomic Layer-Deposited Ti,Si-Doped ZrO2 Antiferroelectric Films for Energy Storage Capacitors. *ACS Appl. Electron. Mater.* **5**, 5907–5915 (2023).

163. Viegas, A. E., Kuehnel, K., Mart, C., Czernohorsky, M. & Heitmann, J. Stabilizing Antiferroelectric-Like Aluminum-Doped Hafnium Oxide for Energy Storage Capacitors. *Advanced Engineering Materials* **25**, 2300443 (2023).

164. Kühnel, K., Czernohorsky, M., Mart, C. & Weinreich, W. High-density energy storage in Si-doped hafnium oxide thin films on area-enhanced substrates. *J. Vac. Sci. Technol. B* **37**, 021401 (2019).

165. Viegas, A., Mart, C. & Czernohorsky, M. Antiferroelectric Si:HfO2 for High Energy Storage using 3D MIM Capacitors. in *2020 Joint Conference of the IEEE International Frequency Control Symposium and International Symposium on Applications of Ferroelectrics (IFCS-ISAF)* 1–3 (2020). doi:10.1109/IFCS-ISAF41089.2020.9234899.

166. Bassyouni, Z., Allagui, A. & Abou Ziki, J. D. Microsized Electrochemical Energy Storage Devices and Their Fabrication Techniques For Portable Applications. *Advanced Materials Technologies* **8**, 2200459 (2023).

167. Zheng, S., Shi, X., Das, P., Wu, Z.-S. & Bao, X. The Road Towards Planar Microbatteries and Micro-Supercapacitors: From 2D to 3D Device Geometries. *Advanced Materials* **31**, 1900583 (2019).

168. Kyeremateng, N. A., Brousse, T. & Pech, D. Microsupercapacitors as miniaturized energy-storage components for on-chip electronics. *Nature Nanotech* **12**, 7–15 (2017).

169. Pech, D. *et al.* Ultrahigh-power micrometre-sized supercapacitors based on onion-like carbon. *Nature Nanotech* **5**, 651–654 (2010).

170. Pech, D. *et al.* Elaboration of a microstructured inkjet-printed carbon electrochemical capacitor. *Journal of Power Sources* **195**, 1266–1269 (2010).

171. Huang, P. *et al.* On-chip and freestanding elastic carbon films for micro-supercapacitors. *Science* **351**, 691–695 (2016).

172. Sun, G. *et al.* A capacity recoverable zinc-ion micro-supercapacitor. *Energy Environ. Sci.* **11**, 3367–3374 (2018).

173. Dinh, T. M., Armstrong, K., Guay, D. & Pech, D. High-resolution on-chip supercapacitors with ultra-high scan rate ability. *J. Mater. Chem. A* **2**, 7170–7174 (2014).

174. Zheng, S. *et al.* Arbitrary-Shaped Graphene-Based Planar Sandwich Supercapacitors on One Substrate with Enhanced Flexibility and Integration. *ACS Nano* **11**, 2171–2179 (2017).

175. Wu, Y. *et al.* Boosting the Electrochemical Performance of Graphene-Based On-Chip Micro-Supercapacitors by Regulating the Functional Groups. *ACS Appl. Mater. Interfaces* **12**, 42933–42941 (2020).

176. Liu, H. *et al.* Laser-induced and KOH-activated 3D graphene: A flexible activated electrode fabricated via direct laser writing for in-plane micro-supercapacitors. *Chemical Engineering Journal* **393**, 124672 (2020).

177. Liu, H. *et al.* Laser-oxidized Fe3O4 nanoparticles anchored on 3D macroporous graphene flexible electrodes for ultrahigh-energy in-plane hybrid micro-supercapacitors. *Nano Energy* **77**, 105058 (2020).

178. Shao, Y. *et al.* 3D Crumpled Ultrathin 1T MoS2 for Inkjet Printing of Mg-Ion Asymmetric Micro-supercapacitors. *ACS Nano* **14**, 7308–7318 (2020).

179. Létiche, M. *et al.* Sputtered Titanium Carbide Thick Film for High Areal Energy on Chip Carbon-Based Micro-Supercapacitors. *Advanced Functional Materials* **27**, 1606813 (2017).

180. Xu, S. *et al.* Screen-printable microscale hybrid device based on MXene and layered double hydroxide electrodes for powering force sensors. *Nano Energy* **50**, 479–488 (2018).

181. Zhang, P. *et al.* Zn-Ion Hybrid Micro-Supercapacitors with Ultrahigh Areal Energy Density and Long-Term Durability. *Advanced Materials* **31**, 1806005 (2019).

182. Durou, H. *et al.* Wafer-level fabrication process for fully encapsulated micro-supercapacitors with high specific energy. *Microsyst Technol* **18**, 467–473 (2012).

183. Shen, C., Wang, X., Zhang, W. & Kang, F. A high-performance three-dimensional micro supercapacitor based on self-supporting composite materials. *Journal of Power Sources* **196**, 10465–10471 (2011).

184. Pu, J. *et al.* High-energy-density, all-solid-state microsupercapacitors with three-dimensional interdigital electrodes of carbon/polymer electrolyte composite. *Nanotechnology* **27**, 045701 (2015).

185. Ning, H. *et al.* Holographic patterning of high-performance on-chip 3D lithium-ion microbatteries. *Proceedings of the National Academy of Sciences* **112**, 6573–6578 (2015).

186. Eustache, E. *et al.* High Areal Energy 3D-Interdigitated Micro-Supercapacitors in Aqueous and Ionic Liquid Electrolytes. *Advanced Materials Technologies* **2**, 1700126 (2017).

187. Dubal, D. P. *et al.* 3D hierarchical assembly of ultrathin MnO2 nanoflakes on silicon nanowires for high performance micro-supercapacitors in Li- doped ionic liquid. *Sci Rep* **5**, 9771 (2015).

188. Beidaghi, M. & Wang, C. Micro-supercapacitors based on three dimensional interdigital polypyrrole/C-MEMS electrodes. *Electrochimica Acta* **56**, 9508–9514 (2011).

189. Orangi, J., Hamade, F., Davis, V. A. & Beidaghi, M. 3D Printing of Additive-Free 2D Ti3C2Tx (MXene) Ink for Fabrication of Micro-Supercapacitors with Ultra-High Energy Densities. *ACS Nano* **14**, 640–650 (2020).

190. Dinh, T. M. *et al.* Hydrous RuO2/carbon nanowalls hierarchical structures for all-solid-state ultrahigh-energy-density micro-supercapacitors. *Nano Energy* **10**, 288–294 (2014).

191. Ferris, A., Garbarino, S., Guay, D. & Pech, D. 3D RuO2 Microsupercapacitors with Remarkable Areal Energy. *Advanced Materials* **27**, 6625–6629 (2015).

192. Wang, X., Yin, Y., Li, X. & You, Z. Fabrication of a symmetric micro supercapacitor based on tubular ruthenium oxide on silicon 3D microstructures. *Journal of Power Sources* **252**, 64–72 (2014).

193. Weng, Q. *et al.* A compact tube-in-tube microsized lithium-ion battery as an independent microelectric power supply unit. *CR-PHYS-SC* **2**, (2021).

194. Laïk, B., Ressejac, I., Venet, C. & Pereira-Ramos, J.-P. Comparative study of electrochemical performance of commercial solid-state thin film Li microbatteries. *Thin Solid Films* **649**, 69–74 (2018).

195. Hur, J. I., Smith, L. C. & Dunn, B. High Areal Energy Density 3D Lithium-Ion Microbatteries. *Joule* **2**, 1187–1201 (2018).

196. Sun, K. *et al.* 3D Printing of Interdigitated Li-Ion Microbattery Architectures. *Advanced Materials* **25**, 4539–4543 (2013).

197. Li, S. *et al.* High-areal-capacity/power lithium metal microbattery configuration based on the mechanically flexible, ultra-lightweight, nanocellulose framework. *Nano Res.* **17**, 8155–8162 (2024).

198. Labyedh, N., Mattelaer, F., Detavernier, C. & Vereecken, P. M. 3D LiMn2O4 thin-film electrodes for high rate all solid-state lithium and Li-ion microbatteries. *J. Mater. Chem. A* **7**, 18996–19007 (2019).

199. Xue, C.-X. *et al.* 16.1 A 22nm 4Mb 8b-Precision ReRAM Computing-in-Memory Macro with 11.91 to 195.7TOPS/W for Tiny AI Edge Devices. in *2021 IEEE International Solid-State Circuits Conference (ISSCC)* vol. 64 245–247 (2021).

200. Yin, S., Sun, X., Yu, S. & Seo, J.-S. High-Throughput In-Memory Computing for Binary Deep Neural Networks With Monolithically Integrated RRAM and 90-nm CMOS. *IEEE Transactions on Electron Devices* **67**, 4185–4192 (2020).

201. Huang, W.-H. *et al.* A Nonvolatile Al-Edge Processor with 4MB SLC-MLC Hybrid-Mode ReRAM Compute-in-Memory Macro and 51.4-251TOPS/W. in *2023 IEEE International Solid-State Circuits Conference (ISSCC)* 15–17 (2023). doi:10.1109/ISSCC42615.2023.10067610.

202. Liu, Q. *et al.* 33.2 A Fully Integrated Analog ReRAM Based 78.4TOPS/W Compute-In-Memory Chip with Fully Parallel MAC Computing. in *2020 IEEE International Solid-State Circuits Conference - (ISSCC)* 500–502 (2020). doi:10.1109/ISSCC19947.2020.9062953.

203. Mochida, R. *et al.* A 4M Synapses integrated Analog ReRAM based 66.5 TOPS/W Neural-Network Processor with Cell Current Controlled Writing and Flexible Network Architecture. in *2018 IEEE Symposium on VLSI Technology* 175–176 (2018). doi:10.1109/VLSIT.2018.8510676.

204. Hung, J.-M. *et al.* An 8-Mb DC-Current-Free Binary-to-8b Precision ReRAM Nonvolatile Computing-in-Memory Macro using Time-Space-Readout with 1286.4-21.6TOPS/W for Edge-AI Devices. in *2022 IEEE International Solid-State Circuits Conference (ISSCC)* vol. 65 1–3 (2022).

205. Correll, J. M. *et al.* An 8-bit 20.7 TOPS/W Multi-Level Cell ReRAM-based Compute Engine. in *2022 IEEE Symposium on VLSI Technology and Circuits (VLSI Technology and Circuits)* 264–265 (2022). doi:10.1109/VLSITechnologyandCir46769.2022.9830490.

206. Wen, T.-H. *et al.* A 28nm Nonvolatile AI Edge Processor using 4Mb Analog-Based Near-Memory-Compute ReRAM with 27.2 TOPS/W for Tiny AI Edge Devices. in *2023 IEEE Symposium on VLSI Technology and Circuits (VLSI Technology and Circuits)* 1–2 (2023). doi:10.23919/VLSITechnologyandCir57934.2023.10185326.

207. Le Gallo, M. *et al.* A 64-core mixed-signal in-memory compute chip based on phase-change memory for deep neural network inference. *Nat. Electron.* **6**, 680–693 (2023).

208. Khwa, W.-S. *et al.* A 40-nm, 2M-Cell, 8b-Precision, Hybrid SLC-MLC PCM Computing-in-Memory Macro with 20.5 - 65.0TOPS/W for Tiny-Al Edge Devices. in *2022 IEEE International Solid-State Circuits Conference (ISSCC)* vol. 65 1–3 (2022).

209. Ambrogio, S. *et al.* An analog-AI chip for energy-efficient speech recognition and transcription. *Nature* **620**, 768–775 (2023).

210. Khaddam-Aljameh, R. *et al.* HERMES-Core—A 1.59-TOPS/mm2 PCM on 14-nm CMOS In-Memory Compute Core Using 300-ps/LSB Linearized CCO-Based ADCs. *IEEE J. Solid-State Circuits* **57**, 1027–1038 (2022).

211. Yang, X.-Y. *et al.* Bayesian Neural Network Realization by Exploiting Random Noise and Mitigating the Impact of Conductance Drift of PCM. *Adv. Mater. Technol.* **11**, e01278 (2026).

212. Jung, S. *et al.* A crossbar array of magnetoresistive memory devices for in-memory computing. *Nature* **601**, 211–216 (2022).

213. Chiu, Y.-C. *et al.* A CMOS-integrated spintronic compute-in-memory macro for secure AI edge devices. *Nat. Electron.* **6**, 534–543 (2023).

214. You, D.-Q. *et al.* A 22nm Nonvolatile AI-Edge Processor with 21.4TFLOPS/W using 47.25Mb Lossless-Compressed-Computing STT-MRAM Near-Memory-Compute Macro. in *2024 IEEE Symposium on VLSI Technology and Circuits (VLSI Technology and Circuits)* 1–2 (2024). doi:10.1109/VLSITechnologyandCir46783.2024.10631408.

215. Kwon, S. *et al.* 37.3 Monolithic in-Memory Computing Microprocessor for End-to-End DNN Inferencing in MRAM-Embedded 28nm CMOS Technology with 1.1Mb Weight Storage. in *2025 IEEE International Solid-State Circuits Conference (ISSCC)* vol. 68 1–3 (2025).

216. Chiu, Y.-C. *et al.* A 22nm 4Mb STT-MRAM data-encrypted near-memory computation macro with a 192GB/s read-and-decryption bandwidth and 25.1-55.1TOPS/W 8b MAC for AI operations. in *2022 IEEE International Solid-State Circuits Conference (ISSCC)* vol. 65 178–180 (2022).

217. Chen, P. *et al.* 7.8 A 22nm Delta-Sigma Computing-In-Memory (Δ∑CIM) SRAM Macro with Near-Zero-Mean Outputs and LSB-First ADCs Achieving 21.38TOPS/W for 8b-MAC Edge AI Processing. in *2023 IEEE International Solid-State Circuits Conference (ISSCC)* 140–142 (2023). doi:10.1109/ISSCC42615.2023.10067289.

218. Tu, F. *et al.* A 28nm 29.2TFLOPS/W BF16 and 36.5TOPS/W INT8 Reconfigurable Digital CIM Processor with Unified FP/INT Pipeline and Bitwise In-Memory Booth Multiplication for Cloud Deep Learning Acceleration. in *2022 IEEE International Solid-State Circuits Conference (ISSCC)* vol. 65 1–3 (2022).

219. Yue, J. *et al.* A 28nm 16.9-300TOPS/W Computing-in-Memory Processor Supporting Floating-Point NN Inference/Training with Intensive-CIM Sparse-Digital Architecture. in *2023 IEEE International Solid-State Circuits Conference (ISSCC)* 1–3 (2023). doi:10.1109/ISSCC42615.2023.10067779.

220. Papistas, I. A. *et al.* A 22 nm, 1540 TOP/s/W, 12.1 TOP/s/mm2 in-Memory Analog Matrix-Vector-Multiplier for DNN Acceleration. in *2021 IEEE Custom Integrated Circuits Conference (CICC)* 1–2 (2021). doi:10.1109/CICC51472.2021.9431575.

221. Parmar, V. *et al.* Demonstration of Differential Mode Ferroelectric Field-Effect Transistor Array-Based in-Memory Computing Macro for Realizing Multiprecision Mixed-Signal Artificial Intelligence Accelerator. *Adv. Intell. Syst.* **5**, 2200389 (2023).

222. Lu, T. *et al.* Two-dimensional fully ferroelectric-gated hybrid computing-in-memory hardware for high-precision and energy-efficient dynamic tracking. *Sci. Adv.* **10**, eadp0174 (2024).

223. Koo, W.-T. *et al.* 224 Tops/W-Level Analog Computation in Memory Cell Using Hybrid Ferroelectric Tunnel Junction Having Enhanced on-State Conductance. in *2025 Symposium on VLSI Technology and Circuits (VLSI Technology and Circuits)* 1–3 (2025). doi:10.23919/VLSITechnologyandCir65189.2025.11074955.

224. Soliman, T. *et al.* First demonstration of in-memory computing crossbar using multi-level cell FeFET. *Nat. Commun.* **14**, 6348 (2023).

225. Wang, X. *et al.* Energy-Efficient Reliable HZO FeFET Computation-in-Memory with Local Multiply & Global Accumulate Array for Source-Follower & Charge-Sharing Voltage Sensing. *Nano Energy* **140**, 111011 (2025).

226. Matsui, C., Toprasertpong, K., Takagi, S. & Takeuchi, K. Energy-efficient reliable HZO FeFET computation-in-memory with local multiply & global accumulate array for source-follower & charge-sharing voltage sensing. in *2021 Symposium on VLSI Technology* 1–2 (2021).

227. Laleni, N., Müller, F., Cuñarro, G., Kämpfe, T. & Jang, T. A High-Efficiency Charge-Domain Compute-in-Memory 1F1C Macro Using 2-bit FeFET Cells for DNN Processing. *IEEE J. Explor. Solid-State Computat. Devices Circuits* **10**, 153–160 (2024).

228. Kim, G. *et al.* Power-Delay Area-Efficient Processing-In-Memory Based on Nanocrystalline Hafnia Ferroelectric Field-Effect Transistors. *ACS Appl. Mater. Interfaces* **15**, 1463–1474 (2022).

229. Li, W. *et al.* A 1763.9TOPS/W 520kb Scalable FeFET-Based Analog Computing-in-Memory Chip Using BEOL HZO Capacitor for On-chip Learning. in *2025 IEEE International Electron Devices Meeting (IEDM)* 1–4 (2025). doi:10.1109/IEDM50572.2025.11353707.

230. Zhang, L., Xu, P., Borggreve, D., Vanselow, F. & Brederlow, R. A FeFET In-memory-computing core with offset cancellation for mitigating computational errors. in *ESSCIRC 2023- IEEE 49th European Solid State Circuits Conference (ESSCIRC)* 29–32 (2023). doi:10.1109/ESSCIRC59616.2023.10268782.